\documentclass[aip,pof,10pt]{revtex4-2}

\usepackage{graphicx,epstopdf}
\usepackage{amsmath,amssymb,amsfonts}

\usepackage[T1]{fontenc}
\usepackage[english]{babel}
\usepackage[table]{xcolor}
\usepackage{rotating}

\usepackage[colorlinks=true,linkcolor=blue,citecolor=blue,urlcolor=blue]{hyperref}
\usepackage{caption,subcaption}
\usepackage{placeins}
\usepackage{float}
\usepackage{enumerate}
\usepackage{setspace}
\usepackage{tikz}
\usepackage{mathptmx} 
\usetikzlibrary{shapes, arrows}

\tikzstyle{terminator} = [rectangle, draw, text centered, rounded corners, minimum height=2em]
\tikzstyle{connector} = [draw, -latex']
\begin{document}

\title{Solute dispersion in magnetically influenced multiphase flow through a porous tube: axial transport and microrotational effects}
\author{Sohel Ahmed}
\affiliation{%
Department of Mathematics, Cooch Behar Panchanan Barma University, Cooch Behar 736101, India}

\author{Nanda Poddar}
\homepage{Email: nandapoddarcr7@gmail.com, nandap@srmist.edu.in (Corresponding author)}
\affiliation{%
Department of Mathematics, SRM Institute of Science and Technology, Kattankulathur, Chennai 603203, India}
\affiliation{%
School of Mathematical and Statistical Sciences, University of Galway, Galway H91 TK33, Ireland}

\author{Jyotirmoy Rana}
\affiliation{%
Department of Mathematics, Indian Institute of Technology Hyderabad, Sangareddy 502284, India}

\author{Kajal Kumar Mondal}
\affiliation{%
Department of Mathematics, Cooch Behar Panchanan Barma University, Cooch Behar 736101, India}

\author{Niall Madden}
\affiliation{%
School of Mathematical and Statistical Sciences, University of Galway, Galway H91 TK33, Ireland}

\begin{abstract}
 This study presents a theoretical investigation of generalized solute dispersion in magnetohydrodynamic multiphase tube flow with porous layers. A two-fluid analytical model is developed for applications in biofluid and environmental fluid dynamics. The model comprises a micropolar (non-Newtonian) fluid core representing the rotational behaviour of red blood cells and a Newtonian plasma periphery embedded with Brinkman and Darcy porous structures, corresponding to the glycocalyx and endothelial layers with distinct permeability characteristics. A transverse magnetic field is incorporated to investigate how magnetic-field-induced modifications of the carrier flow influence solute localisation, with potential relevance to magnetic nanoparticle-mediated drug delivery. Using the generalised dispersion framework of Sankarasubramanian \& Gill \cite{sankara1973royal}, analytical solutions are derived to investigate how the coupled axial velocity field and associated microrotational dynamics influence solute transport. The analytical predictions are independently validated through Brownian dynamics simulations, demonstrating excellent agreement for the temporal evolution of the zeroth and first transport moments. The results reveal the previously unexplored influence of microrotational dynamics on solute concentration, convection coefficients and effective dispersion, providing new insights into the coupled roles of translational and rotational fluid motion in biofluid transport. This work bridges an important gap in the literature and establishes a generalized theoretical framework linking magnetic fields, micropolar fluids and porous arterial structures for biofluid transport, targeted drug delivery and clinical engineering applications.
 \end{abstract}

\maketitle

\section{Introduction}\label{sec:headings}

Recently, the study of solute dispersion in various fluid flow geometries has emerged as a critically important and highly promising research domain. Driven by remarkable advancements in experimental techniques, sophisticated analytical and computational modelling, and state-of-the-art imaging technologies, this field has significantly impacted diverse applications. Notably, precise modelling of solute dispersion is pivotal in enhancing drug mixing within physiological systems, optimising bacterial fermentation processes, improving molecular separation in chemical engineering, and accurately predicting contaminant transport within environmental flows. Thus, the analytical and theoretical exploration of mass transport mechanisms has gained substantial relevance across multiple interdisciplinary fields, including biofluid mechanics, environmental fluid dynamics, and broader engineering sciences.

Taylor \cite{taylor1953royal} initially developed a theoretical framework for the solute dispersion process. He then validated his theory through an experiment using a small-bore tube in a steady laminar viscous fluid flow. Taylor's research clarified the combined effect of axial convection and radial molecular diffusion in dispersing solute throughout the fluid. Later on, Aris \cite{aris1956royal} presented a new analytical technique for Taylor dispersion by removing certain limitations on parameters. These early investigations primarily focused on the dispersion mechanism after the solute had been in the fluid for an extended period of time. However, these studies provided a very limited understanding of the solute's behaviour immediately after it is injected. Gill \& Sankarasubramanian \cite{gill1970royal} suggested a methodology to overcome these constraints by developing a generalised method to study solute dispersion in a circular tube under laminar flow conditions. This approach provides a comprehensive understanding of the dispersion process, beginning from the moment the solute is injected. Sankarasubramanian \& Gill \cite{sankara1973royal} subsequently devised a novel model to investigate the dispersion resulting from certain concentration distributions. They then expanded their research to include solute dispersion with mass transfer between phases via the tube wall, using the generalised dispersion model. The generalised dispersion technique and the Taylor-Aris dispersion model have been extensively used by several researchers to investigate the mass transport processes in various flow geometries. Chatwin \cite{chatwin1975jfm} conducted a study on the longitudinal dispersion of a passive contaminant in a tube under an oscillatory flow, taking into account the existence of an effective molecular diffusion coefficient. Barton \cite{barton1983jfm} resolved many technical challenges while using the separation of variables method to solve Aris' moment equations. Bandyopadhyay \& Mazumder \cite{bandyo1999am} examined the impact of longitudinal dispersion in a channel with pulsatile flow using the Aris-Barton moment technique. By considering an irreversible wall absorption parameter, Nagarani \textit{et al.} \cite{nagarani2004abe} investigated the dispersion of a solute over a large period of time in a tube and channel. Wu \& Chen \cite{wu2014jfm} described the solute dispersion by taking the approach towards the uniformity of concentration distribution. Subsequently, Rana \& Murthy \cite{rana2016jfm} studied the solute dispersion in Casson fluid flow in a tube with wall absorption. In the presence of reversible and irreversible reactions, Das \textit{et al.} \cite{das2022royal} investigated the unsteady solute dispersion by taking a non-Newtonian fluid flowing in a tube with a thin wall. Azari \& Sadeghi \cite{azari2022jfm} theoretically studied the dispersion of a solute band in a semicircular microchannel with irreversible wall reaction by employing the generalised dispersion model. Later, Singh \& Murthy \cite{sing2023jfm} examined the impact of skewness and kurtosis on unsteady solute dispersion in C-Y fluid in a tube. Since the importance of the solute dispersion in different flow geometries has numerous applications in diversified fields, this area has become one of the best platforms for several researchers: Chang \& Santiago \cite{chang2023jfm}; Guan \& Chen \cite{guan2024jfm}; Peng \cite{peng2024jfm}; Aruna \textit{et al}\cite{aruna2024pof}; Poddar \& Dhar \cite{poddar2026arxiv}.

Magnetohydrodynamic (MHD) fluids, which are electrically conductive and magnetically responsive, play a crucial role in various modern technologies and natural phenomena. The control and manipulation of these fluids have numerous applications in fields such as environmental sciences, engineering, and healthcare \cite{annapurna1979royal, saha2024royal, das2024pof}. Among the major phenomena related to MHD fluids, the study of mass transport has significance because it determines the movement of particles, ions, or species throughout the fluid. Recently, Poddar \textit{et al.} \cite{poddar2022pof} investigated the solute dispersion in a magnetohydrodynamic flow through a porous medium under the effect of heterogeneous and bulk chemical reactions. Later, Das \textit{et al.} \cite{das2024pof} investigated solute transport in magnetohydrodynamic pulsatile electroosmotic flow within a microchannel. Due to the significant potential for innovative engineering and medical applications, researchers are highly interested in studying the behaviour of MHD fluids and mass transport phenomena in these streams. In the case of biological systems, this phenomenon is highly important because blood, a naturally conductive fluid, is affected by applied magnetic fields. Therefore, this concept enables the development of various medical applications, including magnetic drug targeting for precise medication delivery to diseased tissues, the creation of magnetic field-guided surgical tools, and innovative techniques for imaging and monitoring cardiovascular conditions. For our study, we considered a transverse magnetic field relative to the direction of blood flow. Naturally, blood tends to transport magnetic nanoparticles downstream, potentially leading to particle loss from the target location. A transverse magnetic field produces a force perpendicular to the flow, attracting magnetic nanoparticles to the blood vessel walls and facilitating their localisation at the target location, which enhances particle retention and precision of drug delivery. Initially, Gupta \& Chatterjee \cite{gupta1968mpcps} applied the transverse magnetic field relative to the flow direction to study the dispersion of a solute between two parallel plates. Later, Soundalgekar \& Gupta \cite{soundalgekar1975ijhmt} investigated how homogeneous and heterogeneous reactions affect solute dispersion in an MHD channel flow. Afterwards, in the presence of a uniform transverse magnetic field, by applying the generalised dispersion model, Annapurna \& Gupta \cite{annapurna1979royal} investigated the solute dispersion in an MHD fluid. Umavathi \textit{et al.} \cite{umavathi2016ijame} investigated the longitudinal dispersion of a solute with and without chemical reaction in an MHD flow between two parallel plates. By applying a magnetic field in the transverse direction of the flow Shah \textit{et al.} \cite{shah2020ichmt} studied the dispersion of a solute in a two-fluid model. Ndenda \textit{et al.} \cite{ndenda2021jap} described the dispersion of a drug carrier in a microvessel by introducing an external magnetic field. Recently, Saha \textit{et al.} \cite{saha2024royal} analysed the convection-diffusion equation under the influence of an induced magnetic field with bulk chemical reactions.

Different researchers employ various fluid models to investigate the solute dispersion process. However, in the case of blood rheology, the micropolar fluid model is one of the most promising models for studying solute transport phenomena. The concept of micropolar fluid was first introduced by Eringen \cite{eringen1966jmm}. Micropolar fluids consider the microstructure and individual rotating motion of fluid particles, in contrast to conventional Newtonian fluids, which are characterised by the Navier-Stokes equations and assume a continuous distribution of fluid particles. Additionally, apart from the axial fluid velocity, there is another velocity known as the micro-rotational velocity. This velocity refers to the rotational motion of the microparticles, which are moving in conjunction with the fluid. In the case of blood, due to the presence of RBCs in the core region, this model is more suitable. For this reason, numerous researchers have discussed and implemented the micropolar model in their work. Boodoo \textit{et al.} \cite{boodoo2013ra} discussed the flow in a porous, layered tube. In the presence of a magnetic field, a micropolar-Newtonian blood flow model was developed by Jaiswal \& Yadav \cite{jaiswal2019pof}. In a circular tube, Vidyanidhi \& Murty \cite{vidyanidhi1976ijes} investigated the dispersion of a solute in a micropolar fluid. Later, Roy \& Beg \cite{roy2021ichmt} discussed the unsteady solute dispersion in a two-fluid blood flow model by taking micropolar fluid at the core region and Newtonian fluid at the periphery region. Although numerous papers have been discussed about solute dispersion in the case of axial velocity of micropolar fluid, the effects of micro-rotational velocity on solute dispersion have received no attention. 

In the present work, the dispersion of a solute in a two-fluid blood flow model, specifically for axial velocity and micro-rotational velocity with an absorptive wall, has been studied. In various fluid systems, where rotational effects profoundly influence mixing, diffusion, and interaction at both micro and macro-scales, the novelty of this work focuses on micro-rotational dynamics and their interplay with axial flow. By exploring this extra velocity field, this study presents a more complete and realistic view of solute dispersion in micropolar fluids. For our study, the micropolar fluid (non-Newtonian) is considered in the core region, and a Newtonian fluid is considered in the peripheral region. Due to the presence of surface layers with resistance to flow, a thin region of the porous layer is considered a Brinkman porous region, and at the outer wall, we have modelled the Darcy porous region. Here, we considered different permeabilities for the two porous regions. Also in these regions, the fluid is taken as Newtonian. A uniform magnetic field was considered in the transverse direction of the flow. Sankarasubramanian \& Gill \cite{sankara1973royal}'s generalised dispersion model was utilised to discuss the transport mechanism of solutes. To independently verify the analytical predictions, the derived dispersion model is further validated using Brownian dynamics simulations. This investigation first examines how the governing flow parameters modify the coupled axial and microrotational velocity fields. Since the axial velocity is intrinsically coupled to the microrotational
motion through the micropolar momentum equations, the influence of microrotation on solute transport is inherently incorporated into the transport coefficients and concentration distributions obtained from
the axial velocity. In addition, a comparative analysis based on the microrotational velocity profile is presented to further characterise
the transport signature associated with the rotational microstructure. It is observed that the micro-rotational velocity of the microparticles plays a vital role in solute dispersion. The rotational effects of the microparticles significantly impact the mean concentration of solute for different values of the absorption parameter. This study, while developed for generalised tube flows with porous layered walls, can also be directly applied to biological tubular systems such as arteries, where fluid exhibits micropolar behaviour and layered porous structures influence transport and flow dynamics in a similar manner. Our work emphasised the biological case of blood flow, which ensures both generality in fluid mechanics modelling and direct relevance to biofluid mechanics applications.

The remainder of this paper is organised as follows. In
$\S \ref{Sec:Mathematical_Formulation}$, we present the mathematical
formulation, including the governing equations for the fluid flow and
solute transport together with the associated boundary and interface
conditions. The analytical solution for the solute concentration is
then developed in $\S \ref{Sec:Method_of_Solution}$ using the
generalised dispersion model of Sankarasubramanian \& Gill
\cite{sankara1973royal}. In $\S \ref{sec:BDS_validation}$, the
analytical framework is independently assessed through Brownian
dynamics simulations by comparing the analytical and stochastic
predictions for the zeroth and first axial moments. In
$\S \ref{Sec:Results_and_Discussion}$, we examine how magnetic
damping, micropolar effects, coupling between translational and
rotational motions, Darcy-layer resistance, and wall absorption modify
the axial and micro-rotational velocity fields, the convection and
dispersion coefficients, and the resulting mean and spatial
concentration distributions. Finally,
$\S \ref{Sec:Conclusion}$ summarises the principal findings and
concluding remarks. In addition,
\hyperref[Sec:Appendix_A]{Appendix A} provides the definitions of the
dimensionless parameters together with the analytical expressions for
the non-dimensional axial and micro-rotational velocity distributions.

\section{Mathematical formulation}\label{Sec:Mathematical_Formulation}
Consider a steady, laminar, axisymmetric, and fully developed fluid flow through a circular cylindrical tube. A cylindrical coordinate system $(r', \theta', z')$ is used, where $r'$ and $z'$ represent the radial and axial directions, respectively. Here we consider the tube with four concentric cylinders with radii $ {r}_{1}'$, $ {r}_{2}'$, $ {r}_{3}'$ and $ {r}_{4}'$, respectively. For the axial symmetry of the circular tube, we divide the regions as follows: $\boldsymbol{\Omega_1}=\big\{(r',z'):~0\leq r'\leq r'_1,~ -\infty<z'<\infty\big\}$, $\boldsymbol{\Omega_2}=\big\{(r',z'):~r_1'\leq r'\leq r'_2,~ -\infty<z'<\infty\big\}$, $\boldsymbol{\Omega_3}=\big\{(r',z'):~r_2'\leq r'\leq r'_3,~ -\infty<z'<\infty\big\}$ and $\boldsymbol{\Omega_4}=\big\{(r',z'):~r_3'\leq r'\leq r'_4,~ -\infty<z'<\infty\big\}$.

Due to the presence of a pressure gradient $ {\bf{\nabla}}p'$, the micropolar-Newtonian fluids are flowing along the $z'$-axis. The model is applicable to flow in a simple tube; however, this study focuses on the biological context of blood flow to establish direct relevance to biofluids. Therefore, blood is assumed as a micropolar fluid (non-Newtonian) which is flowing in the core region $\boldsymbol{\Omega_1}$. Also in this region, the elements of blood are continuously rotating and moving with the flow. In $\boldsymbol{\Omega_2}$, clear cell-free blood plasma is considered, which is treated as a Newtonian fluid. Due to the presence of a thin glycocalyx layer near the tube wall, $\boldsymbol{\Omega_3}$ is assumed as the Brinkman porous region and at the inner wall of the tube, i.e., at the endothelial layer, $\boldsymbol{\Omega_4}$ is considered as the Darcy porous region.
            \begin{figure}
			\centering
			\includegraphics[width=0.8\linewidth, height = 0.5\textwidth]{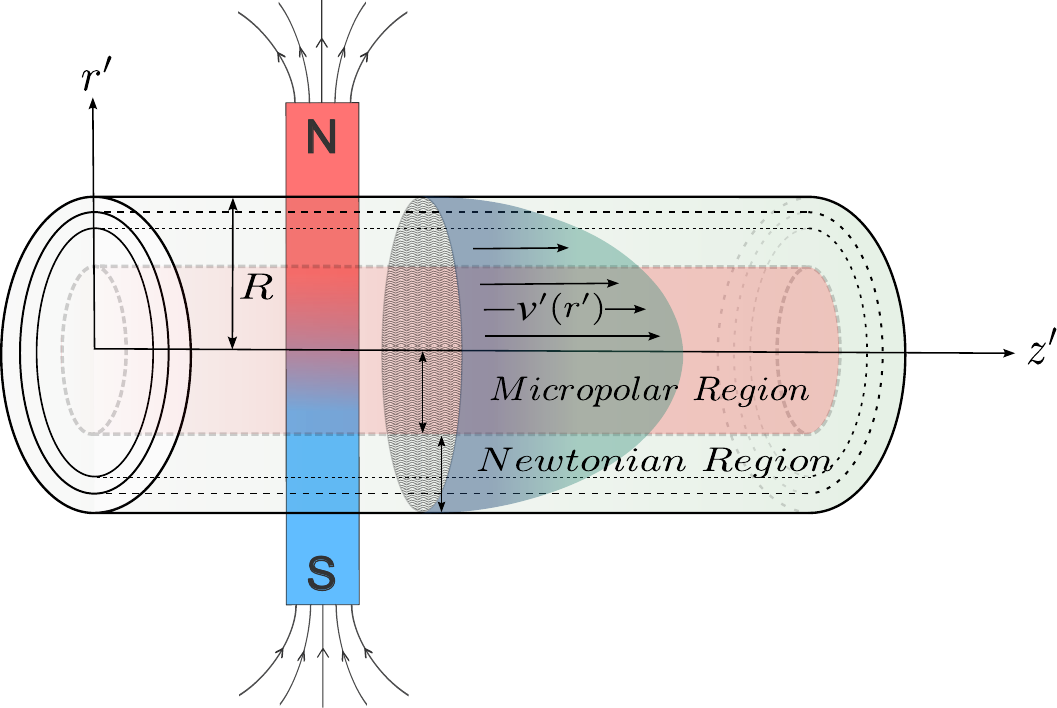}
			\caption{Schematic diagram of the present study.}
			\label{fig:1}
		\end{figure}
In the transverse direction to the flow, an external magnetic field $ \bf{B}$ of uniform strength $ {B}_{0}'$ is applied. Let $(0,0, {v}_i')$ be the velocity vector of the blood flow in $\boldsymbol{\Omega_1}$ to $\boldsymbol{\Omega_4}$, respectively (see figure \ref{fig:1}). In the core region, the velocity vector of the micro-rotational velocity of micro-particles is given by $(0, {w}',0)$.

\subsection{{Velocity distributions}}
The linear and angular momentum equations for the micropolar fluid are coupled, nonlinear differential equations, as given by Eringen \citep{eringen1966jmm}. It is not straightforward to obtain the velocity profiles for micropolar fluids analytically. For this reason, several researchers: Ariman \& Hill \cite{ariman1967pof}, Verma \& Sehgal \cite{verma1968ijes}, Ramkissoon \& Majumdar \cite{ramkissoon1976pof}, Jaiswal \& Yadav \cite{jaiswal2019pof} have applied various simplifying assumptions in order to solve the governing equations analytically. One of these assumptions is Stokes flow, which simplifies the flow equations; it is considered in this study. 

The equations of continuity, linear and angular momentum in select regions are as follows.

The continuity equation for axial velocity can be expressed as
\begin{align}
~~~~~~~&\frac{\partial  {v}_i'}{\partial  {z'}}=0 \mbox{ in } \boldsymbol{\Omega_{i}},~~ i\in \{1,2,3,4\}.\label{eq:2.1}
\end{align}

The linear and angular momentum equations are given by,

\begin{align}
~~~~~~&-\frac{\partial  {p'}}{\partial  {z'}}+ \frac{\mu_{vis}'}{ {r'}}\frac{d}{d {r'}}\left( {r'}\frac{d  {v}_i'}{d  {r'}}\right)+\frac{ {\kappa'}}{ {r'}}\frac{d\left( {r'} {w'}\right)}{d  {r'}} + \Lambda_i- {\sigma}_{ec}' {B}_0'^{2} {v}_i'=0 \mbox{ in } \boldsymbol{\Omega_{i}},~~~i\in \{1,2,3,4\}, \label{eq:2.2}\\
&\mbox{and }{\gamma'}\frac{d}{d  {r'}}\left(\frac{1}{ {r'}}\frac{d}{d  {r'}}\left( {r'} {w'}\right)\right)- {\kappa}'\left(\frac{d  {v}_1'}{d  {r'}}+2 {w'}\right)=0 \mbox{ in } \boldsymbol{\Omega_{1}},\label{eq:2.3}
\end{align}
respectively,\\ 
where ~~ $\mu_{vis}' = 
\begin{cases}
\mu_1' + \kappa', ~\text{in } ~~\boldsymbol{\Omega_1},\\
~~~~\mu', ~~~~~\text{in } ~~\boldsymbol{\Omega_2}, \\
~~~~\mu'_e, ~~~~\text{in } ~~\boldsymbol{\Omega_3},
\end{cases}$
~~~~~~~~~~~~$\Lambda_i = 
\begin{cases}
~~~~0, ~~~~~~~~\text{for } ~~i\in \{1,2\},\\
 -~\frac{v_i'\mu'}{k_{i-2}'}, ~~~~\text{for } ~~i\in \{3,4\},
\end{cases}$\\

~~~~~~~~~~~~~~~~~~~~~~~~~~~and~~ $\sigma_{ec}' = 
\begin{cases}
~\sigma'_1, ~~~\text{in } ~~\boldsymbol{\Omega_1},\\
 ~\sigma', ~~~~\text{in } ~~\boldsymbol{\Omega_2}, \boldsymbol{\Omega_3}~~\text{and} ~~\boldsymbol{\Omega_4}.
\end{cases}$

In this study, $ {\mu}_1'$ is the coefficient of dynamic viscosity and $ {\kappa}'$ is the coefficient of micro-rotation viscosity, 
which only appears in $\boldsymbol{\Omega_1}$ and for other regions, it vanishes. Here $ {\gamma}'$ is the material constant and $ {\sigma}_1'$ is the electrical conductivity of the micropolar fluid; $ {\mu}'$ and $ {\sigma}'$ are viscosity and electrical conductivity of Newtonian fluid, respectively; $ {k}_1'$ is the permeability of the Brinkman porous region and $ {\mu}_e'$ is the effective viscosity; ${k}_2'$ is the permeability of the Darcy porous region. Also, we take $ {k}_1'\ne {k}_2'$ and in case of $\boldsymbol{\Omega_4}$, i.e., for the Darcy porous region, we neglect the second term of L.H.S. of equation \eqref{eq:2.2}.

\begin{subequations}\label{bc_velocity_dimensional}
By assuming the continuous fluid velocities at the interface of select regions, we have the following conditions:
\begin{equation}\label{eq:2.4a}
    v'_i(r'_i) = v'_{i+1}(r'_i) \text{  for  } i\in \{1,2,3\}.
\end{equation}
Also, at the interface of $\boldsymbol{\Omega_1}$ and $\boldsymbol{\Omega_2}$, the continuity of shear stresses leads to the condition
\begin{equation}\label{eq:2.4b}
    (\mu'_1+\kappa')\frac{dv'_1}{dr'_1}+\kappa'w' = \mu'\frac{dv'_2}{dr'} ~~\text{at}~~r'=r'_1.
\end{equation}
The axisymmetry assumption leads to the following condition
\begin{equation}\label{eq:2.4c}
\frac{dv'_1}{dr'}=0 ~~\text{and}~~ w'=0 ~~\text{at}~~ r' = 0.
\end{equation}
In this study, we have considered the micropolar fluid flow in the $\boldsymbol{\Omega_1}$ region. Therefore, the microparticles are only flowing in this zone of radius $r'_1$. Hence, outside of this region, the micro-rotational velocity of the microparticles vanishes. Thus, we can write
\begin{equation}\label{eq:2.4d}
    w' = 0 ~~\text{at}~~ r' = r'_1.
\end{equation}
At the interface of $\boldsymbol{\Omega_2}$ and $\boldsymbol{\Omega_3}$, i.e., between the Newtonian fluid and Brinkman porous regions, owing to the different viscosities of fluid, the shear stresses exerted by the fluids must be different; this leads to the following stress-jump condition Ochoa-Tapia \& Whitaker \cite{ochoa1995ijhmt}
\begin{equation}\label{eq:2.4e}
    \mu'_e \frac{dv'_3}{dr'} - \mu' \frac{dv'_2}{dr'} = \frac{\xi}{\sqrt{k'_1}}\mu' v'_3,
\end{equation}
where $\xi$ is the stress-jump coefficient.
\end{subequations}
The detailed derivation of the velocity field and the corresponding velocity profiles are provided in \hyperref[Sec:Appendix_A]{Appendix A}.

\subsection{{Concentration distribution}}
The present study examines solute transport associated with both the axial translational velocity and the microrotational velocity fields obtained from the micropolar flow model. The axial velocity governs the physical advection of the solute and therefore enters directly into the convection--diffusion equation. To facilitate a direct comparison between translational and rotational transport mechanisms, the same generalised dispersion framework is also applied separately using the microrotational velocity profile. The corresponding results should therefore be interpreted as a comparative theoretical measure of the transport signature associated with the rotational microstructure rather than as an independent physical axial advection.

The following time-dependent convection-diffusion equation controls the concentration of solute:
\begin{equation}\label{eq:2.5}
\frac{\partial {C'}}{\partial {t'}}+{v'}({r'})\frac{\partial {C'}}{\partial {z'}} = D\left[\frac{1}{{r'}}\frac{\partial}{\partial {r'}}\left({r'}\frac{\partial {C'}}{\partial {r'}}\right)+\frac{\partial^2 {C'}}{\partial {z'}^2}\right],
\end{equation}
\begin{subequations}\label{ic_bc_concentration_dimensional}
where, ${C'}$ is the local concentration of the solute, ${t'}$ denotes the time and $D$ is the molecular diffusivity. Following Sankarasubramanian \& Gill \cite{sankara1973royal}, at the initial stage, i.e. at ${t'} = 0$, a radially uniform distribution of solute within a circular cross-section of radius $R$ is considered, where the solute is concentrated at $z'=0$. Therefore, the concentration distribution at the initial time can be interpreted as
\begin{equation}\label{eq:2.6a}
{C'}(0,{z'},{r'}) = C_0\,R\,\delta({{z'}}),
\end{equation}
where $C_0 = \frac{M_0}{\pi R^3}$ is the reference concentration; $M_0$ is the mass of the solute, and $\delta({{z'}})$ is the Dirac-delta function.

Following Sankarasubramanian \& Gill \cite{sankara1973royal}, we have also taken the four boundary conditions. Due to the symmetry of the tube about its axis ${z'}$, the change of concentration along the radial direction is zero, i.e.,
\begin{equation}\label{eq:2.6b}
\frac{\partial {C'}}{\partial {r'}}({t'},{z'},0) = 0.
\end{equation}
 Since there is a first-order heterogeneous irreversible reaction at the tube wall, we have the following condition
\begin{equation}\label{eq:2.6c}
-D \frac{\partial {C'}}{\partial {r'}}({t'},{z'},R) = k \,{C'}({t'},{z'},R),
\end{equation}
where $k$ is taken as the reaction rate constant. For finite quantity of solute, we have
\begin{equation}\label{eq:2.6d}
{C'}({t'},{z'},0) = \text{finite}.
\end{equation}
In addition, the concentration of the solute vanishes at an infinitely large distance owing to its finite amount in the fluid, leading to the conditions given below
\begin{equation}\label{eq:2.6e}
{C'}({t'},\pm\infty,{r'}) = 0.
\end{equation}
\begin{equation}\label{eq:2.6f}
\frac{\partial {C'}}{\partial {z'}}({t'},\pm\infty,{r'}) = 0.
\end{equation}
\end{subequations}

To obtain the non-dimensional form of the concentration equation, the following scaled variables are adopted in this study. 
\begin{equation}\label{eq:2.7}
    C =\frac{{C'}}{C_0},\quad {v} = \frac{{v'}}{v_0},\quad {w} = \frac{{w'}R}{v_0},\quad r = \frac{{r'}}{R},\quad z = \frac{D {z'}}{R^2 v_0},\quad t = \frac{D {t'}}{R^2},
\end{equation}
where $v_0$ is the characteristic velocity. Now, using the above-mentioned scaled variables, the non-dimensional form of the convection-diffusion equation is given by
\begin{equation}\label{eq:2.8}
\frac{\partial C}{\partial t}+v(r)\frac{\partial C}{\partial z} = \frac{1}{r}\frac{\partial }{\partial r}\left(r\frac{\partial C}{\partial r}\right)+\frac{1}{(Pe)^2}\frac{\partial^2 C}{\partial z^2}, ~~~-\infty<z<\infty,~~~0<r<1,
\end{equation}
with the following initial and boundary conditions
\begin{subequations}\label{ic_bc_concentration_non_dimensional}
\begin{equation}\label{eq:2.9a}
C(0,z,r) = \frac{\delta(z)}{Pe},
\end{equation}
\begin{equation}\label{eq:2.9b}
\frac{\partial C}{\partial r}(t,z,0) = 0,
\end{equation}
\begin{equation}\label{eq:2.9c}
\frac{\partial C}{\partial r}(t,z,1) = -\beta\, C(t,z,1),
\end{equation}
\begin{equation}\label{eq:2.9d}
C(t,z,0) = \text{finite},
\end{equation}
\begin{equation}\label{eq:2.9e}
C(t,\pm\infty,r) = 0,
\end{equation}
\begin{equation}\label{eq:2.9f}
\mbox{and }\,\frac{\partial C}{\partial z}(t,\pm\infty,r) = 0,
\end{equation}
where $Pe = {\left(R\, v_0\right)}/{D}$ is the Peclet number and $\beta = {\left(k\, R\right)}/{D}$ is the absorption parameter. 
\end{subequations}

\section{Method of solution for transport coefficients and solute concentration}\label{Sec:Method_of_Solution}
The present study mainly focuses on analysing the transport coefficients as well as the concentration distribution of the solute. 
In this study, we aim to derive the explicit expressions for three transport coefficients, namely, the exchange, convection, and dispersion coefficients, as well as the concentration of the solute. To accomplish this, we follow the analytical procedure proposed by Sankarasubramanian \& Gill \cite{sankara1973royal} and extended by Rana \& Murthy \cite{rana2016jfm}. After plugging the obtained velocity profile into the unsteady convection-diffusion equation \eqref{eq:2.8}, it is solved with the help of initial and boundary conditions \eqref{ic_bc_concentration_non_dimensional}. We formulate the solution of the above equations as a series expansion of the mean concentration. Therefore, the expression of concentration $C(t,z,r)$ is expanded as
\begin{equation}\label{eq:3.1}
C(t,z,r) = \sum_{i=0}^{\infty}f_i(t,r)\frac{\partial^i C_m(t,z)}{\partial z^i},
\end{equation}
where, $f_i(t,r)$ are the unknown functions, for $i = 0,1,2,\ldots~.$ Also, the mean concentration $C_m(t,z)$ in dimensionless form is given by
\begin{align}\label{eq:3.2}
 C_m(t,z) = 2\int_{0}^{1}C(t,z,r)\, r\, dr.
\end{align}
Multiplying both side of equation \eqref{eq:2.8} by $2r$ and then integrating with respect to $r$ from $0$ to $1$, we have
\begin{equation}\label{eq:3.3}
\frac{\partial C_m}{\partial t} = \frac{1}{(Pe)^2}\frac{\partial^2C_m}{\partial z^2}+2\,\frac{\partial C}{\partial r}(t,z,1) - 2\,\frac{\partial}{\partial z}\int_{0}^{1}v(r)\,C\,r\,dr.
\end{equation}
Now, using equation \eqref{eq:3.1} in equation \eqref{eq:3.3}, we have the following equation in terms of $C_{m}(t,z)$ and $f_{i}(t,r)$ as
\begin{align}\label{eq:3.4}
&\frac{\partial C_m}{\partial t} = 2C_m\frac{\partial f_0}{\partial r}(t,1)+ \frac{\partial C_m}{\partial z}\left[2\frac{\partial f_1}{\partial r}(t,1)-2\int_{0}^{1}rv(r)f_0(t,r)dr\right]&\nonumber\\ 
&\hspace{3cm}+ \frac{\partial^2 C_m}{\partial z^2}\left[2\frac{\partial f_2}{\partial r}(t,1)-2\int_{0}^{1}rv(r)f_1(t,r)dr +\frac{1}{(Pe)^2}\right]+\ldots~.&
\end{align}
By taking the coefficients of {\large{$\frac{\partial^i C_m}{\partial z^i}$}} $(i=0,1,2\ldots)$ in R.H.S. of equation \eqref{eq:3.4}, the transport coefficients $K_i(t)$ are obtained by following expressions
\begin{equation}\label{eq:3.5}
K_i(t) = \frac{\delta_{i2}}{(Pe)^2}+2\frac{\partial f_i}{\partial r}(t,1)-2\int_{0}^{1}r\,v(r)\,f_{i-1}(t,r)\,dr,\,\,i = 0,1,2,\ldots,
\end{equation}
where $f_{-1} \equiv0$ and 
$\delta_{ij} = 
\begin{cases}
1, \,\,\text{if }\, i = j,\\
0, \,\,\text{if }\, i \ne j.
\end{cases}$

Here, $K_0(t)$ is known as the exchange coefficient. The solute flux near the tube wall is responsible for the existence of this coefficient. Because of the irreversible reaction at the tube wall, the solute in the system will gradually be depleted over time, as indicated by the negative value of this coefficient.
Furthermore, $K_1(t)$ is the convection coefficient, which arises due to the velocity of the fluid. It mainly tells us how the solute moves with the fluid. $K_2(t)$ is the dispersion coefficient, which results from both the molecular diffusion and the velocity gradient of the fluid over the cross section of the tube. The primary purpose of obtaining it is to explain the solute's dispersion within the fluid.

Therefore, using \eqref{eq:3.5}, equation \eqref{eq:3.4} becomes,
\begin{equation}\label{eq:3.6}
\frac{\partial C_m}{\partial t} = K_0(t)C_m + K_1(t)\frac{\partial C_m}{\partial z}+K_2(t)\frac{\partial^2 C_m}{\partial z^2}+\ldots\,.
\end{equation}
To obtain the solution for mean concentration $C_m$, we need to solve equations \eqref{eq:3.5} and \eqref{eq:3.6} and also, we need to find the unknown functions $f_k(k = 0,1,2)$ with respect to appropriate initial and boundary conditions. For this, we substitute equation \eqref{eq:3.1} into \eqref{eq:2.8} for $(i = 0,1,2,\ldots)$. Now, equating the coefficients of {\Large$\frac{\partial^n C_m}{\partial z^n}$}$(n = 0,1,2,\ldots)$, we have
\begin{equation}\label{eq:3.7}
\frac{\partial f_n}{\partial t} = \frac{1}{r}\frac{\partial}{\partial r}\left(r\frac{\partial f_n}{\partial r}\right) - f_{n-1}v(r)-\sum_{i=0}^{n}f_{n-i}K_i(t) + \frac{f_{n-2}}{(Pe)^2},
\end{equation}
where, $n=0,1,2,\ldots$ and $f_{-1}=f_{-2}=0$.
Using the initial condition, $C_m$ and $f_n$ are expressed as,

\begin{subequations}
\begin{equation}\label{eq:3.8a}
C_m(0,z) = \frac{\delta(z)}{Pe},
\end{equation}
\begin{equation}\label{eq:3.8b}
f_0(0,r) = 1,
\end{equation}
\begin{equation}\label{eq:3.8c}
f_n(0,r) = 0,~~~~n=1, 2, \dots,
\end{equation}
and using the boundary conditions for $f_n$ and $C_m$ from equations \eqref{eq:2.9b} to \eqref{eq:2.9f}, we have,
\begin{equation}\label{eq:3.8d}
\frac{\partial f_n}{\partial r}(t,0) = 0,
\end{equation}
\begin{equation}\label{eq:3.8e}
\frac{\partial f_n}{\partial r}(t,1) = -\beta f_n(t,1),
\end{equation}
\begin{equation}\label{eq:3.8f}
\text{and}~~~f_n(t,0) = \mbox{finite}.
\end{equation}
\begin{equation}\label{eq:3.8g}
\mbox{Also,}~~~C_m(t,\infty) = 0,
\end{equation}
\begin{equation}\label{eq:3.8h}
\frac{\partial C_m}{\partial z}(t,\infty) = 0.
\end{equation}
Now, using equation \eqref{eq:3.2} into \eqref{eq:3.1}, we have one more additional condition as
\begin{equation}\label{eq:3.8i}
\int_{0}^{1}f_n(t,r)rdr =\frac{1}{2} \delta_{n0},~~~~n=0,1,2.
\end{equation}
\end{subequations}
The transport coefficients $K_{n}(t)$ and the unknown functions $f_{n}(t,r)$ are obtained by solving the coupled equations \eqref{eq:3.5} and \eqref{eq:3.7}. We have adopted the eigenfunction expansion method to evaluate these coefficients $K_{n}(t)$ and functions $f_{n}(t,r)$.

\subsection{{Solution for $f_0(t,r)$ and $K_0(t)$}}
From \eqref{eq:3.7} for $n=0$, we have the following differential equation
\begin{equation}\label{eq:3.9}
\frac{\partial f_0}{\partial t} = \frac{1}{r}\frac{\partial}{\partial r}\left(r\frac{\partial f_0}{\partial r}\right)-f_0K_0.
\end{equation}

The solution for $f_0(t,r)$ is determined by solving equation \eqref{eq:3.9} along with the conditions \eqref{eq:3.8b}, \eqref{eq:3.8d}, \eqref{eq:3.8e}, \eqref{eq:3.8f} and \eqref{eq:3.8i}. The final solution for $f_0(t,r)$ is derived to
\begin{equation}\label{eq:3.10}
f_0(t,r) = \frac{\sum\limits_{n=0}^{\infty}A_nJ_0(\mu_n r) e^{-\mu_n^2t}}{2\sum\limits_{n = 0}^{\infty}\left(\frac{A_n}{\mu_n}\right)J_1(\mu_n) e^{-\mu_n^2t}},
\end{equation}
where $J_0$ and $J_1$ are the Bessel functions of the first kind of order zero and one, respectively. Also, $\mu_n$ are the roots of the transcendental equation
\begin{equation}\label{eq:3.11}
\mu_n J_1(\mu_n) = \beta J_0(\mu_n).
\end{equation}
From this transcendental equation, we can find the roots $\mu_n$ by any suitable numerical method. The term $A_n$ in equation \eqref{eq:3.10}, which is dependent on the initial solute distribution, can be expressed as
\begin{equation}\label{eq:3.12}
A_n = \frac{2\mu_nJ_1(\mu_n)}{(\mu_n^2+\beta^2)[J_0(\mu_n)]^2}.
\end{equation}
Now from equation \eqref{eq:3.5} for $i=0$, we have the expression for exchange coefficient as
\begin{equation}\label{eq:3.13}
K_0(t) = -\frac{ \sum\limits_{n = 0}^{\infty}A_n \mu_n J_1(\mu_n) e^{-\mu_n^2t} }{ \sum\limits_{n = 0}^{\infty}\left(\frac{A_n}{\mu_n}\right)J_1(\mu_n) e^{-\mu_n^2t} }.
\end{equation}
It is clearly seen that the velocity is not present in the expression of the exchange coefficient, resulting in it being independent of the fluid velocity (Rana \& Murthy \cite{rana2016jfm}). However, $K_0(t)$ depends on the initial distribution of the solute via the term $A_n$ given in equation \eqref{eq:3.12} and the rate of the wall absorption via the transcendental equation \eqref{eq:3.11}.

\subsection{{Solution for $f_1(t,r)$ and $K_1(t)$}}
From \eqref{eq:3.7} for $n=1$, we have the differential equation for $f_1(t,r)$ as
\begin{equation}\label{eq:3.14}
\frac{\partial f_1}{\partial t} = \frac{1}{r}\frac{\partial}{\partial r}\left(r\frac{\partial f_1}{\partial r}\right)-f_1K_0-f_0\,\Big(K_1+v(r)\Big).
\end{equation}

Using the eigenfunction expansion method, equation \eqref{eq:3.14} is solved along with the conditions \eqref{eq:3.8c}-\eqref{eq:3.8f} and \eqref{eq:3.8i}. The final expression of $f_1(t,r)$ is obtained as 
\begin{equation}\label{eq:3.15}
f_1(t,r) = -\frac{\sum\limits_{n=0}^{\infty}\left[B_{1n} T_1(t) + \sum\limits_{m=0}^{\infty}B_{1mn} I_{mn} \int_{0}^{t}e^{-s(\mu_m^2-\mu_n^2)} ds\right]J_0(\mu_n r)e^{-\mu_n^2t}}{2\sum\limits_{n = 0}^{\infty}\left(\frac{A_n}{\mu_n}\right)J_1(\mu_n) e^{-\mu_n^2t}}.
\end{equation}
Therefore, we have,
\begin{equation}\label{eq:3.16}
    \frac{\partial f_1}{\partial r}(t,1) = \frac{\sum\limits_{n=0}^{\infty}\left[B_{1n} T_1(t)+\sum\limits_{m=0}^{\infty}B_{1mn} I_{mn} \int_{0}^{t}e^{-s(\mu_m^2-\mu_n^2)} ds\right]\mu_n J_1(\mu_n)e^{-\mu_n^2t}}{2\sum\limits_{n = 0}^{\infty}\left(\frac{A_n}{\mu_n}\right)J_1(\mu_n) e^{-\mu_n^2t}}.
\end{equation}
Using \eqref{eq:3.16}, the expression for $K_1(t)$ from \eqref{eq:3.5} for $i=1$ becomes,
\begin{align}\label{eq:3.17}
& K_1(t) = \frac{\sum\limits_{n=0}^{\infty}\left[B_{1n} T_1(t) + \sum\limits_{m=0}^{\infty}B_{1mn} I_{mn} \int_{0}^{t}e^{-s(\mu_m^2-\mu_n^2)} ds\right]\mu_n J_1(\mu_n)e^{-\mu_n^2t}}{\sum\limits_{n = 0}^{\infty}\left(\frac{A_n}{\mu_n}\right)J_1(\mu_n) e^{-\mu_n^2t}}\nonumber\\
&\hspace{5cm}-2\int_{0}^{1}r\left[\frac{\sum\limits_{n = 0}^{\infty}A_nJ_0(\mu_n r) e^{-\mu_n^2t}}{2\sum\limits_{n = 0}^{\infty}\left(\frac{A_n}{\mu_n}\right)J_1(\mu_n) e^{-\mu_n^2t}}\right] v(r)dr,
\end{align}
where the expression for $T_1(t)$ is given by
\begin{subequations}  
\begin{align}\label{eq:3.18}
    &T_1(t) =- \frac{\sum\limits_{n=0}^{\infty}\sum\limits_{m=0}^{\infty}B_{1mn}\left(\frac{J_1(\mu_n)}{\mu_n}\right)e^{-\mu_n^2t}I_{mn} \int_{0}^{t}e^{-s(\mu_m^2-\mu_n^2)} ds}{\sum\limits_{n=0}^{\infty}B_{1n}\left(\frac{J_1(\mu_n)}{\mu_n}\right)e^{-\mu_n^2t}},\\
    &A_n' = \frac{\sqrt{2}}{\sqrt{[J_0(\mu_n)]^2+[J_1(\mu_n)]^2}},\,\,\,B_{1n} = A_n A_n'^2 \left(\frac{[J_0(\mu_n)]^2+[J_1(\mu_n)]^2}{2}\right),\\
  &B_{1mn} = A_m A_n'^2, \,\,\,\,\,\,I_{mn} = \int_{0}^{1}r v(r)J_0(\mu_m r)J_0(\mu_n r) dr.\label{integral_I_mn}
  \end{align}
\end{subequations}
We can clearly see that $K_1(t)$ depends on the fluid velocity via the integral $I_{mn}$ given in equation \eqref{integral_I_mn}. This coefficient mainly describes the convection of solute flowing in the fluid. In addition, it depends on the initial distribution of the solute as well as the rate of wall absorption via the terms as discussed earlier for $K_0(t)$. 
\subsection{{Solution for $f_2(t,r)$ and $K_2(t)$}}
From \eqref{eq:3.7}, we have the following equation for $f_2(t,r)$ as
\begin{equation}\label{eq:3.19}
\frac{\partial f_2}{\partial t} = \frac{1}{r}\frac{\partial}{\partial r}\left(r\frac{\partial f_2}{\partial r}\right)-f_2\,K_0+f_0\left(\frac{1}{(Pe)^2}-K_2\right)-f_1\Big(v(r)+K_1\Big).
\end{equation}

By applying the eigenfunction expansion method, the differential equation \eqref{eq:3.19} is solved along with the conditions \eqref{eq:3.8c}-\eqref{eq:3.8f} and \eqref{eq:3.8i}. We have the final solution for $f_2(t,r)$ as 
\begin{align}\label{eq:3.20}
&f_2(t,r)\nonumber\\
&=\frac{\sum\limits_{n=0}^{\infty}\left[\sum\limits_{m=0}^{\infty}B_{2mn}\left\{F_{1mn}(t)+\sum\limits_{l=0}^{\infty}F_{2mnl}(t)\right\}-B_{5n}T_2(t)+\sum\limits_{m=0}^{\infty}B_{2n}F_{3mn}(t)\right]{J_0(\mu_nr)}e^{-\mu_n^2t}}{2\sum\limits_{n = 0}^{\infty}\left(\frac{A_n}{\mu_n}\right)J_1(\mu_n) e^{-\mu_n^2t}}.
\end{align}
Now, we have
\begin{align}\label{eq:3.21}
&\frac{\partial f_2}{\partial r}(t,1)\nonumber\\&=\frac{\sum\limits_{n=0}^{\infty}\left[\sum\limits_{m=0}^{\infty}B_{2mn}\left\{F_{1mn}(t)+\sum\limits_{l=0}^{\infty}F_{2mnl}(t)\right\}-B_{5n}T_2(t)+\sum\limits_{m=0}^{\infty}B_{2n}F_{3mn}(t)\right]\mu_n{J_1(\mu_n)}e^{-\mu_n^2t}}{2\sum\limits_{n = 0}^{\infty}\left(\frac{A_n}{\mu_n}\right)J_1(\mu_n) e^{-\mu_n^2t}}.
\end{align}
Using \eqref{eq:3.21}, the expression for $K_2(t)$ from \eqref{eq:3.5} for $i=2$  becomes,
\begin{multline}\label{eq:3.22}
K_2(t) = \frac{1}{(Pe)^2} +\\
\hspace{0.4cm}
2\left[\frac{\sum\limits_{n=0}^{\infty}\left[\sum\limits_{m=0}^{\infty}B_{2mn}\left\{F_{1mn}(t)+\sum\limits_{l=0}^{\infty}F_{2mnl}(t)\right\}-B_{5n}T_4(t)+\sum\limits_{m=0}^{\infty}B_{2n}F_{3mn}(t)\right]\mu_n{J_1(\mu_n)}e^{-\mu_n^2t}}{2\sum\limits_{n = 0}^{\infty}\left(\frac{A_n}{\mu_n}\right)J_1(\mu_n) e^{-\mu_n^2t}}\right] \\
\hspace{0.4cm}
+ 2\int_{0}^{1}rv(r)\left[\frac{\sum\limits_{n=0}^{\infty}\left[B_{1n} T_1(t) + \sum\limits_{m=0}^{\infty}B_{1mn}I_{mn} \int_{0}^{t}e^{-s(\mu_m^2-\mu_n^2)} ds\right]J_0(\mu_n r)e^{-\mu_n^2t}}{2\sum\limits_{n = 0}^{\infty}\left(\frac{A_n}{\mu_n}\right)J_1(\mu_n) e^{-\mu_n^2t}}\right]dr,
\end{multline}
where the expression for $T_2(t)$ is given by
\begin{subequations}
\begin{equation}\label{eq:3.23a}
T_2(t) =\frac{\sum\limits_{n=0}^{\infty}\left[\sum\limits_{m=0}^{\infty}B_{2mn}F_{1mn}(t)+\sum\limits_{m=0}^{\infty}\sum\limits_{l=0}^{\infty}B_{2mn}F_{2mnl}(t)+\sum\limits_{m=0}^{\infty}B_{2n}F_{3mn}(t)\right]\left(\frac{J_1(\mu_n)}{\mu_n}\right)e^{-\mu_n^2t}}{\sum\limits_{n=0}^{\infty}B_{3n}\left(\frac{J_1(\mu_n)}{\mu_n}\right)e^{-\mu_n^2t}},
\end{equation}
with the following integrals and coefficients
\begin{equation}\label{eq:3.23b}
    F_{1mn}(t) = A_m A_m'\frac{[J_0(\mu_m)]^2+[J_1(\mu_m)]^2}{2}I_{mn}\int_{0}^{t}e^{-s(\mu_m^2 -\mu_n^2)}\left(\int_{0}^{s}K_1(s') ds'\right)ds,
\end{equation}
\begin{equation}\label{eq:3.23c}
    F_{2mnl}(t) = A_l A_m' I_{mn}I_{lm}\int_{0}^{t}\left[e^{-s(\mu_m^2-\mu_n^2)}\left(\int_{0}^{s}e^{- s'(\mu_l^2-\mu_m^2)} ds'\right)\right]ds,
\end{equation}
\begin{equation}\label{eq:3.23d}
    F_{3mn}(t) = \int_{0}^{t}\left[\int_{0}^{s}\left(G_{mn}(s')+A_n A_n'\frac{[J_0(\mu_n)]^2+[J_1(\mu_n)]^2}{2}K_1(s')\right)ds'\right]K_1(s)ds,
\end{equation}
\begin{equation}\label{eq:3.23e}
    G_{mn}(s') = A_m A_n'e^{-s'(\mu_m^2-\mu_n^2)}I_{mn},
\end{equation}
\begin{align}\label{eq:3.23f}
B_{2mn} = A_m'A_n''^2,\,\,\,\,\,\,B_{2n} = A_n' A_n''^2\left(\frac{[J_0(\mu_n)]^2+[J_1(\mu_n)]^2}{2}\right),\\B_{3n} = A_n A_n''^2\left(\frac{[J_0(\mu_n)]^2+[J_1(\mu_n)]^2}{2}\right) ~\mbox{ and }~ A''^2_{n} = \frac{2\mu_n^2}{(\mu_n^2+\beta^2)J_0^2(\mu_n)}.
\end{align}
\end{subequations}
From equation \eqref{eq:3.22}, $K_2(t)$ is the effective diffusivity or the dispersion coefficient, which is associated with both the fluid velocity as well as the molecular diffusivity. The first term $1/Pe^2$ in the given equation \eqref{eq:3.22} indicates the diffusive component of $K_2(t)$; the second term in the given equation \eqref{eq:3.22} denotes the convective component of $K_2(t)$, and it is shear-enhanced diffusivity, occurring owing to the shearing of the fluid motion. Note that this coefficient also depends on the initial distribution of the solute as well as the rate of wall absorption via the terms as discussed earlier for $K_0(t)$. 

To get the high-order transport coefficients $K_{n}(t)\,(n\geq3)$, a similar methodology can be applied to the differential equation \eqref{eq:3.7} for $n\geq3$ along with the conditions \eqref{eq:3.8c}-\eqref{eq:3.8f} and \eqref{eq:3.8i}, and $K_{n}(t)$ can be determined from equation \eqref{eq:3.5} for $i\geq3$.

\subsection{{Solution for mean concentration $C_m(t,z)$ and local concentration of solute $C(t,r,z)$}}
After determining the transport coefficients $K_{n}(t)$ at the transient time, we are focused on finding out the mean and local concentration of the solute. To accomplish this, we solve equation \eqref{eq:3.6} in the presence of absorption at the tube wall along with initial and boundary conditions \eqref{eq:3.8a}, \eqref{eq:3.8g} and \eqref{eq:3.8h}. Note that $K_{n}(t)\,(n\geq3)$ are negligibly small in magnitude at large times and their contributions in finding solute concentration are neglected in this study. Consequently, we restrict the present study to the second-order dispersion model for finding the mean concentration of the solute $C_m (t,z)$, which results in the following solution for $C_m$
\begin{equation}\label{eq:3.24}
C_m(t,z) = \frac{1}{2Pe\sqrt{\pi T}}\exp\left[\zeta-\frac{z_1^2}{4T}\right],
\end{equation}
~~~~~~~~~~~~~~~~~~~~~~~~~~~~~~~~\text{where}
\begin{equation}\label{eq:3.25}
\zeta(t) = \int_{0}^{t}K_0(\eta)d\eta,
\end{equation}
\begin{equation}\label{eq:3.26}
z_1(t,z) = z + \int_{0}^{t}K_1(\eta)d\eta,
\end{equation}
~~~~~~~~~~~~~~~~~~~~~~~~~~~~~~~~\text{and}
\begin{equation}\label{eq:3.27}
T(t) = \int_{0}^{t}K_2(\eta)d\eta.
\end{equation}

Accordingly, the local concentration $C(t,z,r)$ of solute can be written utilizing \eqref{eq:3.1} as
\begin{equation}\label{eq:3.28}
    C(t,z,r) = \frac{1}{2Pe\sqrt{\pi T}} \left[f_0 - \frac{z_1}{2T}f_1 + \left(\frac{z_1^2}{4T^2} - \frac{1}{2T}\right)f_2\right]\exp\left[\zeta-\frac{z_1^2}{4T}\right].
\end{equation}

\section{Validation of the analytical results against Brownian dynamics simulations}
\label{sec:BDS_validation}

To validate the analytical solution, an independent Brownian dynamics
simulation (BDS) (Jiang \textit{et al.} \cite{jiang2022jfm}) is constructed such that the probability density of the
stochastic particle ensemble satisfies the same governing transport
equation. The deterministic concentration field
\(C(t,z,r)\) is governed by equation~\eqref{eq:2.8}. The corresponding
stochastic formulation is obtained by matching the drift and diffusion
coefficients in equation~\eqref{eq:2.8} with those of the associated
Fokker--Planck equation.

The axial and radial particle positions therefore satisfy the It\^o
stochastic differential equations
\begin{align}
\mathrm{d}z
&=
v(r)\,\mathrm{d}t
+
\frac{\sqrt{2}}{Pe}\,\mathrm{d}W_{z},
\label{eq:bds_sde_z}
\\
\mathrm{d}r
&=
\frac{1}{r}\,\mathrm{d}t
+
\sqrt{2}\,\mathrm{d}W_{r},
\label{eq:bds_sde_r}
\end{align}
where \(W_z\) and \(W_r\) are independent standard Wiener processes.
The drift term \(v(r)\) represents advection by the prescribed axial
velocity profile, whereas the diffusion coefficient
\(\sqrt{2}/Pe\) produces the required axial diffusivity
\(1/Pe^2\). The geometric drift term \(1/r\) in
equation~\eqref{eq:bds_sde_r} arises naturally from the cylindrical
coordinate system and represents the stochastic counterpart of the
radial diffusion operator
\[
\frac{1}{r}
\frac{\partial}{\partial r}
\left(
r
\frac{\partial C}{\partial r}
\right).
\]

For numerical implementation,
equations~\eqref{eq:bds_sde_z}--\eqref{eq:bds_sde_r}
are discretized using the Euler--Maruyama scheme. For a uniform time
increment \(\Delta t\), the particle positions are updated according to
\begin{align}
z_{k+1}
&=
z_k
+
v(r_k)\Delta t
+
\frac{\sqrt{2\Delta t}}{Pe}\,\xi_{z,k},
\label{eq:bds_discrete_z}
\\
r_{k+1}
&=
r_k
+
\frac{\Delta t}{r_k}
+
\sqrt{2\Delta t}\,\xi_{r,k},
\label{eq:bds_discrete_r}
\end{align}
where
\[
\xi_{z,k},\,\xi_{r,k}
\sim
\mathcal{N}(0,1)
\]
are independent standard normal random variables generated at each time
step.

The deterministic initial condition is incorporated into the stochastic
formulation through the initial particle distribution. This corresponds
to an instantaneous line source located at \(z=z_s\)
(\(z_s=0\) in the present study), with the released solute uniformly
distributed over the circular cross-section. Accordingly, all particles
are initially released from the prescribed axial source location,
\begin{equation}
z_0=z_s,
\label{eq:bds_initial_z}
\end{equation}
where \(z_s\) denotes the injection location.

In the radial direction, particles are distributed uniformly with
respect to the cross-sectional area over the interval
\(r\in[r_0,1]\), where \(r_0\) is a small positive cutoff introduced solely to avoid the
singularity of the geometric drift term during numerical integration. Since the differential area element in cylindrical coordinates is proportional to \(r\,dr\), a uniform distribution over the
cross-sectional area yields the probability density
\begin{equation}
f_{R_0}(r)
=
\frac{2r}{1-r_0^2},
\qquad
r_0\le r\le1,
\label{eq:bds_initial_pdf}
\end{equation}
with cumulative distribution function
\[
F(r)
=
\frac{r^2-r_0^2}{1-r_0^2}.
\]
Applying inverse-transform sampling yields
\begin{equation}
R_0
=
\sqrt{
r_0^2+
(1-r_0^2)\eta},
\qquad
\eta\sim\mathcal{U}(0,1),
\label{eq:bds_initial_r}
\end{equation}
where \(\mathcal U(0,1)\) denotes the continuous uniform distribution on
the interval \((0,1)\), which is precisely the initialization employed
in the Brownian dynamics simulation.

A sufficiently large ensemble of particles is evolved according to
equations~\eqref{eq:bds_discrete_z}--\eqref{eq:bds_discrete_r}. At each
recorded time level, the macroscopic concentration field is reconstructed
from normalized two-dimensional histograms of the particle positions
over the computational domain. The reconstructed concentration is then
used to evaluate the same statistical quantities as those derived from
the analytical solution.

To validate the analytical model, we compare the temporal evolution of
the zeroth and first axial moments obtained from the analytical solution
and the Brownian dynamics simulation. The zeroth moment,
\begin{equation}
M_0(t)
=
\int_{-\infty}^{\infty}
\int_0^1
C(t,z,r)\,
r\,dr\,dz,
\label{eq:validation_M0}
\end{equation}
represents the total solute mass remaining within the domain and therefore provides a direct measure of the mass evolution and wall-
induced depletion during the transport process. The first moment,
\begin{equation}
M_1(t)
=
\int_{-\infty}^{\infty}
\int_0^1
z\,C(t,z,r)\,
r\,dr\,dz,
\label{eq:validation_M1}
\end{equation}
is the first axial moment and characterizes the downstream transport of the solute distribution. Its temporal evolution reflects the cumulative effects of advection and diffusion on the axial migration of the solute
cloud. The Brownian dynamics moments are evaluated using the same definitions as those employed in the analytical solution,
allowing a direct quantitative comparison.

\begin{figure}
\centering
\begin{minipage}{0.48\textwidth}
\centering
\includegraphics[width=\linewidth]{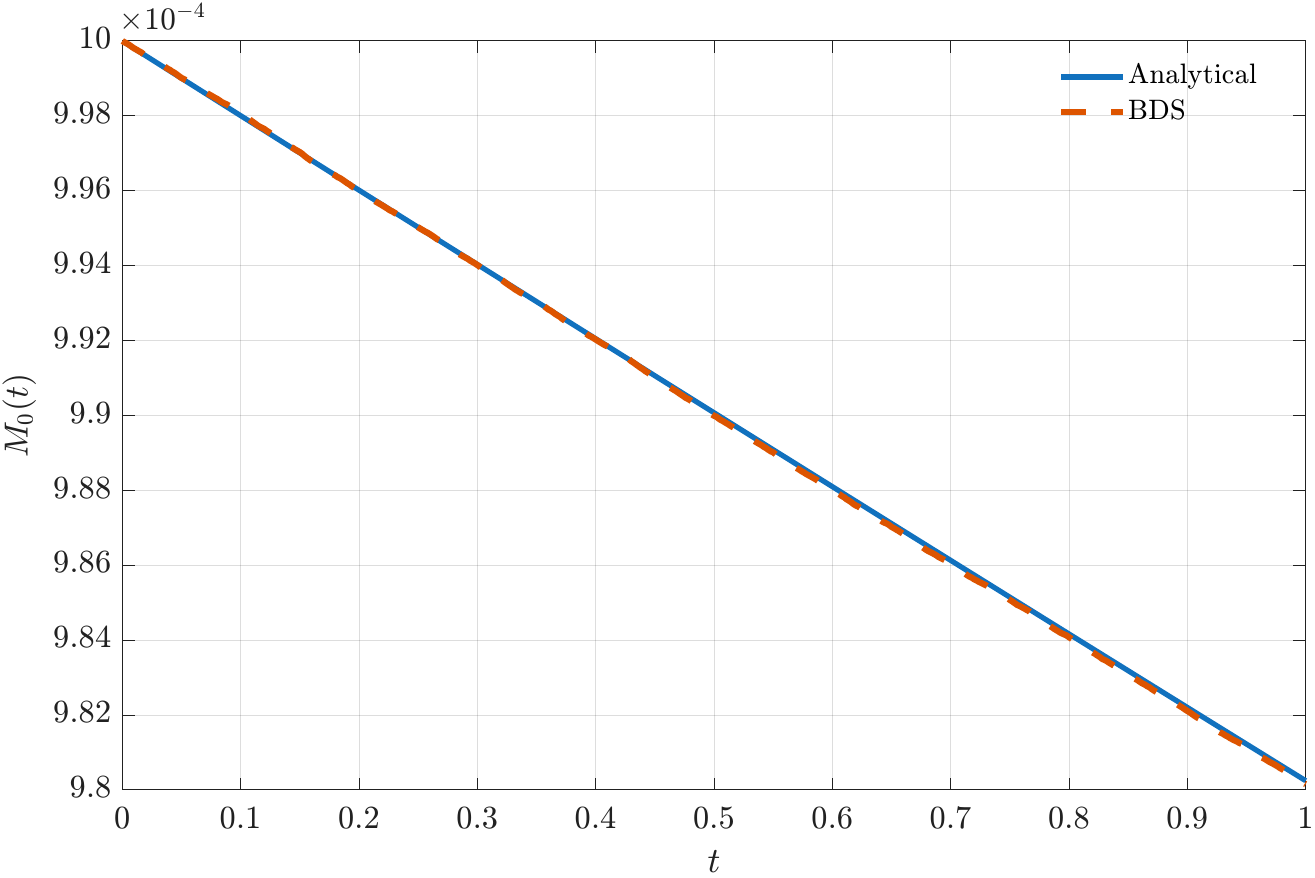}
\caption*{(a)}
\end{minipage}
\hfill
\begin{minipage}{0.48\textwidth}
\centering
\includegraphics[width=\linewidth]{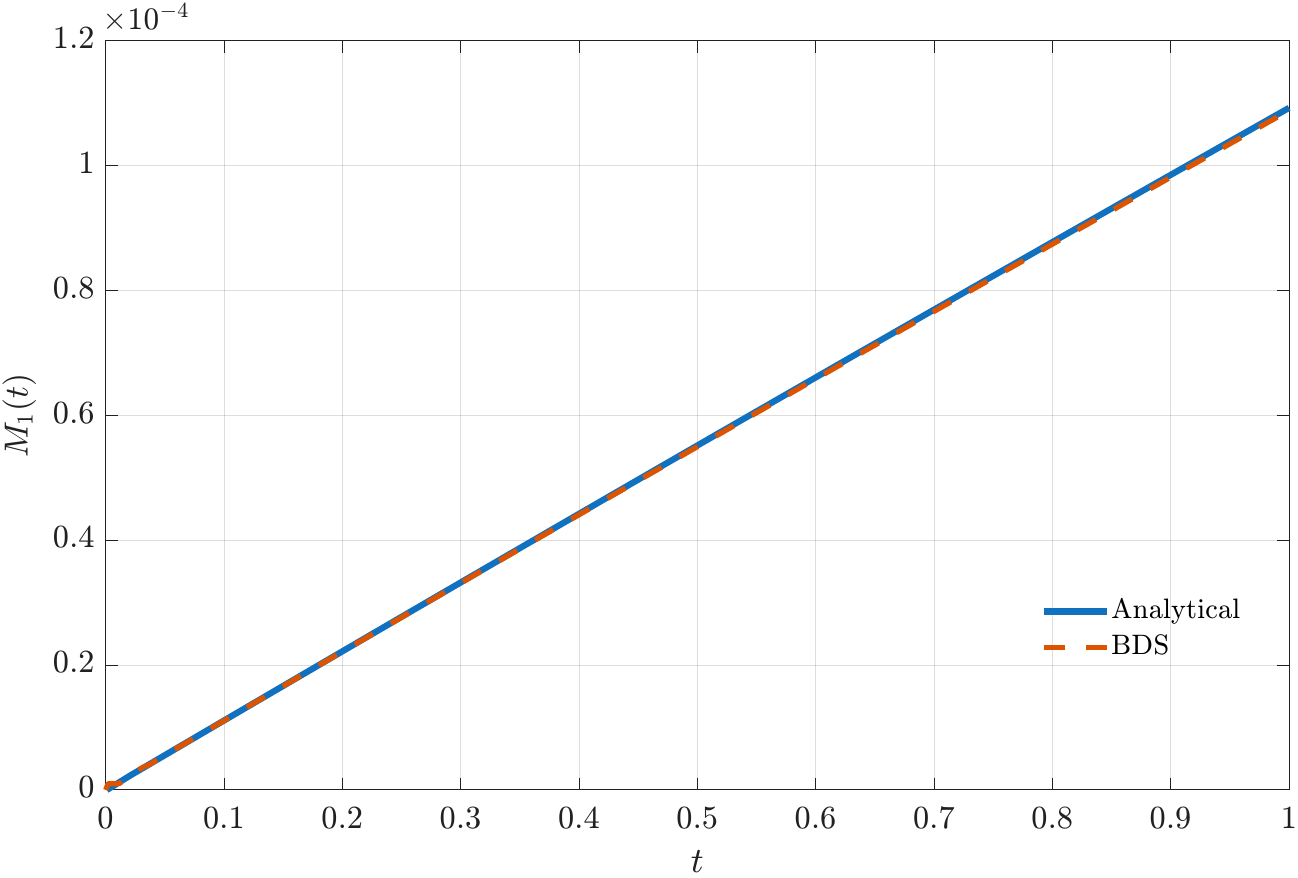}
\caption*{(b)}
\end{minipage}
\caption{
Comparison between the analytical solution and the Brownian dynamics
simulation for (a) the zeroth axial moment \(M_0(t)\) and (b) the first axial moment
\(M_1(t)\).
}
\label{fig:analytical_BDS_moments}
\end{figure}

Figure~\ref{fig:analytical_BDS_moments} compares the analytical solution
with the Brownian dynamics simulation for the temporal evolution of the
zeroth and first moments for a Peclet number of \(Pe=1000\). As shown in
figure~\ref{fig:analytical_BDS_moments}(a), the Brownian dynamics simulation accurately reproduces the analytically predicted temporal evolution of \(M_0(t)\) and gradual depletion of the total solute mass caused by wall absorption. The close agreement confirms that the stochastic formulation accurately reproduces the mass evolution prescribed by the reactive-wall analytical model.

Figure~\ref{fig:analytical_BDS_moments}(b) shows that the first moment
\(M_1(t)\) increases monotonically with time as the solute cloud is
advected downstream by the background flow. The Brownian dynamics
results closely follow the analytical solution over the entire
simulation, accurately reproducing both the magnitude and temporal
evolution of the first axial moment for \(Pe=1000\).

\section{Results and discussion}\label{Sec:Results_and_Discussion}
This study examines the influence of magnetically induced multiphase fluid flow on solute dispersion phenomena in a porous, layered tube. Our approach can be applied to a flow through different tubes; however, we centre around blood flow to ensure its practical relevance in biofluid studies.
We have examined how MHD fluid flow parameters, such as the micropolar parameter, coupling number, Hartmann number, porosity parameter, and effective viscosity ratio parameter, affect the axial velocity of fluids in various areas of a porous layered blood vessel, as well as the micro-rotational velocity of microparticles in the core region. This study mainly analyses the role of the axial and micro-rotational velocities on solute transport. To study the dispersion mechanism, we have employed the generalised dispersion model proposed by Sankarasubramanian \& Gill \cite{sankara1973royal}. Three transport coefficients, like the exchange, convection, and dispersion coefficients, are obtained at small and large times to see how the solute transports within the fluid in different zones. The authors have examined the impact of the absorption parameter on the solute's mean concentration for different values, ranging from small to high. We have considered the appropriate ranges of the parameters given in Table \ref{tab:table1}. The selection of parameter ranges in this study is based on the physiological and physical aspects of blood flow through arterial segments with porous walls. The Hartmann number ($M$) is considered in the range $0 \leq M \leq 4$, which represents both the non-magnetic baseline and moderate magnetic field conditions typically encountered in biomedical magnetohydrodynamic applications, such as magnetically influenced micro-arterial flows (Tiwari \textit{et al.}\cite{tiwari2020epj}). The micropolar parameter ($m$), representing the microrotation of suspended microparticles in the core region, is varied from $0$ to $\infty$, ensuring that both Newtonian and highly non-Newtonian regimes are addressed, consistent with previous micropolar blood flow studies (Eringen \cite{eringen1966jmm}). The coupling number ($N_c$) is varied between $0$ and $1$, corresponding respectively to the limits of no coupling (Newtonian flow) and complete coupling between the rotational and translational motions of the micropolar medium (Tiwari \textit{et al.}\citep{tiwari2020epj}). These parameter ranges comprehensively represent the interactions across the four zones of the arterial segment: the central \textit{micropolar core}, the intermediate \textit{Newtonian layer}, the thin \textit{Brinkman transitional region} (characterised by resistivity $n_1$), and the outer \textit{Darcy porous region} (parametrised by $n_2$). Such zoning and parametric coverage reflect the realistic physical behaviour of physiological flows, and align with well-established models for micropolar and porous arterial environments (Jaiswal \& Yadav \cite{jaiswal2019pof}). The viscosity ratio parameter $\lambda$ is taken as the ratio of the viscosity of the Newtonian fluid and the non-Newtonian fluid and ranges from $0.5$ to $1.0$; the effective viscosity ratio parameter $\lambda_e$ which is the ratio of the viscosity of Newtonian fluid in the Brinkman porous region and the clear Newtonian region is ranges from 1 to 1.5; and the porosity parameter $\phi=\frac{1}{\lambda_e^2}$ (Whitaker \cite{whitaker1999kap}) is taken from $0.4$ to $1$. Also, from the study of (Ochoa-Tapia \& Whitaker \cite{ochoa1995ijhmt}), the stress-jump coefficient $(\xi)$ is from $-1$ to $1.47$. The explicit forms of these parameters, together with the velocity fields, are provided in \hyperref[Sec:Appendix_A]{Appendix A}.

\begin{table}
	\begin{center}
		\begin{tabular}{l|l|l} 
			\bf{Parameters} & \textbf{~~~Values~~~} & \textbf{~~~~~~Source} \\
			\hline
			Micropolar parameter $(m)$ & ~~~[0, $\infty$)~~~ &~~~~~~Eringen \cite{eringen1966jmm}\\
			Coupling number $(N_c)$ &~~~[0, 1)~~~ &~~~~~~Tiwari \textit{et al.} \cite{tiwari2020epj}\\
			Hartmann number $(M)$ &~~~[0, 4]~~~ & ~~~~~~Tiwari \textit{et al.} \cite{tiwari2020epj}\\
			Effective viscosity ratio $(\lambda_e)$ & ~~~[1, 1.5]~~~ & ~~~~~~Tiwari \textit{et al.} \cite{tiwari2020epj}\\
			Viscosity ratio $(\lambda)$ & ~~~[0.5, 1.0]~~~ & ~~~~~~Tiwari \textit{et al.} \cite{tiwari2020epj}\\
			Porosity $(\phi)$ &~~~[0.4, 1]~~~ & ~~~~~~Tiwari \textit{et al.} \cite{tiwari2020epj}\\
			Stress-jump parameter $(\xi)$ & ~~~[-1, 1.47]~~~ & ~~~~~~Ochoa-Tapia \& Whitaker \cite{ochoa1995ijhmt}\\
		\end{tabular}
		\caption{Ranges of the given parameters in the study.}\label{tab:table1}
	\end{center}
\end{table}

\subsection{Effects of flow parameters on axial and micro-rotational velocity}
\begin{figure}
\centering
 \begin{subfigure}{0.49\linewidth}
  \centering
  \includegraphics[width=\linewidth]{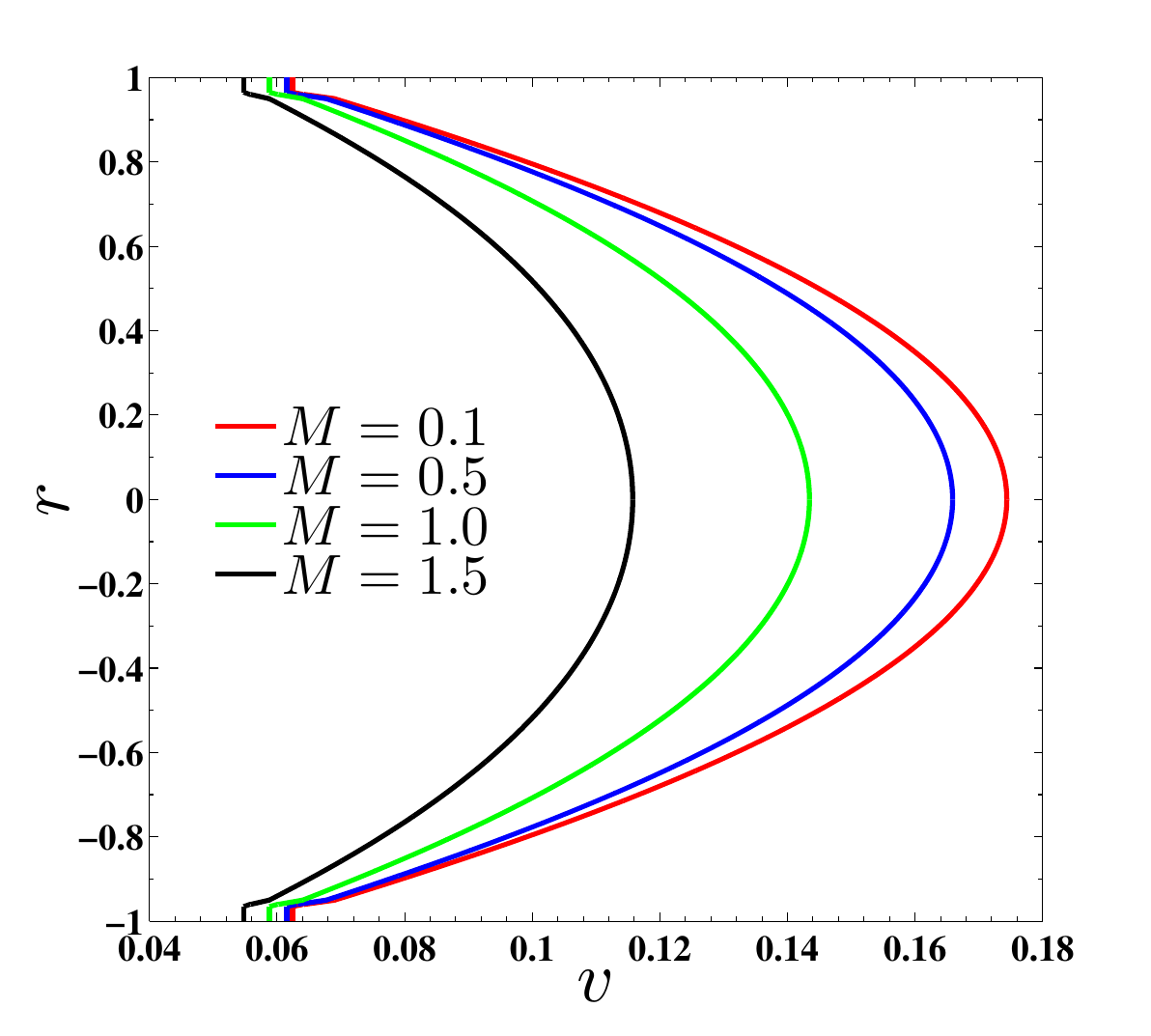}
  \caption{}
  \label{fig:2a}
  \end{subfigure}
 \begin{subfigure}{0.49\linewidth}
     \centering
     \includegraphics[width=\linewidth]{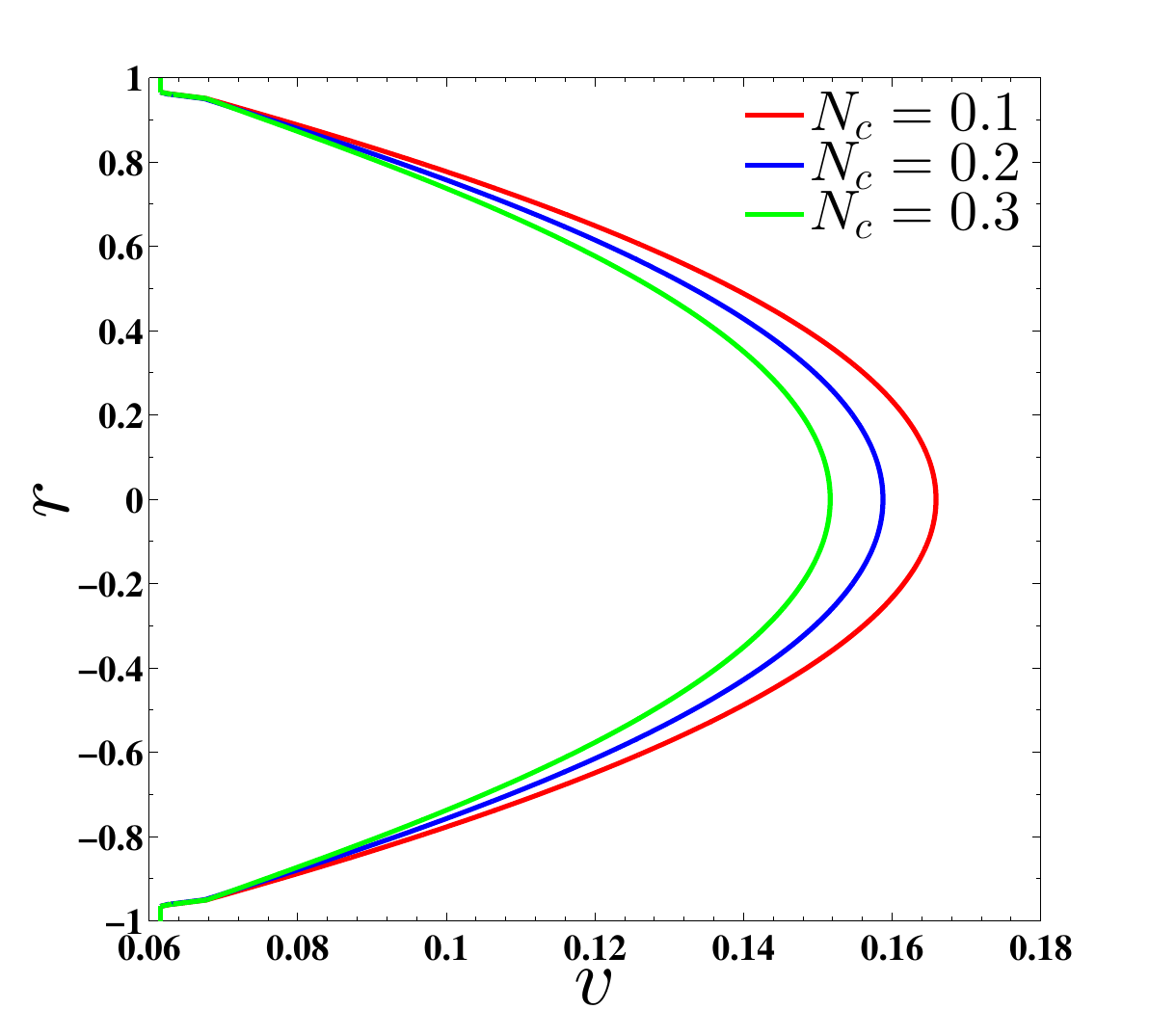}
     \caption{}
     \label{fig:2b}
     \end{subfigure}
    \begin{subfigure}{0.48\linewidth}
     \centering
     \includegraphics[width=\linewidth]{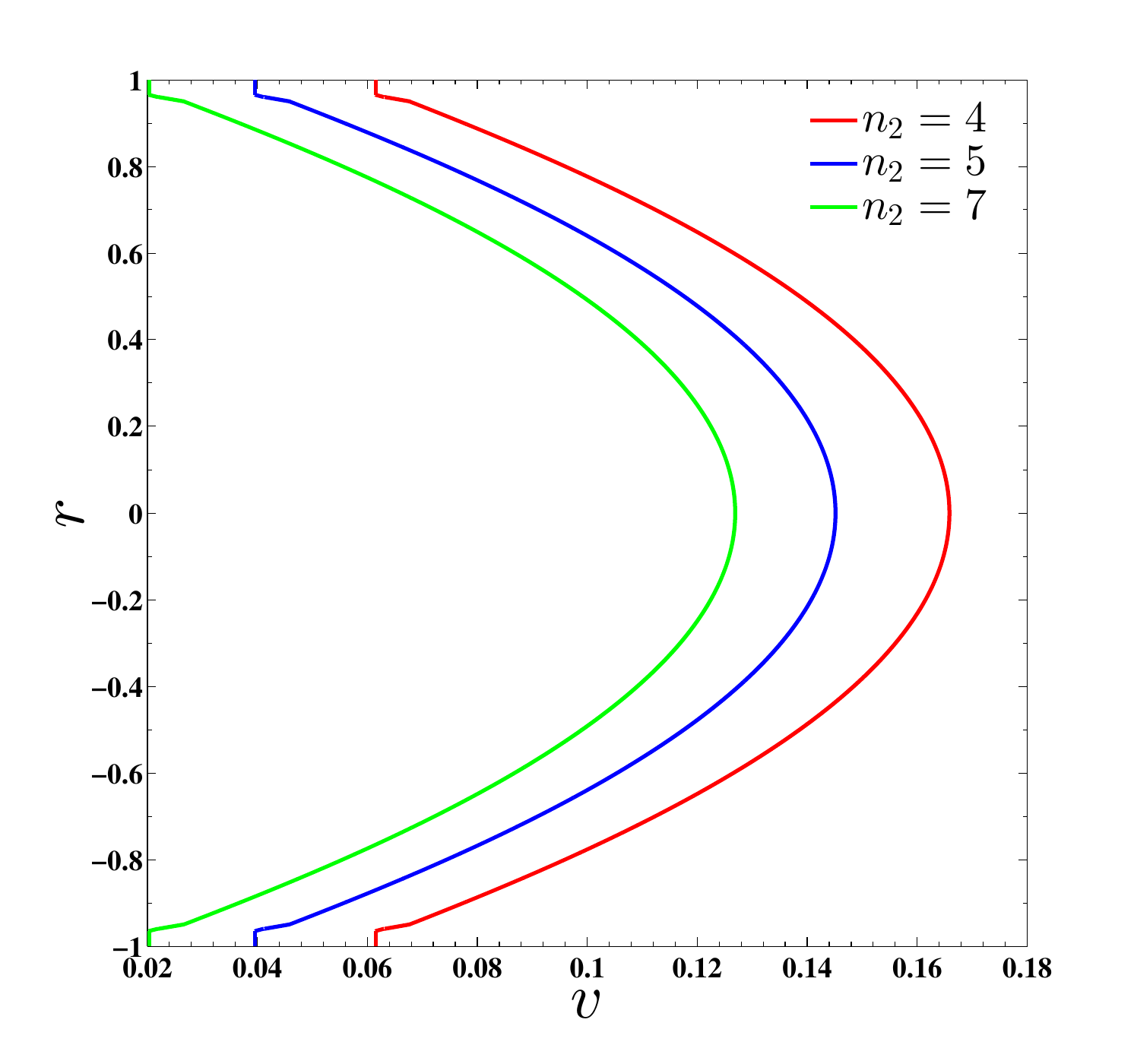}
     \caption{}
     \label{fig:2c}
     \end{subfigure}
  \caption{ Velocity distribution for various values of (a) Hartmann number $M$ with $m=10$, $N_c=0.1$, $n_1=0.5$, $n_2=4$; (b) coupling number $N_c$ with $m=10$, $n_2=4$, $M=0.5$, $n_1=0.5$ and (c) resistivity of Darcy porous region $n_2$ with $m=10$,  $N_c=0.1$, $n_1=1$, $M=0.5$ with $\phi=0.69$, $\xi=0.5$, $\lambda_e=1.2$ $\gamma=0.5$, $\lambda=0.5$ and $P=-1$.}
  \label{fig:2}
\end{figure}

The effect of various parameters, such as Hartmann number $M$, micropolar parameter $m$, coupling number $N_c$, and resistivity parameter of Darcy porous region $n_2$, on the fluid velocities in select regions of the porous layered artery, has been displayed in Figure \ref{fig:2}. From figure \ref{fig:2a}, as $M$ increases, the fluid velocity decreases in each region. This suggests that the flow of fluid in select regions is hindered by the applied transverse magnetic field. The opposing force generated by the magnetic field is referred to as the Lorentz force, which reduces the axial velocity. From figure \ref{fig:2a}, the fluid velocity in the core region $(0\le r \le0.95)$ drops significantly compared to that for other regions with increasing $M$. The velocity profile undergoes a transition from a parabolic form to a blunt one. Furthermore, the velocity profiles in the Brinkman porous region $(0.961\le r \le 0.965)$ and the Darcy porous region $(0.965 \le r\le 1)$ exhibit minor influences. For visualization over the entire tube diameter, the radial profiles are mirrored about the centerline and plotted over $-1 \le r \le 1$; whereas, the governing equations are solved only within $0 \le r \le 1$.

The impact of $N_c$ on the axial fluid velocity is illustrated in figure \ref{fig:2b}. By increasing the value of $N_c$, axial fluid velocity falls significantly. This happens because the coupling number quantifies the interaction between the translational and rotational motions of the microparticles. A stronger coupling between these movements is indicated by a larger coupling number, which results in a decrease in axial velocity. Figure \ref{fig:2c} demonstrates the effect of $n_2$ on the axial velocity. Higher resistivity means lower permeability, which is a characteristic of the porous medium that determines how easily a fluid can flow through it. Therefore, higher flow resistance is correlated with smaller permeability values (higher resistivity), which allows the fluid to pass through the porous medium more slowly. In figure \ref{fig:2c}, we can clearly see that if we increase $n_2$, the flow resistance increases, making it more difficult for the fluid to flow.

\begin{figure}
	\centering
    	\begin{subfigure}{0.49\linewidth}
		\centering
		\includegraphics[width=\linewidth]{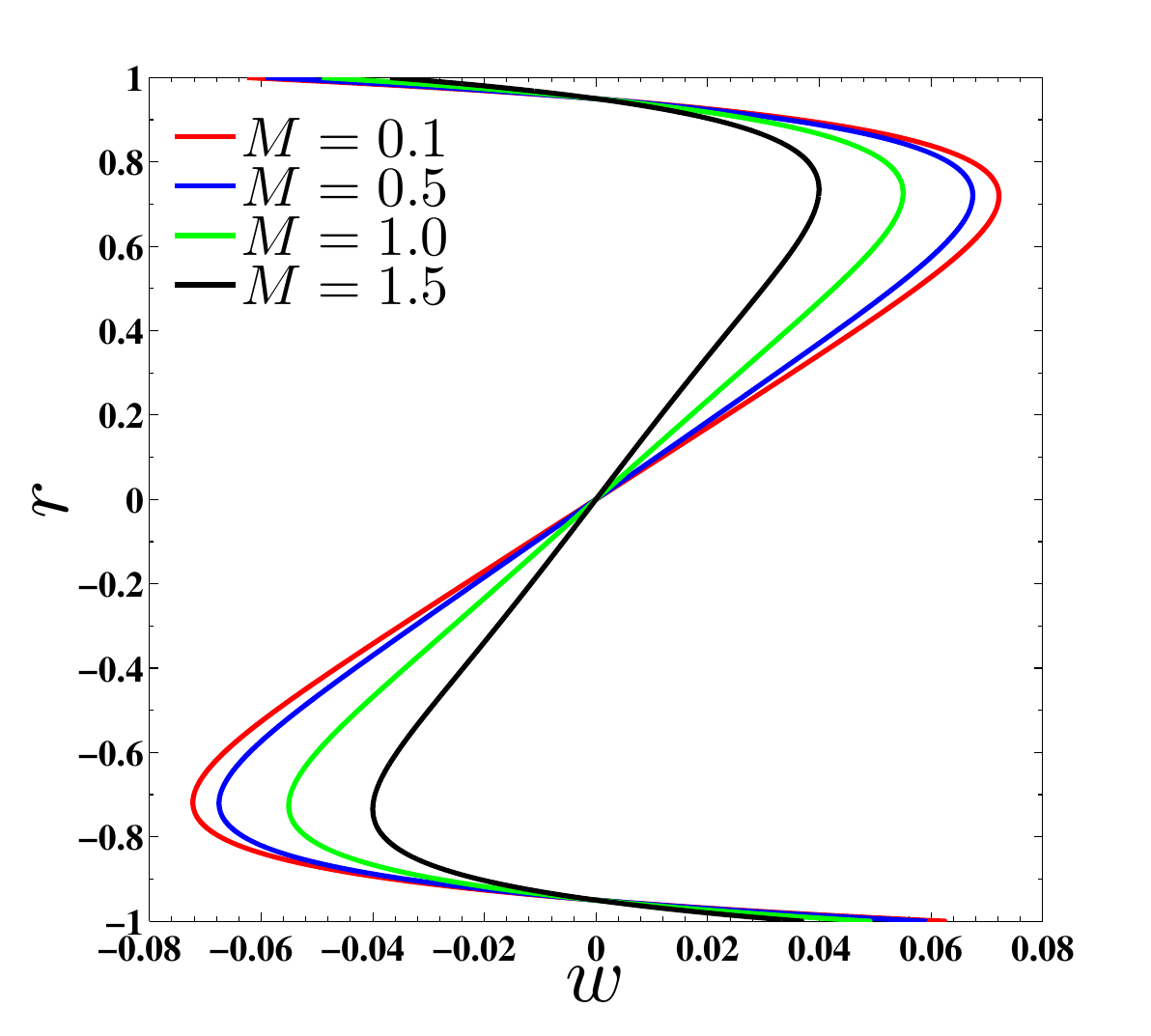}
		\caption{}
		\label{fig:3a}
	\end{subfigure}
	\begin{subfigure}{0.48\linewidth}
		\centering
		\includegraphics[width=\linewidth]{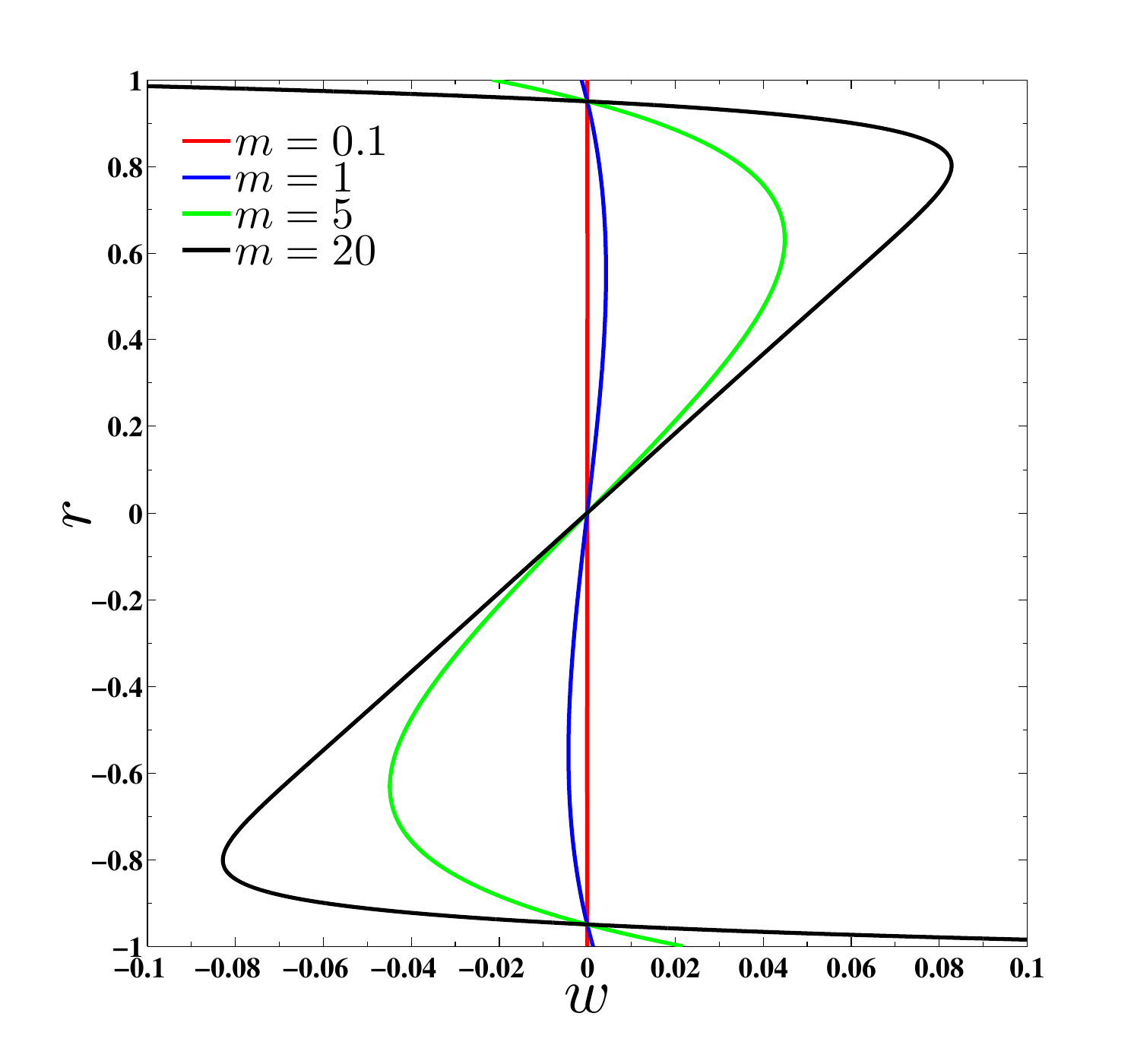}
		\caption{}
		\label{fig:3b}
	\end{subfigure}
	\begin{subfigure}{0.49\linewidth}
		\centering
		\includegraphics[width=\linewidth]{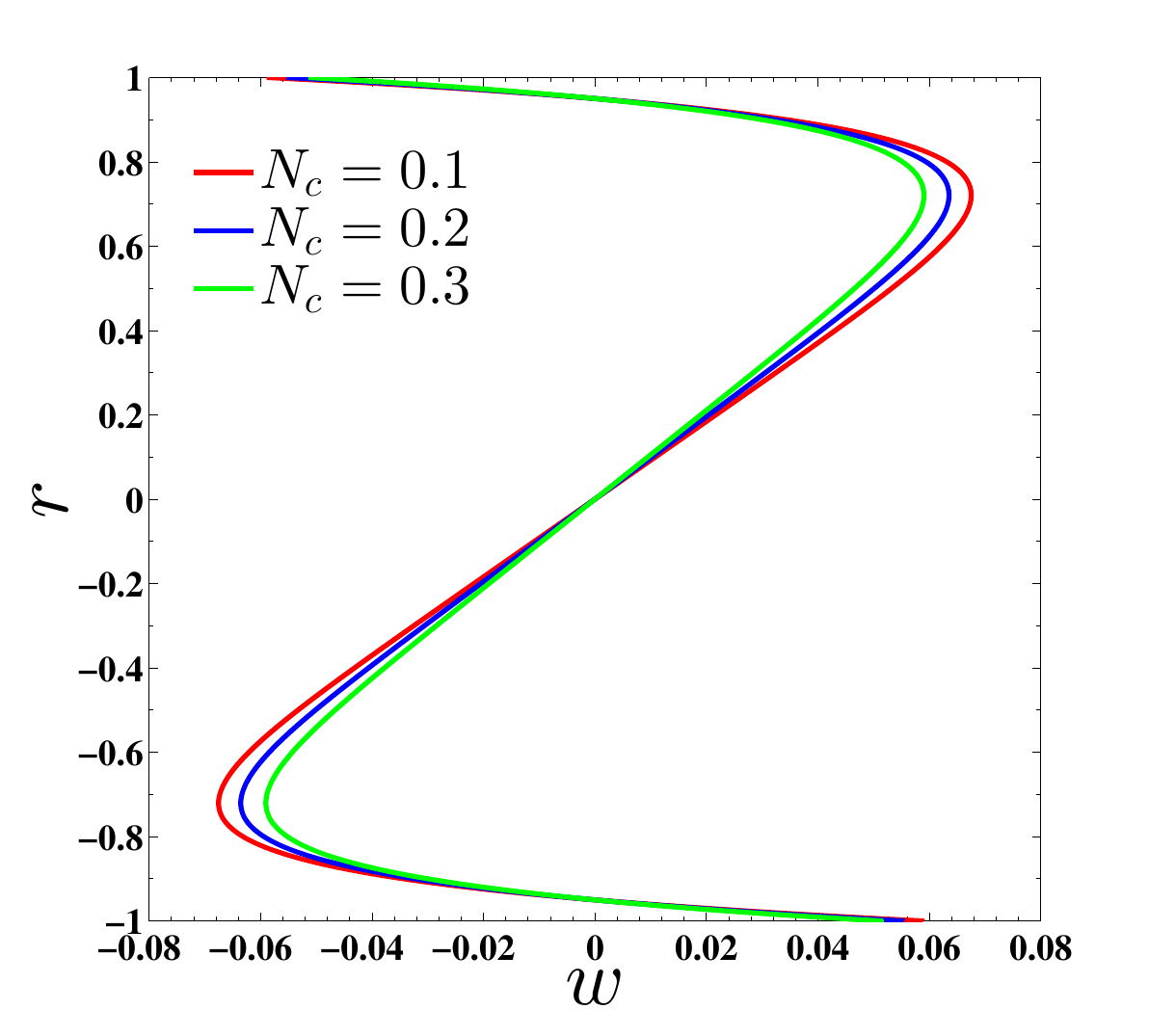}
		\caption{}
		\label{fig:3c}
	\end{subfigure}
	\caption{ Variation of micro-rotational velocity for different values of (a) Hartmann number $M$ with $m=10$, $N_c=0.1$; (b)  micropolar parameter $m$ with $N_c=0.1$, $M=0.5$ and (c) coupling number $N_c$ with $m=10$, $M=0.5$ with $\gamma=0.5$, $\lambda=0.5$, $\lambda_e=1.2$, $n_2=4$, $\xi=0.5$, $n_1=0.5$  and $P=-1$.}
	\label{fig:3}
\end{figure}

The effect of the micropolar parameter $m$, coupling number $N_c$, and Hartmann number $M$ on the micro rotational velocity has been investigated and is displayed in Figure \ref{fig:3}. From figure \ref{fig:3a}, an increase in $M$ leads to a reduction in the micro-rotational velocity of the micropolar fluid. The physical reason is explained as follows. The magnetic field creates an opposing force, known as the Lorentz force, which counteracts the fluid flow and reduces the fluid velocity. Note that the magnetic field intensity enhances with $M$. The high intensity of the magnetic field significantly diminishes the fluid velocity due to the activation of a strong Lorentz force, resulting in a substantial drop in micro-rotational velocity.

From figure \ref{fig:3b}, as $m$ increases, the micro-rotational velocity reaches its highest value close to the core region's boundary. This result suggests that the micropolar parameter stimulates the fluid particles to rotate, particularly in the direction of the boundary of the core region. Figure \ref{fig:3c} describes the effect of $N_c$ on the micro-rotational velocity. The translational motion resists the rotating motion more in the presence of a higher coupling effect. This means that as the particles attempt to rotate, their ability to do so freely is impeded by the forces associated with their movement through the fluid. This interaction decreases the micro-rotational velocity of the particles. 

The temporal behaviour of the exchange coefficient $K_0(t)$ with the effect of the wall absorption parameter $\beta$ is observed in this study, and a similar behaviour of $K_0$ is predicted as described in the study of Sankarasubramanian \& Gill \cite{sankara1973royal}; Rana \& Murthy \cite{rana2016jfm}. To avoid repetition, we have skipped the discussion of $K_0$ in this study. Note that $K_0$ does not depend on the flow behaviour of the fluid (Rana \& Murthy \cite{rana2016jfm}). 

Unless otherwise stated, two complementary sets of transport results
are presented throughout the next subsections. The first employs the coupled axial velocity profile and therefore represents the physical solute transport predicted by the convection--diffusion model. The second uses the microrotational velocity profile within the same analytical framework to quantify the corresponding rotational transport signature, thereby providing a systematic comparison between the translational and rotational kinematic fields.

\subsection{Convection coefficient $K_1(t)$}
The convection coefficient depends on the prescribed kinematic profile and therefore on the governing flow parameters. The axial-profile-based coefficient represents physical solute convection, whereas the microrotational-profile-based coefficient is used as a comparative transport measure. In this study, we have discussed the effect of flow parameters on the convection coefficient for select values of the wall absorption parameter $\beta$.

\begin{figure}
    \centering
    \begin{subfigure}{0.49\linewidth}
    \centering
    \includegraphics[width=\linewidth]{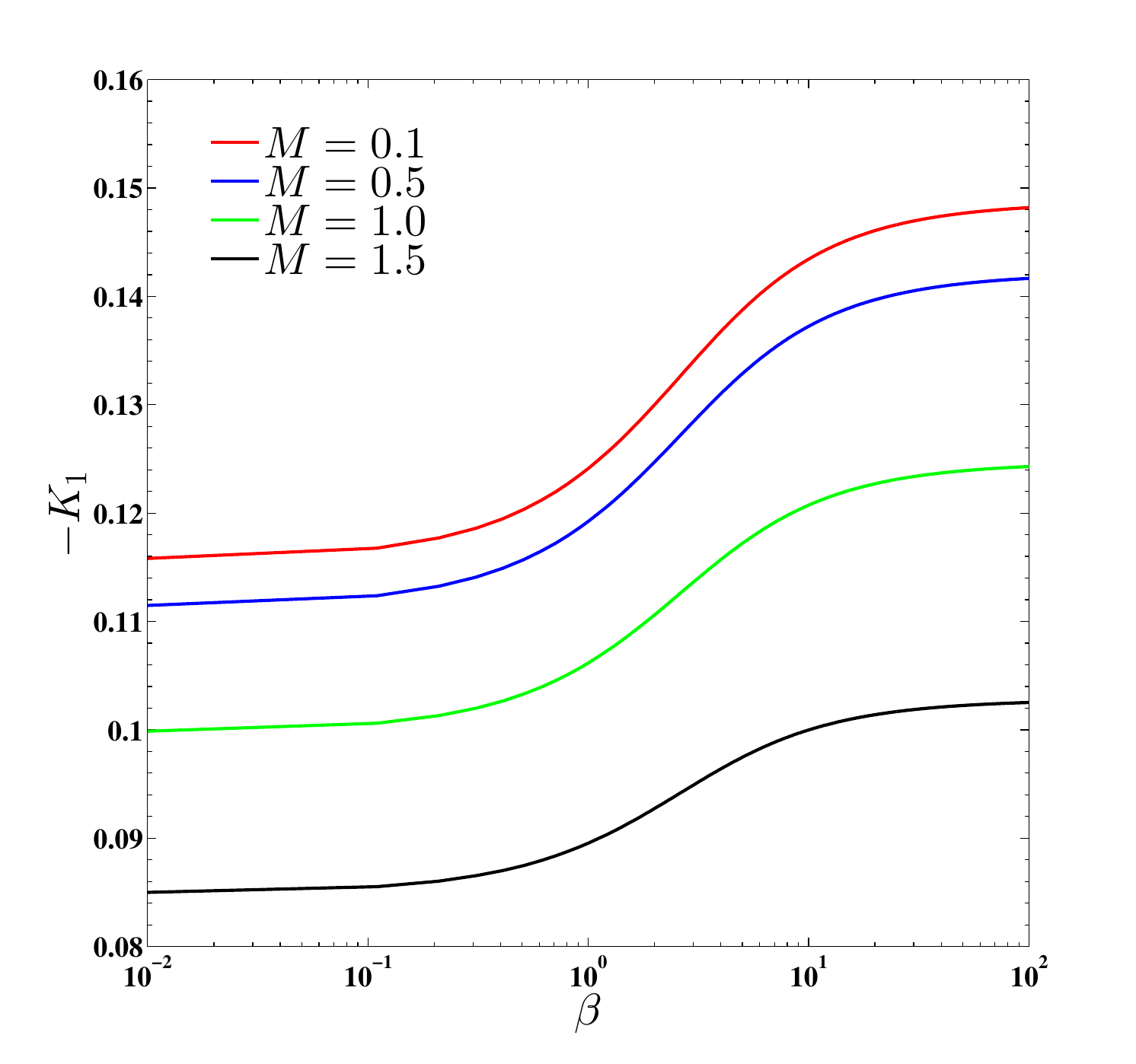}
    \caption{}
    \label{fig:4a}
  \end{subfigure}
\begin{subfigure}{0.49\linewidth}
    \centering
    \includegraphics[width=\linewidth]{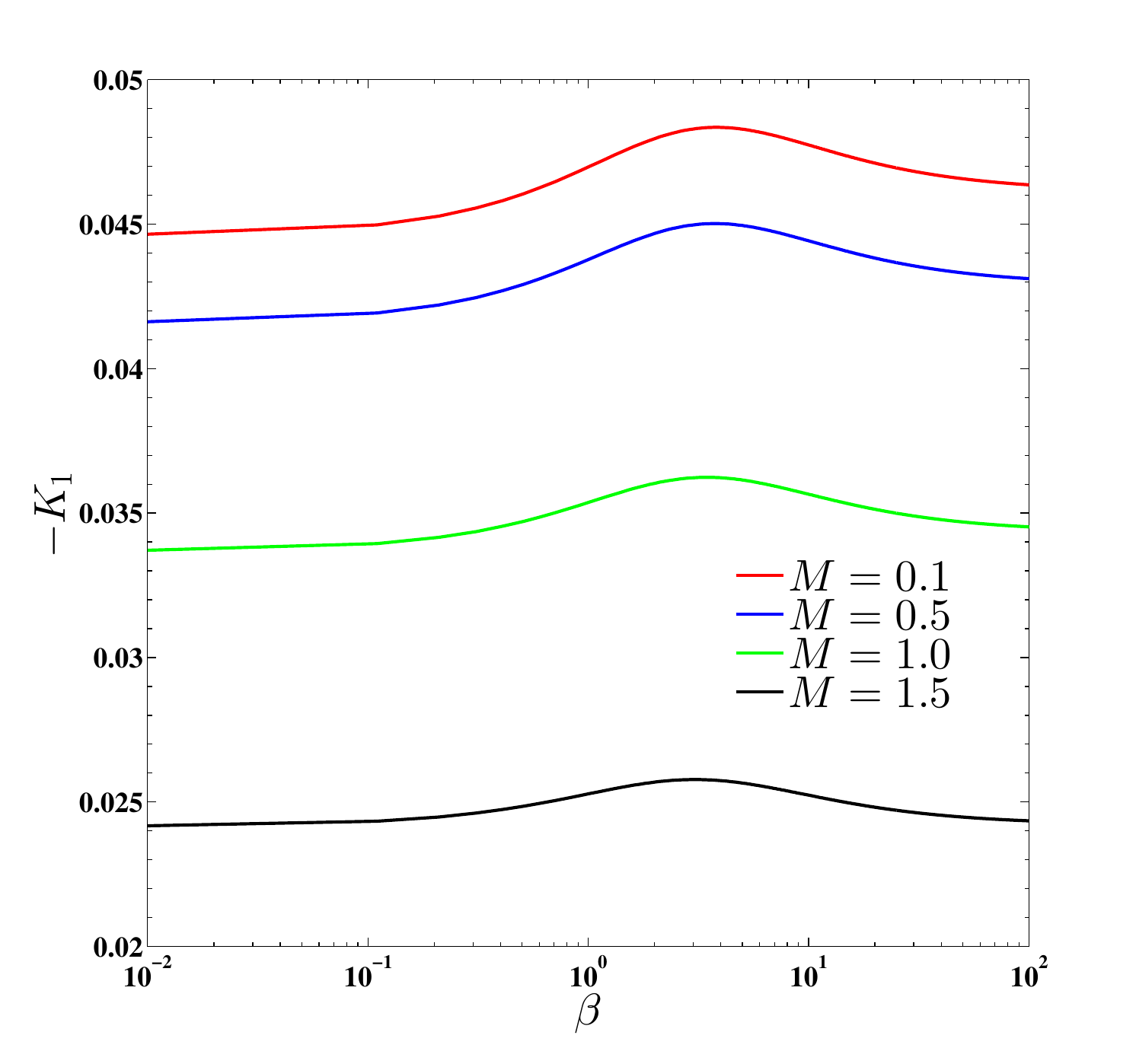}
    \caption{}
    \label{fig:4b}
  \end{subfigure}
     \caption{ Variation of negative convection coefficient $-K_1$ with absorption parameter $\beta$ at time $t=0.5$ for the (a) axial velocity and (b) micro-rotational velocity with $m=10$, $\gamma=0.5$, $\lambda=0.5$, $N_c=0.1$, $n_1=0.5$, $n_2=4$, $\phi=0.69$, $\xi=0.5$, $\lambda_e=1.2$ and $P=-1$.}
   \label{fig:4}
\end{figure}

\begin{figure}
    \centering
    \begin{subfigure}{0.49\linewidth}
    \centering
    \includegraphics[width=\linewidth]{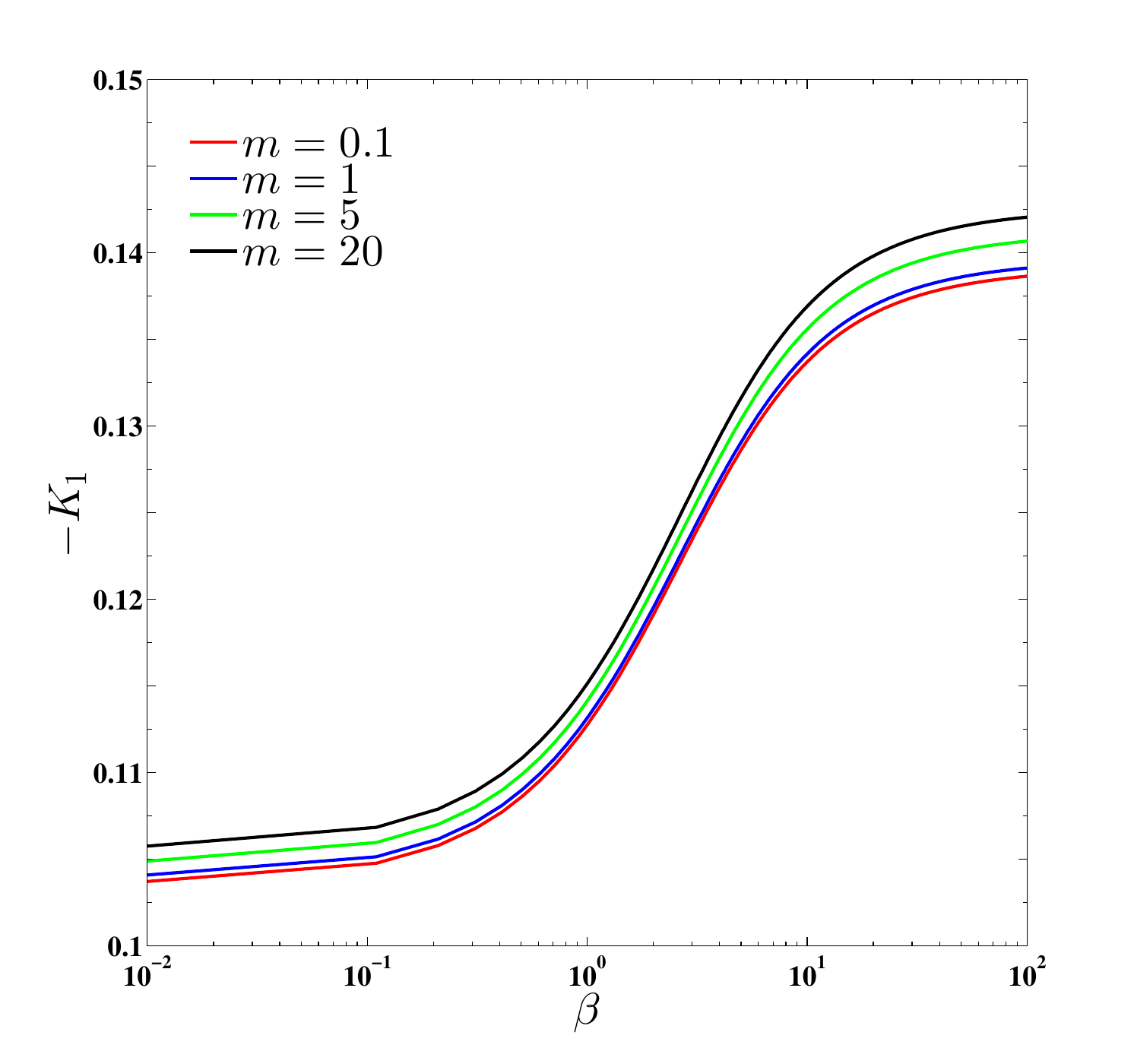}
    \caption{}
    \label{fig:5a}
  \end{subfigure}
\begin{subfigure}{0.49\linewidth}
    \centering
    \includegraphics[width=\linewidth]{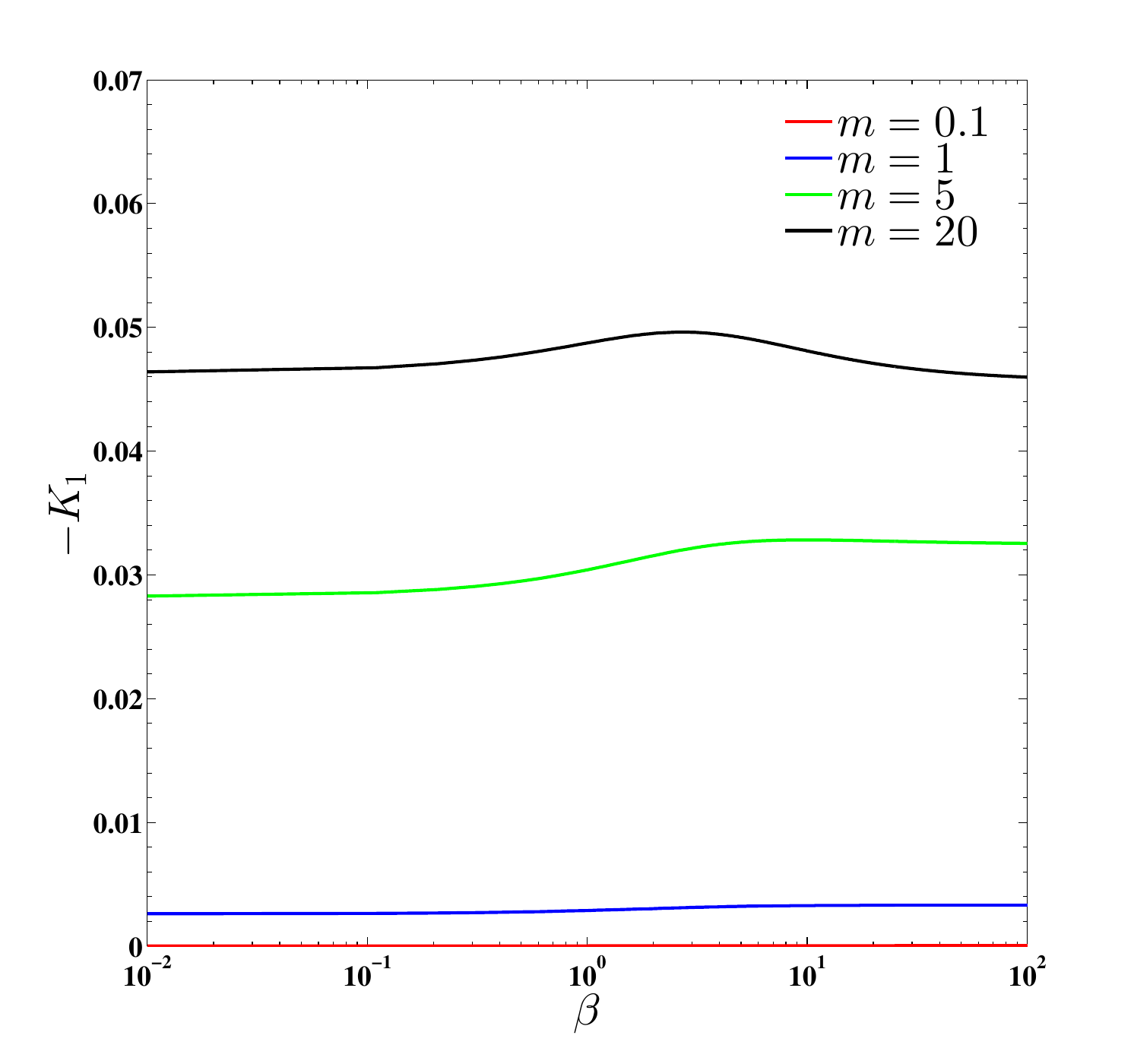}
    \caption{}
    \label{fig:5b}
  \end{subfigure}
    \caption{Variation of negative convection coefficient $-K_1$ with micropolar parameter $m$ at all times for (a) axial velocity and (b) micro-rotational velocity with $\gamma=0.5$, $\lambda_e=1.2$, $\lambda=0.5$, $N_c=0.1$, $n_2=4$, $M=0.5$, $\phi=0.69$, $\xi=0.5$, $n_1=0.5$ and $P=-1$.}
    \label{fig:5}
\end{figure}

\begin{figure}
    \centering
    \begin{subfigure}{0.49\linewidth}
    \centering
    \includegraphics[width=\linewidth]{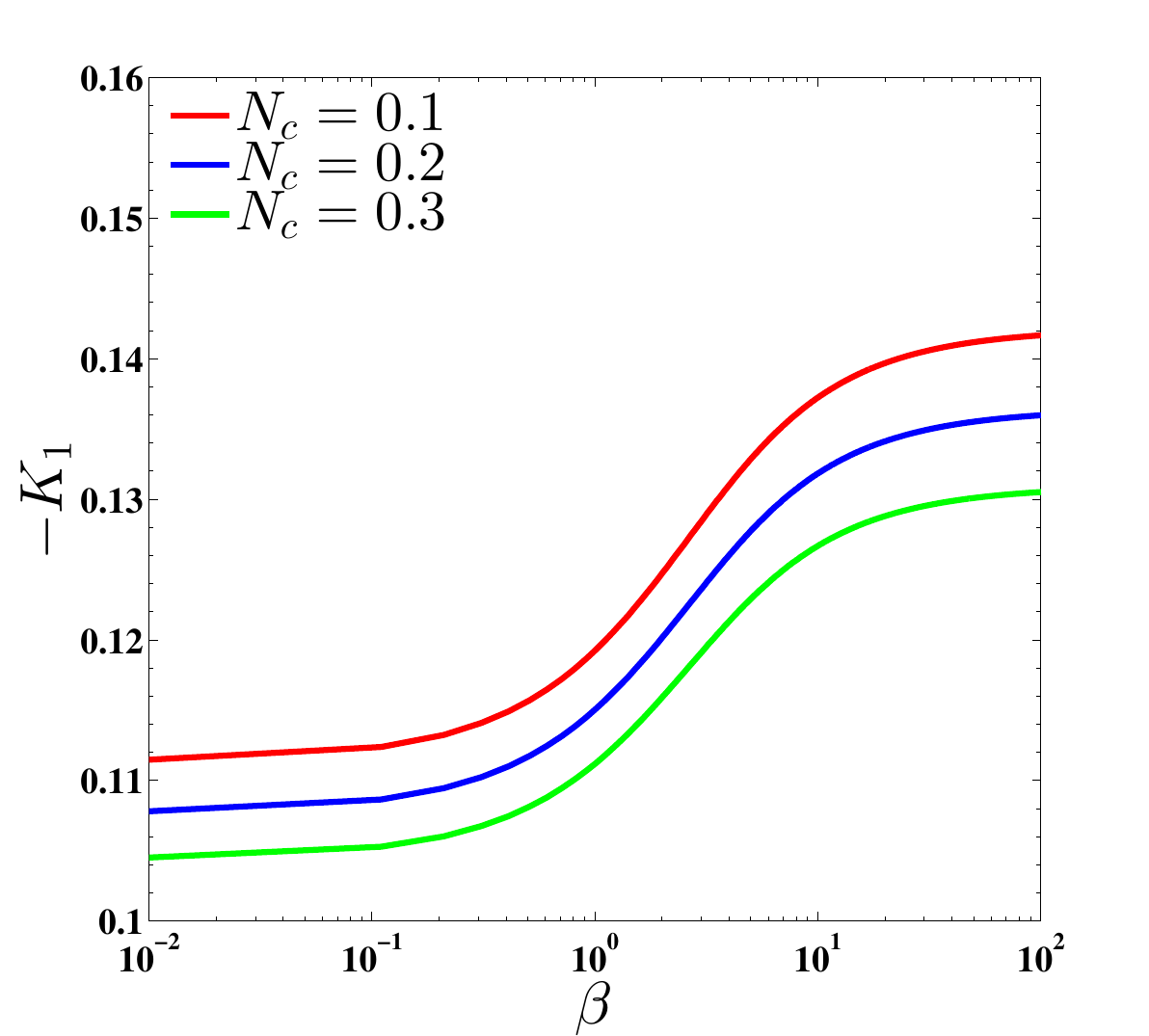}
    \caption{}
    \label{fig:6a}
  \end{subfigure}
\begin{subfigure}{0.49\linewidth}
    \centering
    \includegraphics[width=\linewidth]{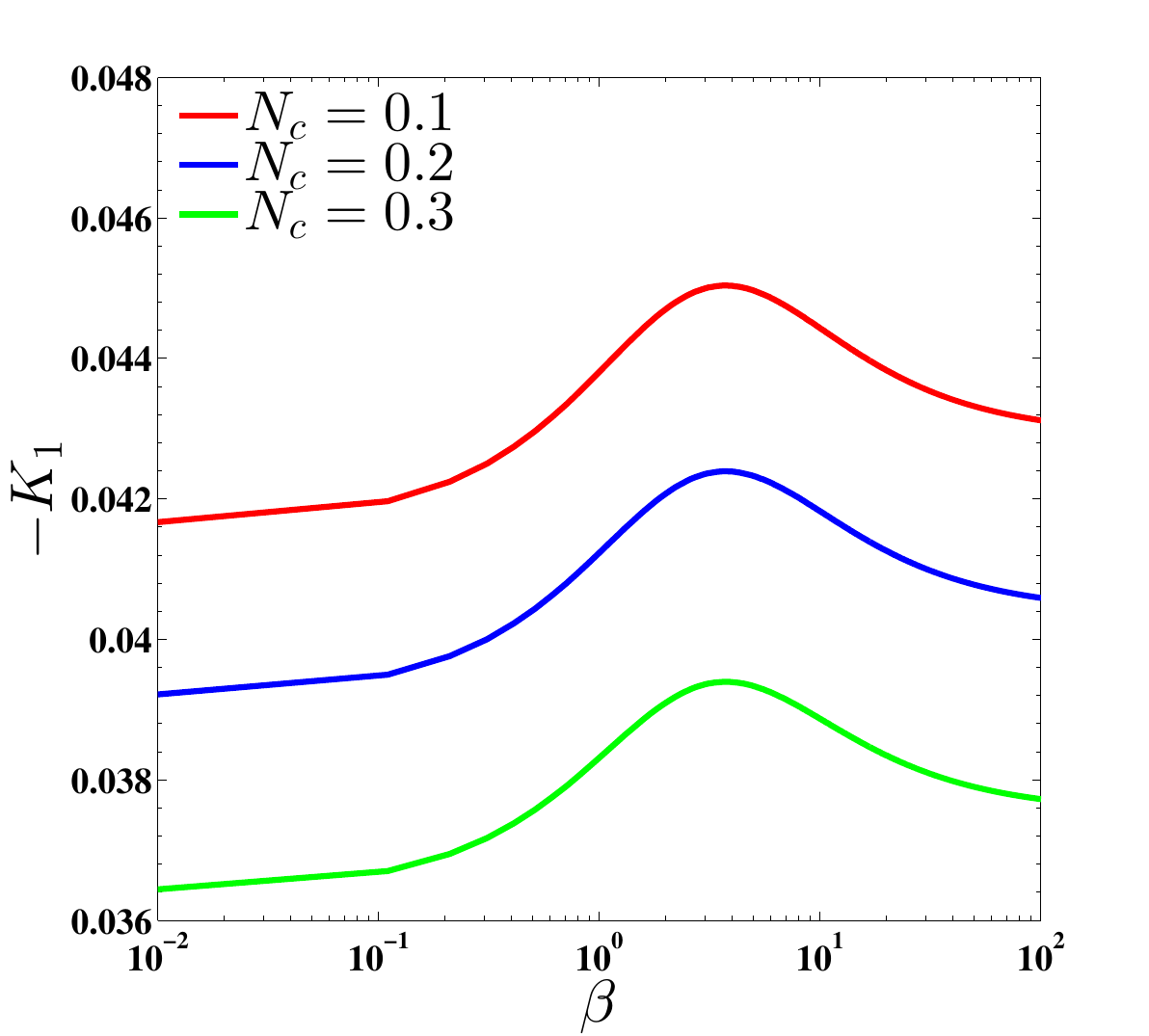}
    \caption{}
    \label{fig:6b}
  \end{subfigure}

 \caption{ Variation of negative convection coefficient $-K_1$ with coupling number $N_c$ at all time for (a) axial velocity and (b) micro-rotational velocity with $\gamma=0.5$, $\lambda_e=1.2$, $\lambda=0.5$, $m=10$, $n_2=4$, $M=0.5$, $\phi=0.69$, $\xi=0.5$, $n_1=0.5$ and $P=-1$.}
   \label{fig:6}
\end{figure}

\begin{figure}
    \centering
    \begin{subfigure}{0.49\linewidth}
    \centering
    \includegraphics[width=\linewidth]{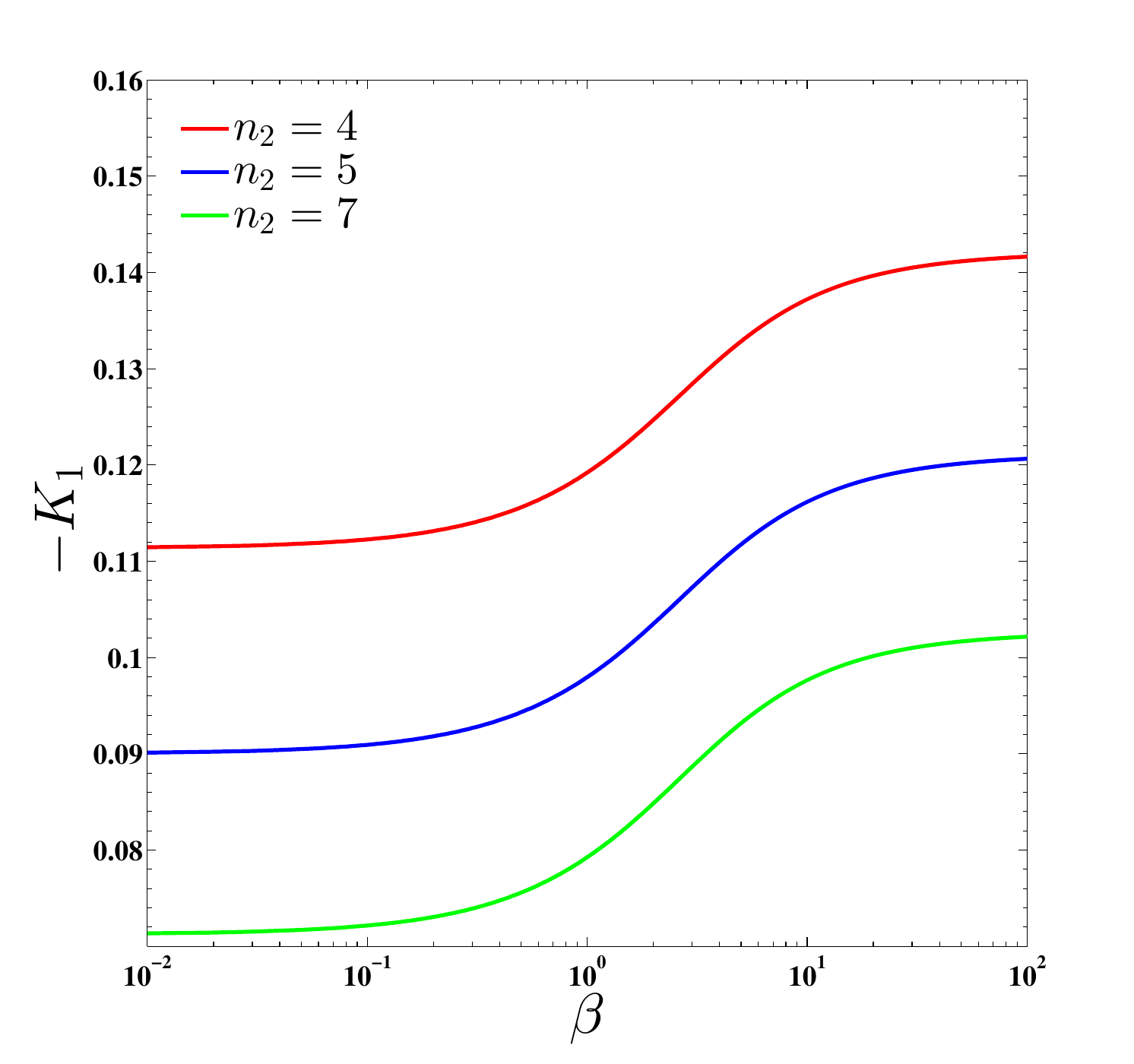}
    \caption{}
    \label{fig:7a}
  \end{subfigure}
\begin{subfigure}{0.49\linewidth}
    \centering
    \includegraphics[width=\linewidth]{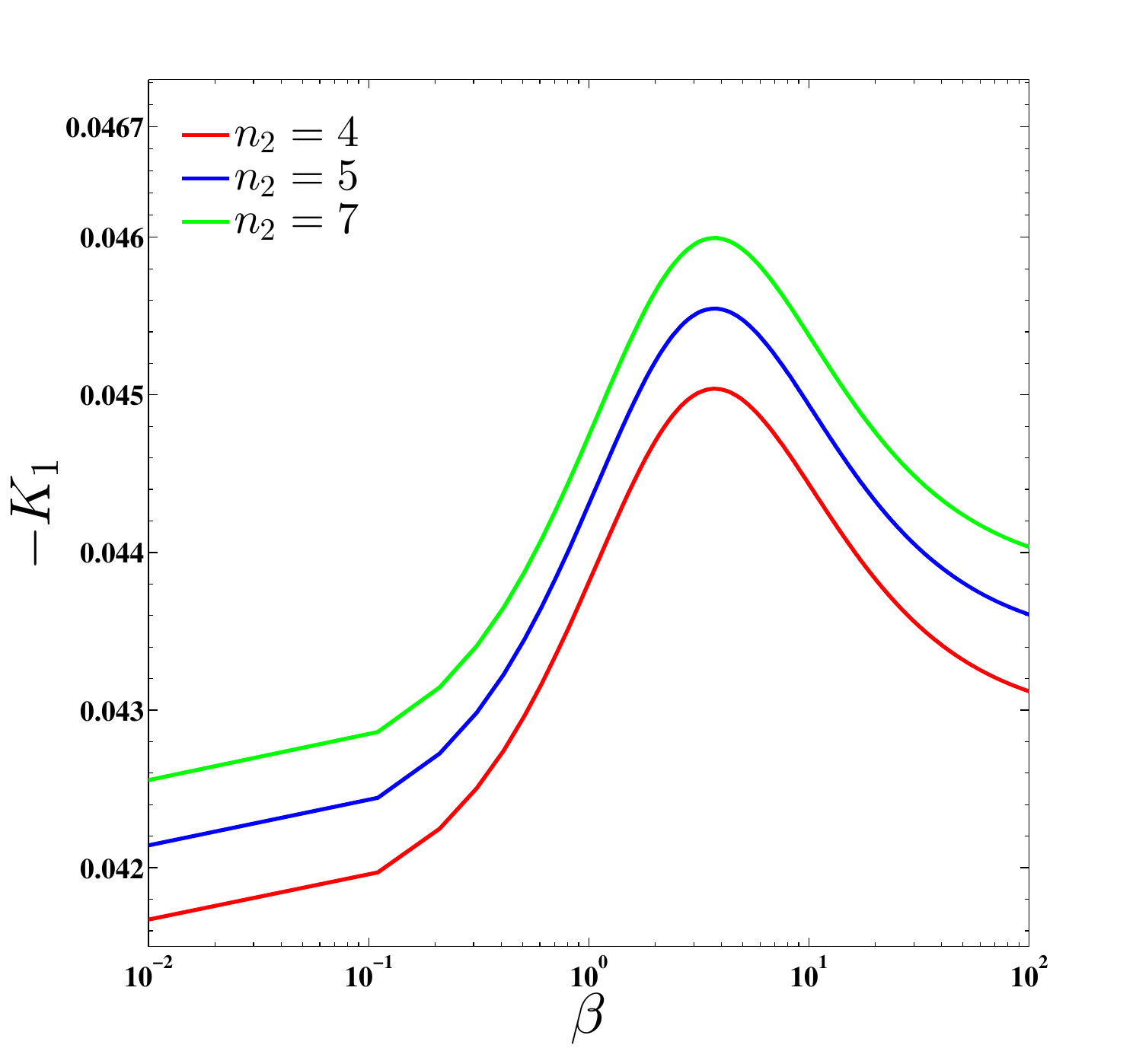}
    \caption{}
    \label{fig:7b}
  \end{subfigure}

 \caption{ Variation of negative convection coefficient $-K_1$ with permeability parameter $n_2$ at all time for (a) axial velocity and (b) micro-rotational velocity with $m=10$, $\gamma=0.5$, $\lambda=0.5$, $N_c=0.1$, $n_1=1$, $M=0.5$, $\phi=0.69$, $\xi=0.5$, $\lambda_e=1.2$ and $P=-1$. }
   \label{fig:7}
\end{figure}

Figure \ref{fig:4} displays the effect of the absorption parameter $\beta$ on the convection coefficient $-K_1$ for different Hartmann numbers $M$ for the axial and micro-rotational velocities. From figure \ref{fig:4a}, for the axial velocity, the convection coefficient shows a slow increase for small $\beta$, then it sharply rises for both moderate and large $\beta$. This pattern arises due to the higher values of $\beta$ where the solute is rapidly consumed by the wall, resulting from an increased absorption rate at the wall. As a result, the concentration of solute in the tube's wall region is lower than that in the centre. Thus, the solute exhibits faster movement in the surrounding areas of the centre compared to the wall region. The convection coefficient shows a similar pattern for various values of $M$ in different regions. When $M$ increases, a Lorentz force emerges that minimises the fluid velocity. As a result, the movement of solute particles within the fluid is decelerated, leading to much slower solute convection. $-K_1$ exhibits lower magnitudes with increasing $M$. When $M=0.1$, for small $\beta$, $-K_1$ starts at $0.1158$ and goes to $0.1482$ for large $\beta$, while for $M = 1.5$, at small $\beta$, $-K_1$ sharply decreases and starts at $0.08501$ and then gradually reaches to $0.1025$ for large $\beta$.

In the case of the micro-rotational velocity (figure \ref{fig:4b}), the convection coefficient shows different behaviour. Since the micro-rotational velocity is influenced by $M$, the convection process is also hampered by $M$. By increasing the value of $M$, a similar decreasing trend for $-K_1$ is predicted, as shown for the case of axial velocity. Moreover, for small values of $\beta$, the convection coefficient slowly increases to a maximum peak value $0.04835$ for $M=0.1$ and then decreases suddenly and reaches a constant value $0.04636$. A similar behaviour has been seen for $M=1.5$ where the coefficient $-K_1$ takes the maximum peak value $0.02578$ at $\beta = 3.2$.

The effect of the micropolar parameter $m$ on $-K_1$ is shown in Figure \ref{fig:5}. From figure \ref{fig:5}, as $m$ increases, $-K_1$ increases. Note that $m$ describes the rotational effect applied by microparticles. As $m$ increases, the micro-rotational velocity enhances, which speeds up the movement of the solute. As a result, the solute convection related to the micro-rotational velocity increases (see figure \ref{fig:5b}). For $m=0.1,1,5$ and $20$, the magnitude of $-K_1$ becomes $2.78\times10^{-5}, 2.623\times10^{-3}, 2.828\times10^{-2}$ and $4.597\times10^{-2}$, respectively, when $\beta=10^{-2}$. Since the parameter $m$ is involved in the rotational movement of the fluid, it does not influence the axial velocity. Consequently, with increasing $m$, the minor changes are predicted in solute convection related to the axial velocity (see figure \ref{fig:5a}). For the axial velocity, the magnitude of $-K_1$ takes values $0.1037, 0.1041, 0.1049$ and $0.1058$ for small $\beta$ and gradually increases to the values $0.1387, 0.1391, 0.1407$ and $0.1421$ for large $\beta$ for $m=0.1,1,5$ and $20$, respectively. In the rest of our investigation, we have skipped the discussion about the influence of $m$ on the solute transport for the axial velocity.

Figure \ref{fig:6} shows the effect of $\beta$ on $-K_1$ for select values of the coupling number $N_c$. From figure \ref{fig:6}, as $N_c$ increases, solute convection decreases. In figure \ref{fig:6a}, we have observed the effect of the axial fluid velocity on solute convection. The axial velocity of the fluid diminishes with increasing $N_c$, which reflects the same in solute convection. For small values of $\beta$, $-K_1$ exhibits the magnitudes $0.1115, 0.1078,$ and $0.1045$ for $N_c = 0.1, 0.2,$ and $0.3$, respectively. From figure \ref{fig:6b}, $-K_1$ gradually increases for low to moderate values of $\beta$ but suddenly decreases for high $\beta$. As $N_{c}$ increases, the micro-rotational velocity decreases, which leads to the reduction of solute convection. For $N_c=0.1, 0.2,$ and $0.3$, the peaks of $-K_1$ are at $0.04504,0.04239,$ and $0.0394$ for $\beta = 3.914,3.813,$ and $3.713$, respectively.

The impact of $\beta$ on $-K_1$ for various values of the resistivity parameter $n_2$ is displayed in Figure \ref{fig:7}. We have seen the effect of the axial fluid velocity on solute convection in Figure \ref{fig:6a}. As $n_2$ increases, the solute is convected with a lower axial velocity of the fluid.
When $\beta$ is small, $-K_1$ exhibits the magnitudes $0.1115, 0.1078,$ and $0.1045$ for $n_2 = 4, 5,$ and $7$, respectively. However, the micro-rotational velocity shows a different trend. Figure \ref{fig:7b} illustrates that $-K_1$ increases with $n_2$. The solute is convected with the higher micro-rotation velocity of the fluid, resulting in enhanced solute convection. As discussed before for other parameters, a similar behaviour of $\beta$ on $-K_1$ is observed. When $\beta = 3.713$, for $n_2 = 4, 5,$ and $7$, the maxima of $-K_1$ is achieved at $0.04504,0.0455,$ and $0.046$, respectively. However, the effect of $n_2$ on solute convection related to the micro-rotational velocity is not significant. 

\subsection{Dispersion coefficient $K_2(t)$}
The dispersion coefficient $K_2(t)$, which exists owing to the fluid movement and the molecular diffusivity of the solute, is obtained to investigate the solute dispersion process throughout the system. Since the axial and micro-rotational velocities of the fluid in different layers are affected by the above-mentioned parameters $M$, $m$, $N_c$ and $n_2$, they also affect the dispersion coefficient. This study describes the effect of these parameters on solute dispersion presented below. 

From figures \ref{fig:8} - \ref{fig:11}, we can see that when time $t$ increases, the dispersion of solute also increases. Solute particles transfer from areas of higher concentration to areas of lower concentration due to a fundamental process called molecular diffusion. In addition, owing to the fluid movement in the system, the convective flux contributes significantly to the transfer of the solute. The combined actions intensify with time, making it easy for solute particles to disperse more widely throughout the fluid. This spreading contributes to an increase in the dispersion coefficient for both the axial and micro-rotational velocities.

The dispersion of solute in fluid flow is greatly influenced by axial velocity, especially in processes where the fluid is convected at a faster velocity. Higher axial velocity gradually improves the longitudinal dispersion of solutes over time. Solutes usually stay concentrated close to where they started in the early phases of flow. As time progresses, the axial velocity stretches the solute concentration profile along the flow direction, leading to increased diffusion of solutes across the length of the tube. In contrast, micro-rotational velocity has a more complex and localised effect on solute dispersion. The rotating motion of microparticles or microstructures within the fluid generates micro-rotational effects, enhancing solute mixing. Unlike axial velocity, which determines the bulk motion of solutes, micro-rotational velocity affects the local dynamics of rotation, which in turn affects the dispersion of solutes over time.

\begin{figure}
\centering
 \begin{subfigure}{0.49\linewidth}
   \centering
   \includegraphics[width=\linewidth]{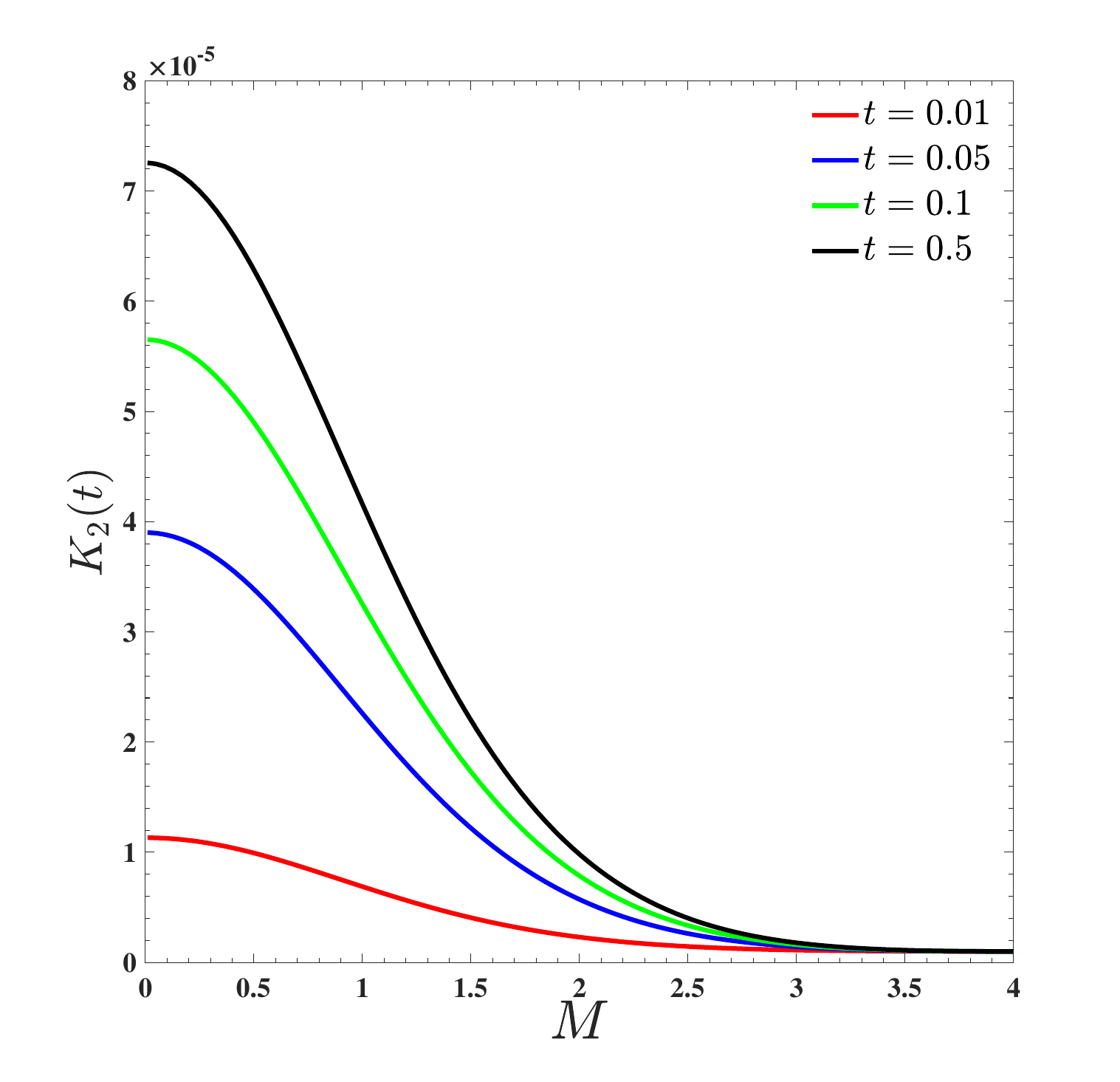}
   \caption{}
   \label{fig:8a}
 \end{subfigure}
 \begin{subfigure}{0.49\linewidth}
   \centering
   \includegraphics[width=\linewidth]{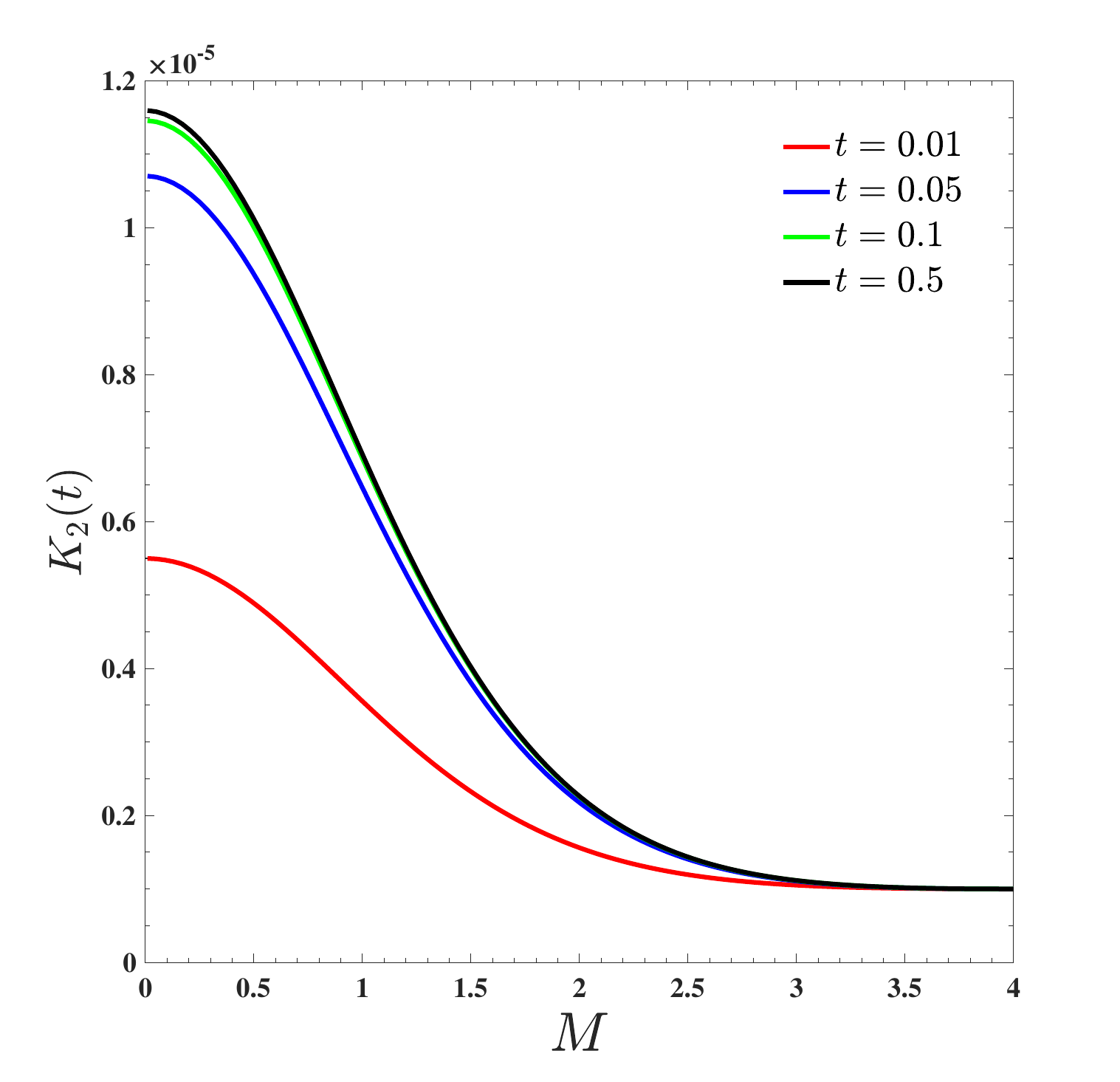}
   \caption{}
   \label{fig:8b}
 \end{subfigure}
\caption{Variation of dispersion coefficient $K_2$ with Hartmann number $M$ for different times for (a) axial velocity and (b) micro-rotational velocity with $m=10$, $\lambda_e = 1.2$, $\gamma=0.5$, $\lambda=0.5$, $N_c=0.1$, $n_2=4$, $\xi=0.5$, $n_1=0.5$, $\beta = 0.01$, $P=-1$ and $Pe = 1000$.}
   \label{fig:8}
\end{figure}

\begin{figure}
\centering
   	\includegraphics[width=0.49\linewidth]{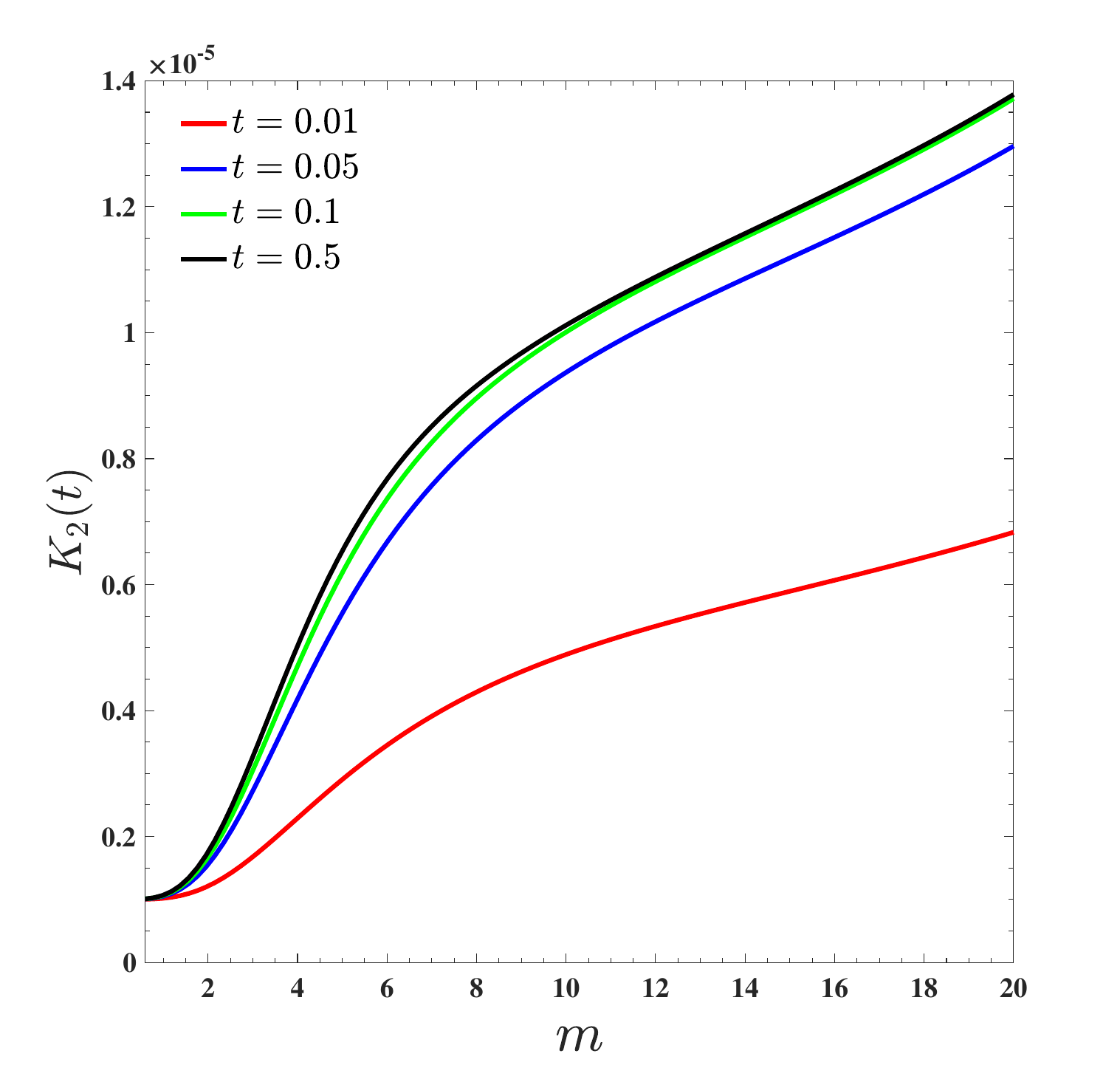}   
\caption{Variation of dispersion coefficient $K_2$ with micropolar parameter $m$ for different times for micro-rotational velocity with $M=0.5$, $\lambda_e = 1.2$, $\gamma=0.5$, $\lambda=0.5$, $N_c=0.1$, $n_2=4$, $\xi=0.5$, $n_1=0.5$, $\beta = 0.01$, $P=-1$ and $Pe = 1000$.}
   \label{fig:9}
\end{figure}

\begin{figure}
\centering
 \begin{subfigure}{0.49\linewidth}
   \centering
   \includegraphics[width=\linewidth]{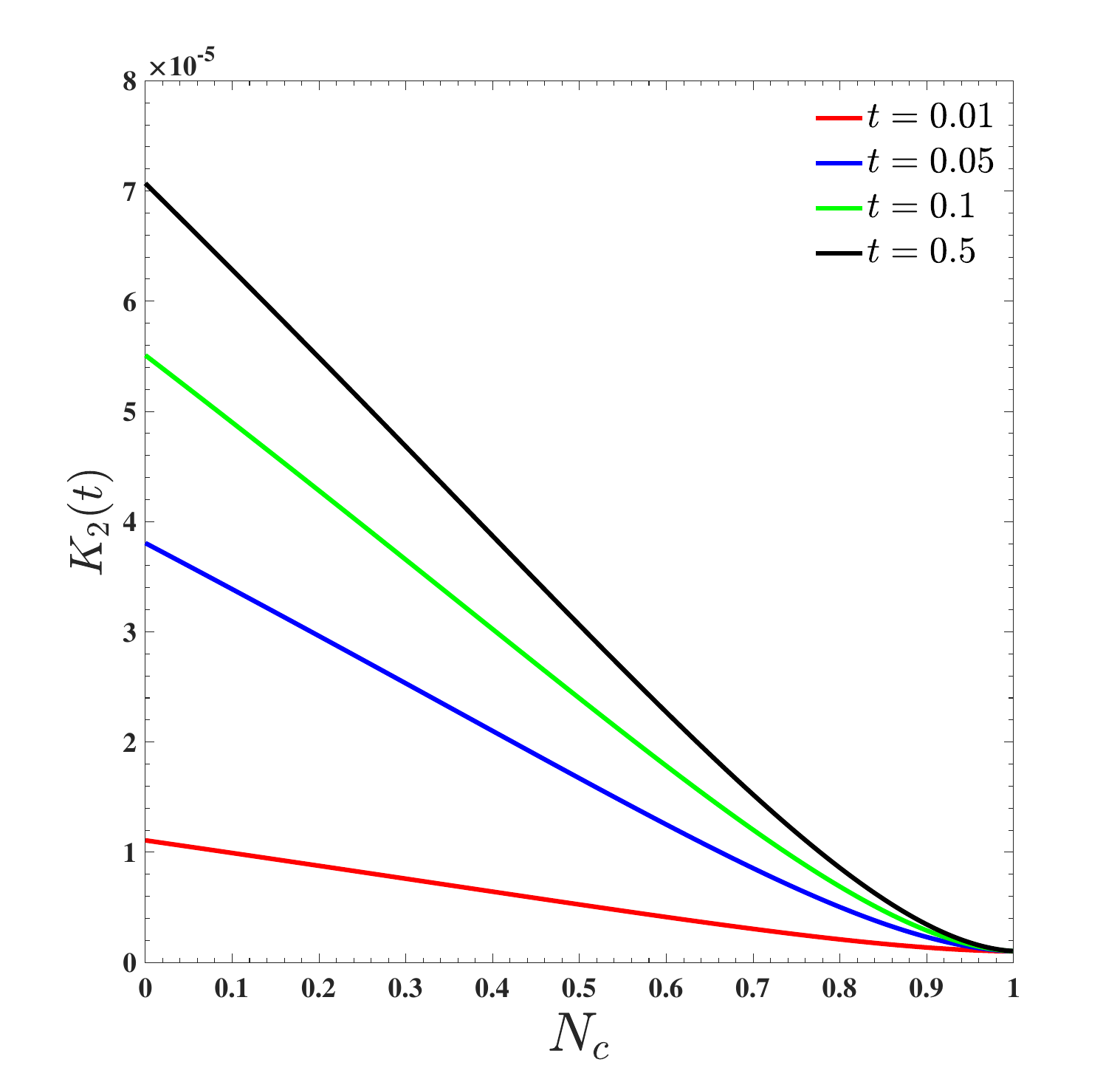}
   \caption{}
   \label{fig:10a}
 \end{subfigure}
 \begin{subfigure}{0.49\linewidth}
   \centering
   \includegraphics[width=\linewidth]{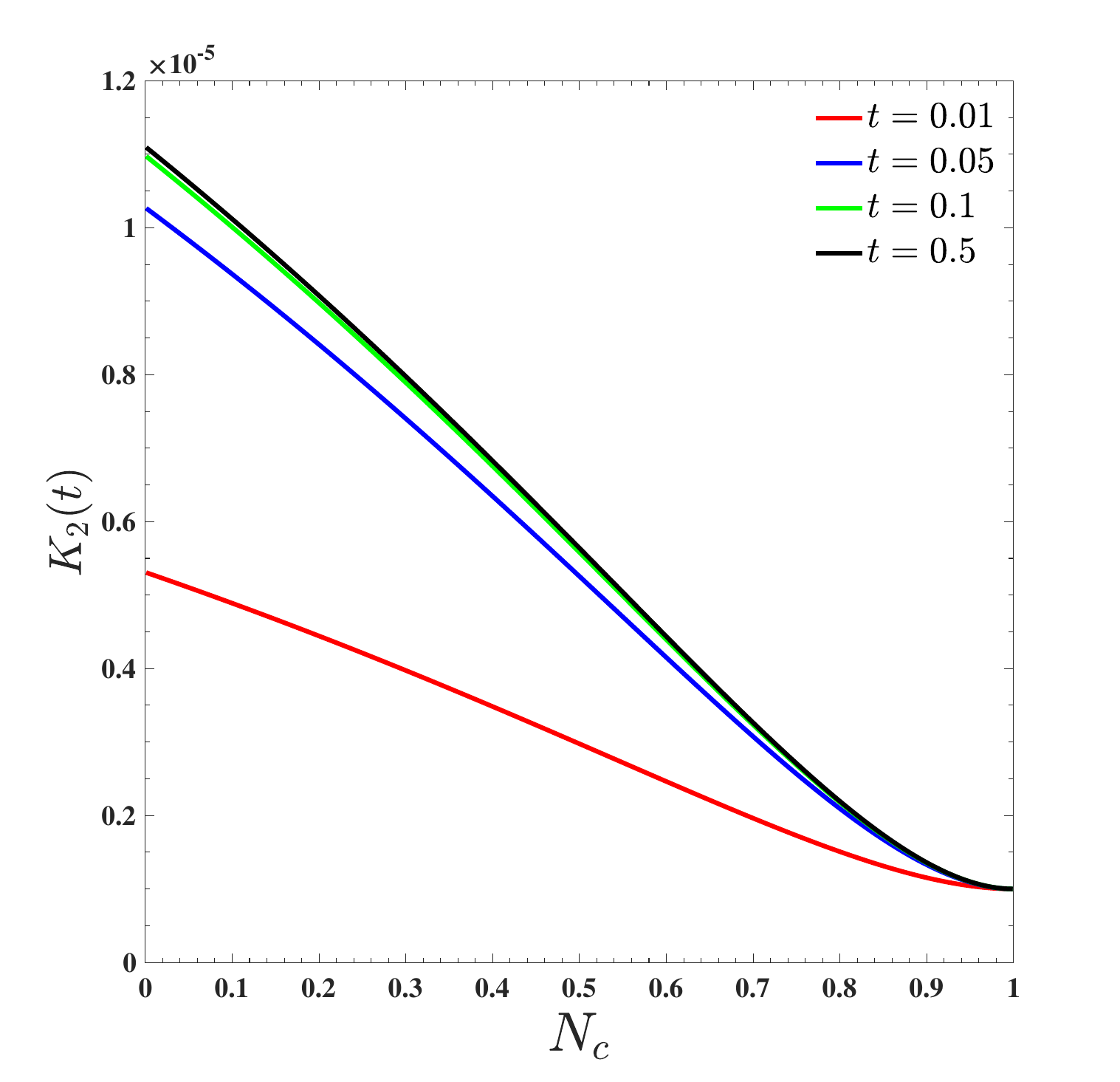}
   \caption{}
   \label{fig:10b}
 \end{subfigure}
\caption{Variation of dispersion coefficient $K_2$ with coupling number $N_c$ for different times for  (a) axial velocity and (b) micro-rotational velocity with $m=10$, $\lambda_e = 1.2$, $\gamma=0.5$, $\lambda=0.5$, $M=0.5$, $n_2=4$, $\xi=0.5$, $n_1=0.5$, $\beta = 0.01$, $P=-1$ and $Pe = 1000$.}
   \label{fig:10}
\end{figure}

\begin{figure}
\centering
   \includegraphics[width=0.49\linewidth]{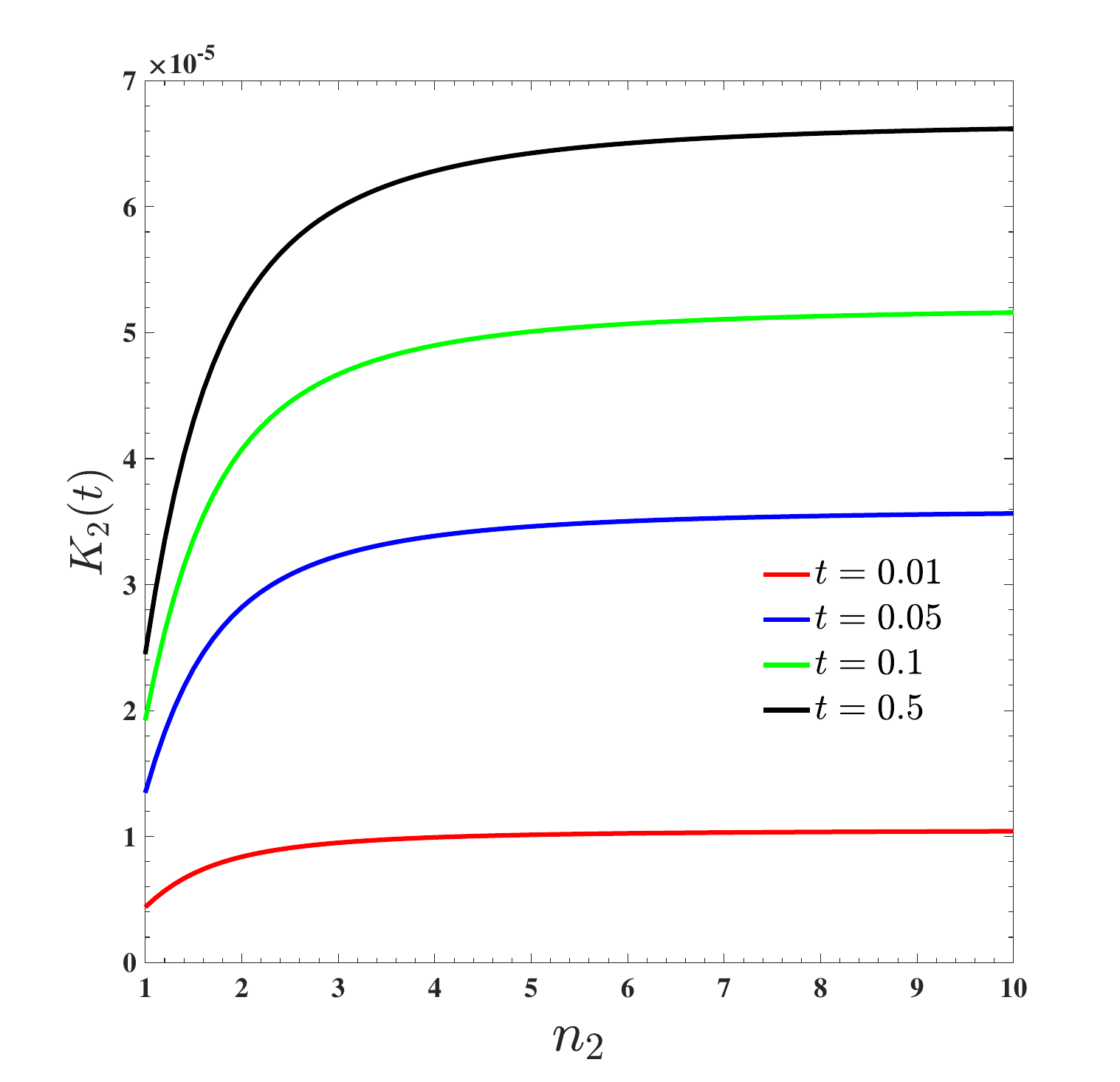}
\caption{Variation of dispersion coefficient $K_2$ for resistivity parameter $n_2$ of the Darcy porous region with time for axial velocity with $m=10$, $\lambda_e = 1.2$, $\gamma=0.5$, $\lambda=0.5$, $M=0.5$, $N_c=0.1$, $\xi=0.5$, $n_1=1$, $\beta = 0.01$, $P=-1$ and $Pe = 1000$.}
   \label{fig:11}
\end{figure}

The effect on $K_2(t)$ for various values of time $t$ with Hartmann number $M$ is shown in figure \ref{fig:8}. The graphs show an initial rise at $M \approx 0$, drop to near zero as $M$ increases (see figure \ref{fig:8a}). This reflects the transition from weak to strong electromagnetic damping of fluid motion. At small $M$, magnetic effects enhance mixing, increasing the dispersion coefficient, but beyond that, at a higher magnitude, the Lorentz force dominates and suppresses both flow and dispersion. The dispersion coefficient decreases as the Hartmann number increases. Then, as it moves past approximately $M \approx 3$, the curves for all time instances approach a steady, near-zero value $(10^{-6})$, showing that the system transitions to a magnetically dominated regime where further increases in field strength produce negligible change in solute dispersion, signifying steady-state suppression of mixing.

In case of microrotational velocity, $K_2(t)$ also exhibits a similar behaviour (see figure \ref{fig:8b}) with respect to the Hartmann number $M$. An initial maximum is observed at $M \approx 0$, followed by a sharp decline in $K_2(t)$ for larger $M$. This behaviour closely parallels the axial velocity section, with Lorentz force first facilitating mixing up to a critical $M$, then suppressing both axial and microrotational contributions to dispersion as electromagnetic dominance sets in. Beyond $M \approx 3.3$, further increases have only a minimal effect on $K_2(t)$, marking the magnetic steady-state regime.

Figure \ref{fig:9} illustrates the influence of micropolar parameter $m$ on solute dispersion. The $K_2(t)$ profile for microrotational velocity displays a pronounced increase as the micropolar parameter $m$ rises, mirroring the enhanced micro-particle rotation and the resultant mixing in the fluid. Initially, growth is rapid, then transitions to a less steep slope, and eventually the $K_2(t)$ curves begin to increase, reflecting a higher value at higher $m$ values for all times. This trend reveals how greater micro-rotation fosters increased dispersion through more particle transport, in contrast to the damping experienced with strong coupling or high Hartmann number.

\figurename~\ref{fig:10} demonstrates the influence on solute dispersion for different values of coupling numbers $N_c$.
The dispersion coefficient $K_2(t)$ decreases linearly with increasing coupling number $N_c$, for all time values considered.
Physically, a higher $N_c$ signifies a stronger coupling between the axial fluid velocity and the microrotational velocity in a micropolar fluid, leading to suppression of solute dispersion due to reduced velocity gradients. This effect diminishes the fluid's ability to transport solute molecules axially. As $N_c$ approaches 1 (specifically at $N_c \approx 0.96$), the value of $K_2(t)$ flattens toward zero (see figure \ref{fig:10a}), indicating that the system reaches a near steady state with minimal further mixing or dispersion as coupling saturates.

Similar to the axial velocity case, the dispersion coefficient $K_2(t)$ associated with the microrotational velocity decreases with increasing coupling number $N_c$ for all considered $t$. The coupling between the axial and micro-rotational motions acts to suppress variation in both flow components, reducing overall solute mixing as $N_c$ increases. As $N_c$ exceeds $0.98$, the curves approach a steady regime (see figure \ref{fig:10b}), indicating saturation in the suppressive coupling effects. This consistency with the axial velocity analysis underlines the fundamental role of $N_c$ in damping dispersion via coupled microstructural and fluid phenomena. 

The impact of the resistivity parameter $n_2$ on $K_2$ for the axial velocity is presented in Figure \ref{fig:11}. We can see that, $K_2(t)$ increases rapidly with $n_2$ for small values at different times, but after a certain point (typically beyond $n_2 \approx 2, 3, 3.5 ~\mbox{and}~ 4$), the curves level out and reach an asymptotic steady-state value $1.04\times10^{-5}, 3.56\times10^{-5}, 5.16\times10^{-5} ~\mbox{and}~ 6.62\times10^{-5}$ for $t = 0.01, 0.05, 0.1 ~\mbox{and}~ 0.5$ respectively. This behaviour represents the effect of increasing resistance from the porous medium: initially, higher resistivity amplifies dispersion through local mixing. When the resistivity becomes very large, further increases cease to affect fluid transport, and the dispersion coefficient remains constant, resulting in a diffusion-limited steady state. Thus, for large $n_2$, the system becomes insensitive to further changes in resistivity.

\subsection{Mean Concentration $C_m$}
The mean concentration $C_m(t,z)$ depends on the transport coefficients $K_0(t),K_1(t)$ and $K_2(t)$. By using these transport coefficients in equations \eqref{eq:3.25}- \eqref{eq:3.27}, we have solved equation (\ref{eq:3.24}) to get the expression for $C_m$. In this study, the axial distribution of $C_m$ is analysed with the effect of the absorption parameter $(\beta)$, the Hartmann number $(M)$, the coupling number $(N_c)$, the resistivity parameter $n_2$, and the micropolar parameter $(m)$.

\begin{figure}
	\centering
	\begin{subfigure}{0.49\linewidth}
		\centering
		\includegraphics[width=\linewidth]{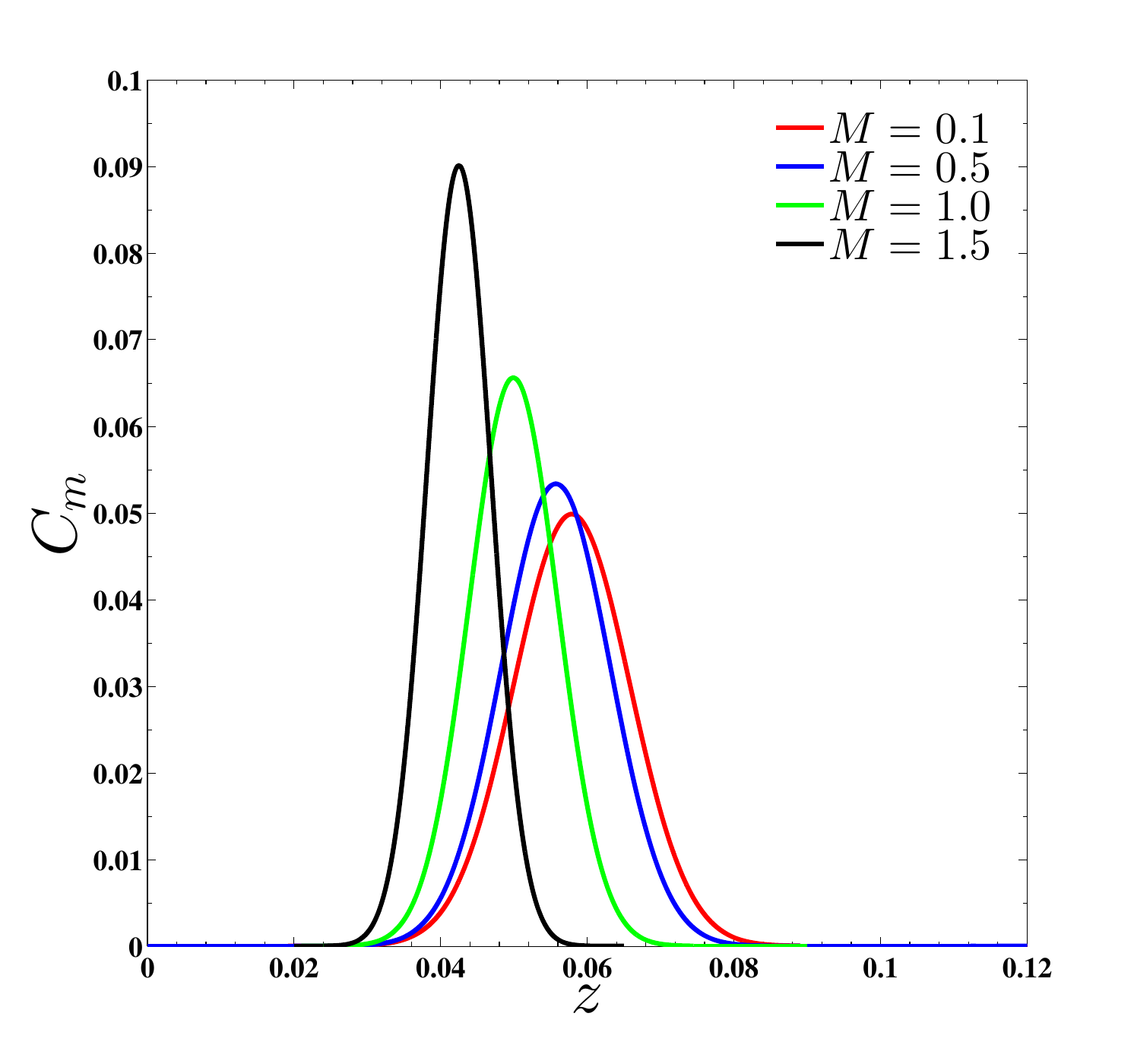}
		\caption{}
		\label{fig:12a}
	\end{subfigure}
		\begin{subfigure}{0.49\linewidth}
		\centering
		\includegraphics[width=\linewidth]{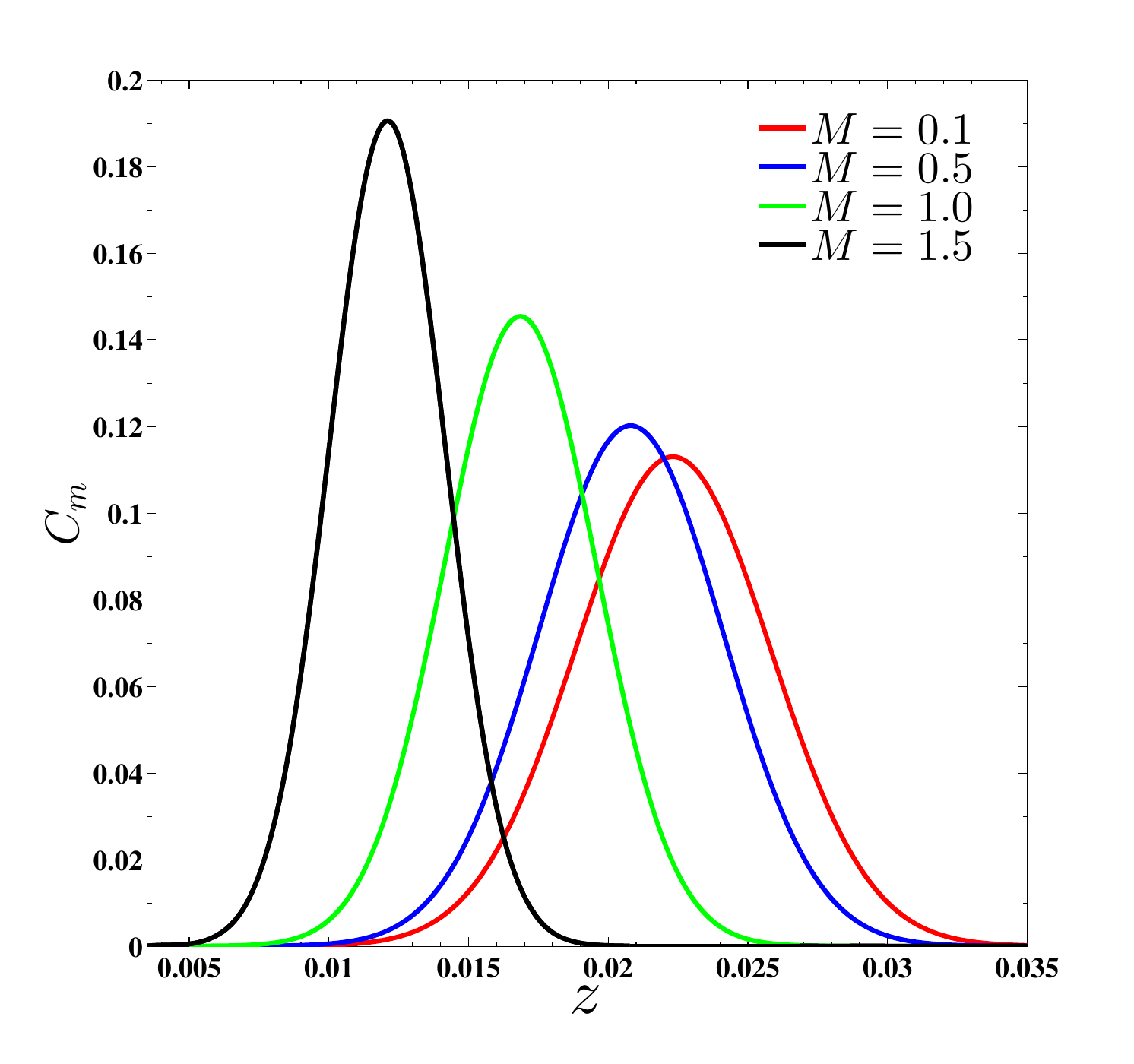}
		\caption{}
		\label{fig:12b}
	\end{subfigure}
	
\caption{ Variation of axial mean concentration $C_m(t,z)$ for Hartmann number $M$ with axial distance $z$ for (a) axial and (b) micro-rotational velocity with $t=0.5$, $\beta=0.01$, $m=10$, $\gamma=0.5$, $\lambda=0.5$, $N_c=0.1$, $n_1=0.5$, $n_2=4$, $\phi=0.69$, $\xi=0.5$, $\lambda_e=1.2$, $P=-1$ and $Pe = 1000$.}
\label{fig:12}
\end{figure}

Figure \ref{fig:12} displays the effect of $M$ on the mean concentration of the solute. As we increase the value of $M$, the peak of the mean concentration $C_m$ increases significantly. As we discussed earlier, the fluid velocity drops with increasing $M$. A reduction in fluid velocity results in less solute movement, thereby decreasing solute convection. Less convection results in a reduction in solute dispersion, which implies that the solute is dispersed through the fluid less effectively. This reduction in both convection and dispersion causes solute to accumulate in certain regions, leading to an increase in the mean concentration of solute in those areas. On the other hand, in the case of the micro-rotational velocity, a similar kind of behaviour is observed. From figure \ref{fig:12}, the peak of $C_m$ is much greater in the case of the micro-rotational velocity than that for axial velocity. However, the axial spreading of concentration is more downstream for the axial velocity than the micro-rotational velocity.

\begin{figure}
	\centering
		\begin{subfigure}{0.5\linewidth}
		\centering
		\includegraphics[width=\linewidth]{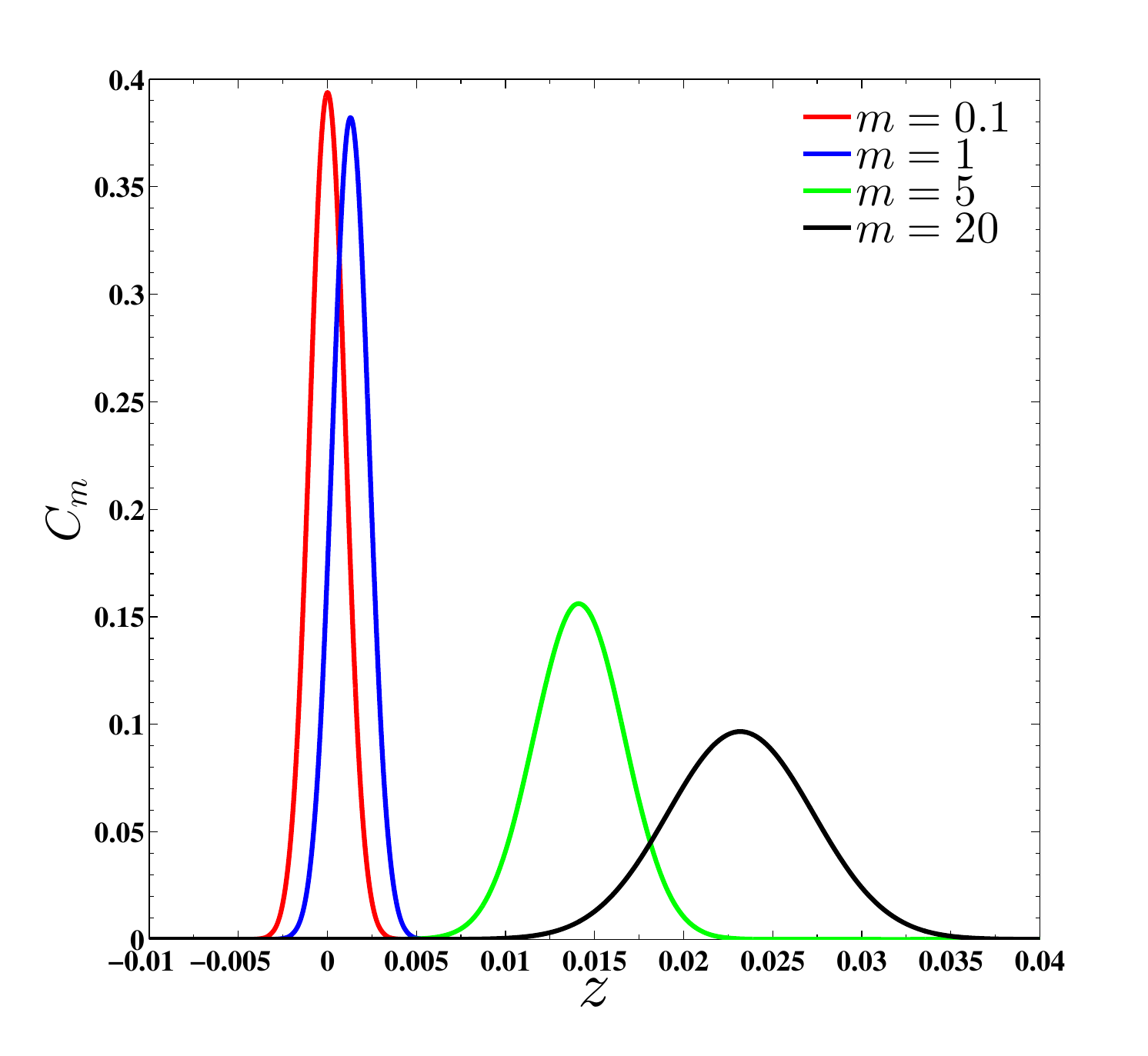}
		\caption{}
	\end{subfigure}
\caption{Variation of axial mean concentration $C_m(t,z)$ for micropolar parameter $m$ with axial distance $z$ for micro-rotational velocity with $t=0.5$, $\beta=0.01$, $M=0.5$, $\gamma=0.5$, $\lambda=0.5$, $N_c=0.1$, $n_1=0.5$, $n_2=4$, $\phi=0.69$, $\xi=0.5$, $\lambda_e=1.2$, $P=-1$ and $Pe = 1000$.}
\label{fig:13}
\end{figure}

From figure \ref{fig:13}, it is observed that as we increase the value of $m$, the mean concentration is greatly decreased by $m$ for the micro-rotation velocity. The solute is convected with the faster velocity, which leads to a more downstream spreading of the solute, resulting in a lower peak of $C_{m}$. For small $m$, both the upstream and downstream distributions are predicted, whereas it is only the downstream for large $m$.

\begin{figure}
	\centering	
		\begin{subfigure}{0.49\linewidth}
		\centering
		\includegraphics[width=\linewidth]{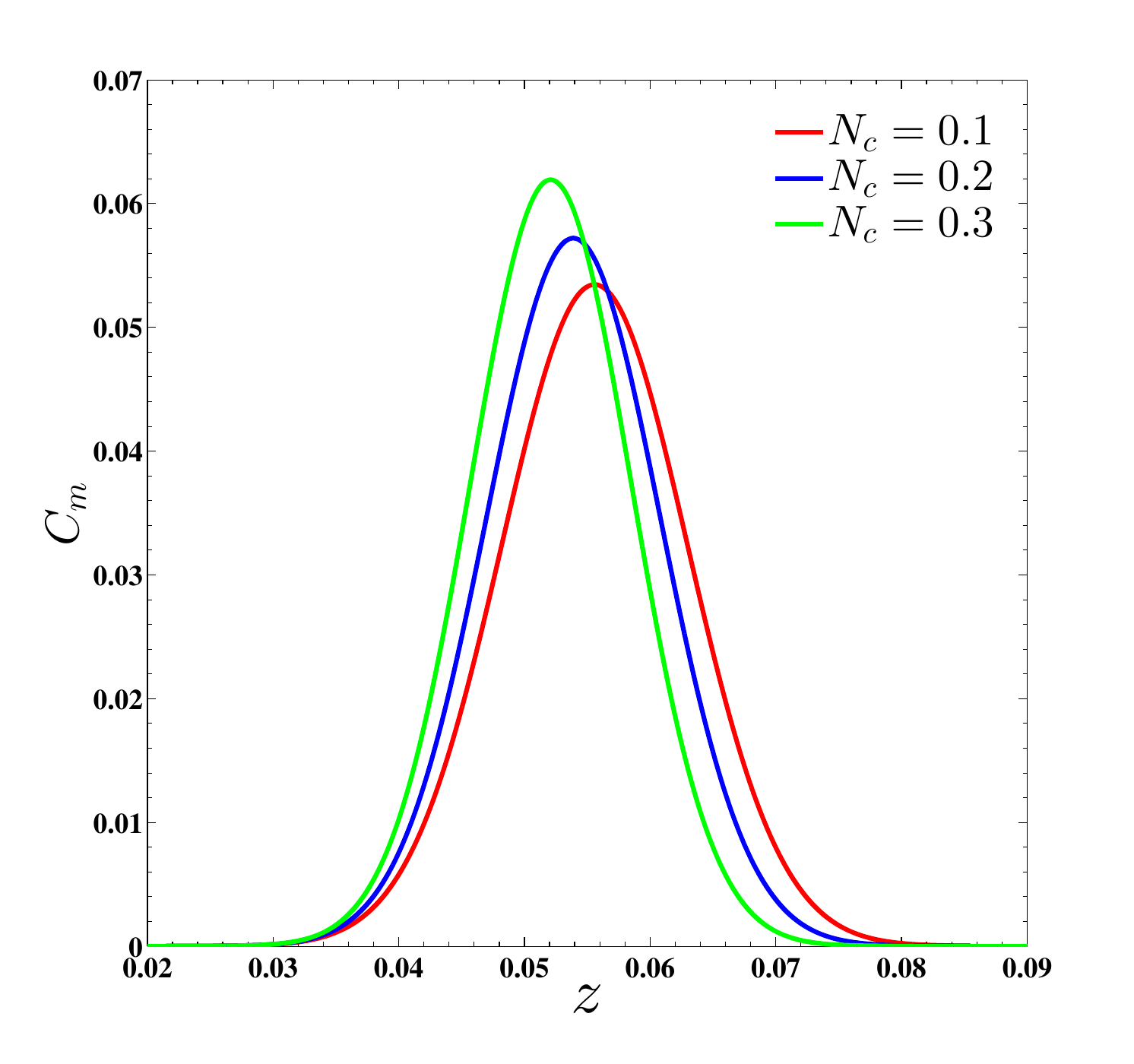}
		\caption{}
		\label{fig:14a}
	\end{subfigure}
	\begin{subfigure}{0.49\linewidth}
		\centering
		\includegraphics[width=\linewidth]{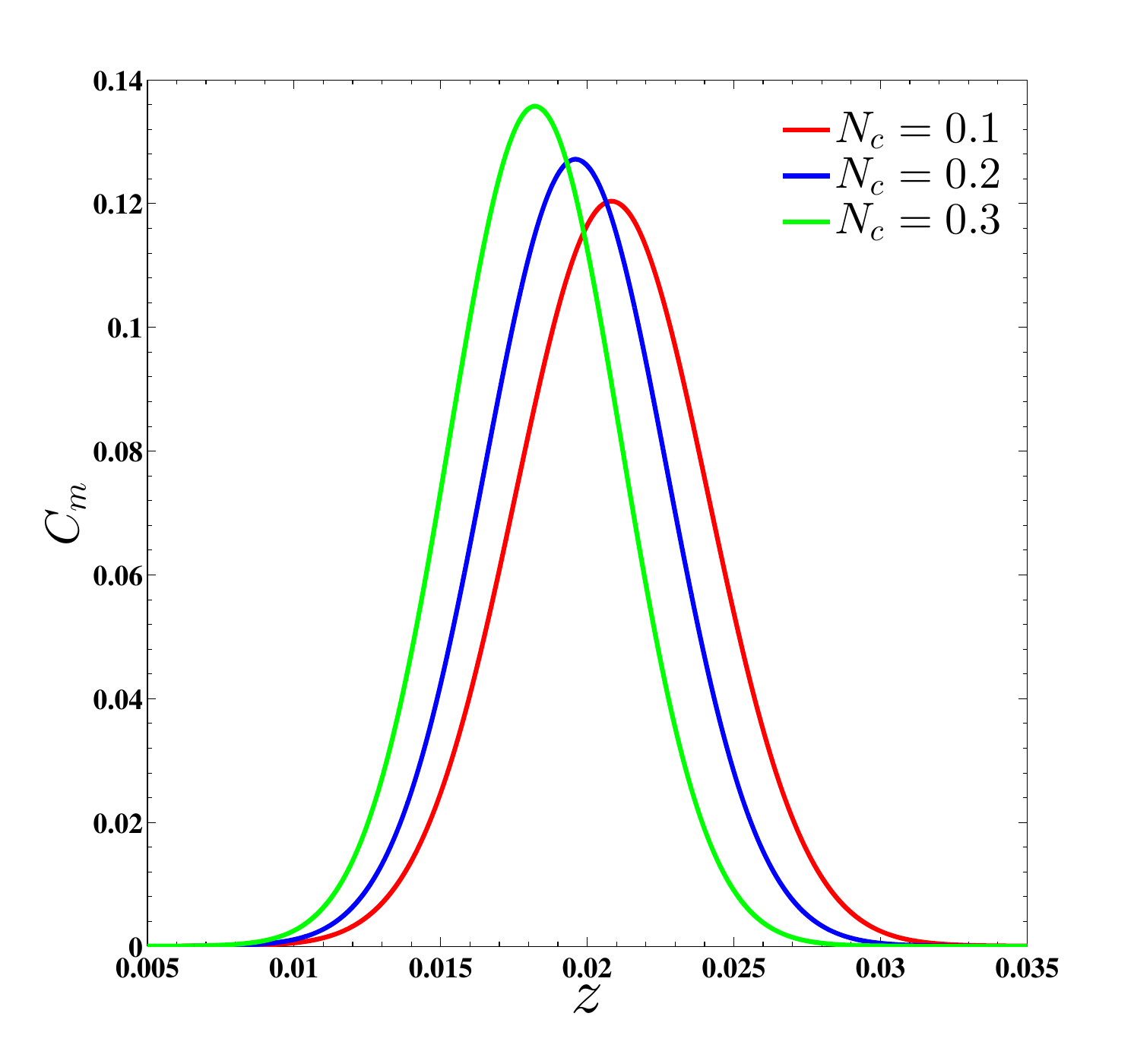}
		\caption{}
		\label{fig:14b}
	\end{subfigure}
\caption{ Variation of axial mean concentration $C_m(t,z)$ for coupling number $N_c$ with axial distance $z$ for (a) axial and (b) micro-rotational velocity with $t=0.5$, $\beta=0.01$, $m=10$, $\gamma=0.5$, $\lambda=0.5$, $M=0.5$, $n_1=0.5$, $n_2=4$, $\phi=0.69$, $\xi=0.5$, $\lambda_e=1.2$, $P=-1$ and $Pe = 1000$.}
\label{fig:14}
\end{figure}

The axial distribution profiles of $C_{m}$ for select values of $N_c$ have been shown in Figure \ref{fig:14}. As $N_c$ increases, the interaction between the fluid's translational and rotational motions becomes stronger. As a result, the fluid's capacity to flow freely is restricted by this increased coupling, which lowers the axial as well as the micro-rotational velocity. Therefore, the solute is convected with the lower velocity of the fluid, which causes a reduction in the dispersion. Therefore, the solute is not evenly distributed throughout the flow, and it tends to concentrate more in specific regions. Because of the more concentrated solute accumulation carried by the decreased dispersion, the mean concentration of the solute rises.

\begin{figure}
	\centering	
		\includegraphics[width=0.49\linewidth]{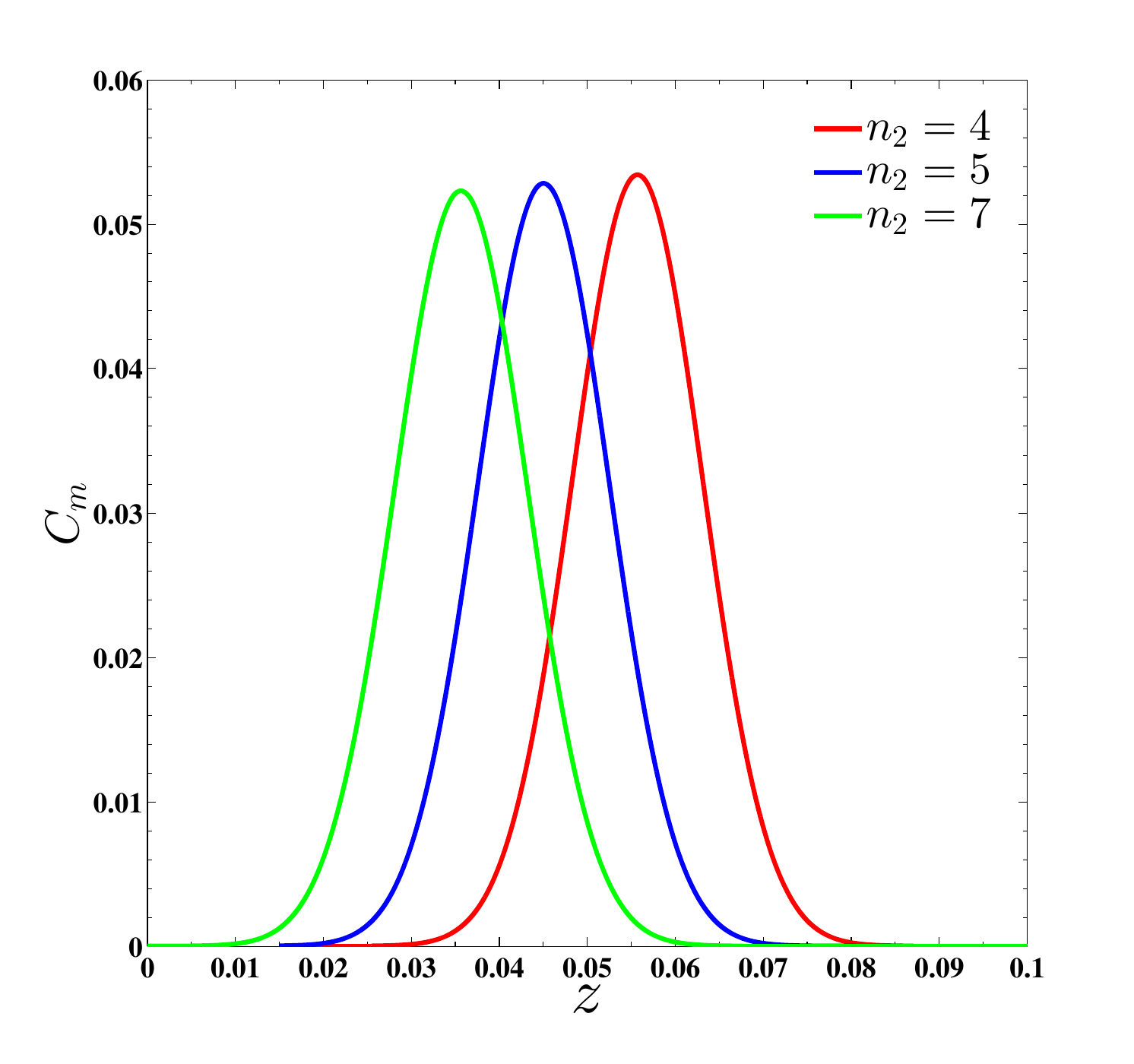}
	\caption{ Variation of axial mean concentration $C_m(t,z)$ for resistivity parameter of Darcy porous region $n_2$ with axial distance $z$ for axial velocity with $t=0.5$, $\beta=0.01$, $m=10$, $\gamma=0.5$, $\lambda=0.5$, $N_c=0.1$, $n_1=1$, $M=0.5$, $\phi=0.69$, $\xi=0.5$, $\lambda_e=1.2$, $P=-1$ and $Pe = 1000$.}
	\label{fig:15}
\end{figure}

Figure \ref{fig:15} demonstrates the effect of the resistivity parameter $n_2$ of the Darcy porous region on $C_m$ along the axial distance $z$. From figure \ref{fig:15}, as $n_2$ increases, solute concentration is less distributed downstream. The physical reason is explained below. A rise in the resistivity parameter in the Darcy porous region slows down fluid velocity, which in turn reduces solute convection and dispersion, resulting in less spreading of the mean solute concentration. These results can be physically interpreted in the context of blood flow through vascular structures. Resistance to blood flow arises due to the presence of the endothelial and glycocalyx layer along the inner lining of the blood vessels. This resistance decreases the downstream distribution of the solute (e.g. drugs, nutrients) in the blood flow and controls the transport of the solute across the blood vessel wall.

\subsection{Spatial concentration distribution}
The local concentration of solute $C(t,r,z)$, which depends on all transport coefficients $K_0(t), K_1(t)$ and $K_2(t)$, is analysed to interpret both longitudinal and transverse spreading of the solute under the influence of the absorption parameter $(\beta)$, the Hartmann number $(M)$, the coupling number $(N_c)$, the resistivity parameter $n_2$, and the micropolar parameter $(m)$.

\begin{figure}
	\centering
	\begin{subfigure}{0.4\linewidth}
		\raggedright
        \text{(a)}
		\includegraphics[width=\linewidth]{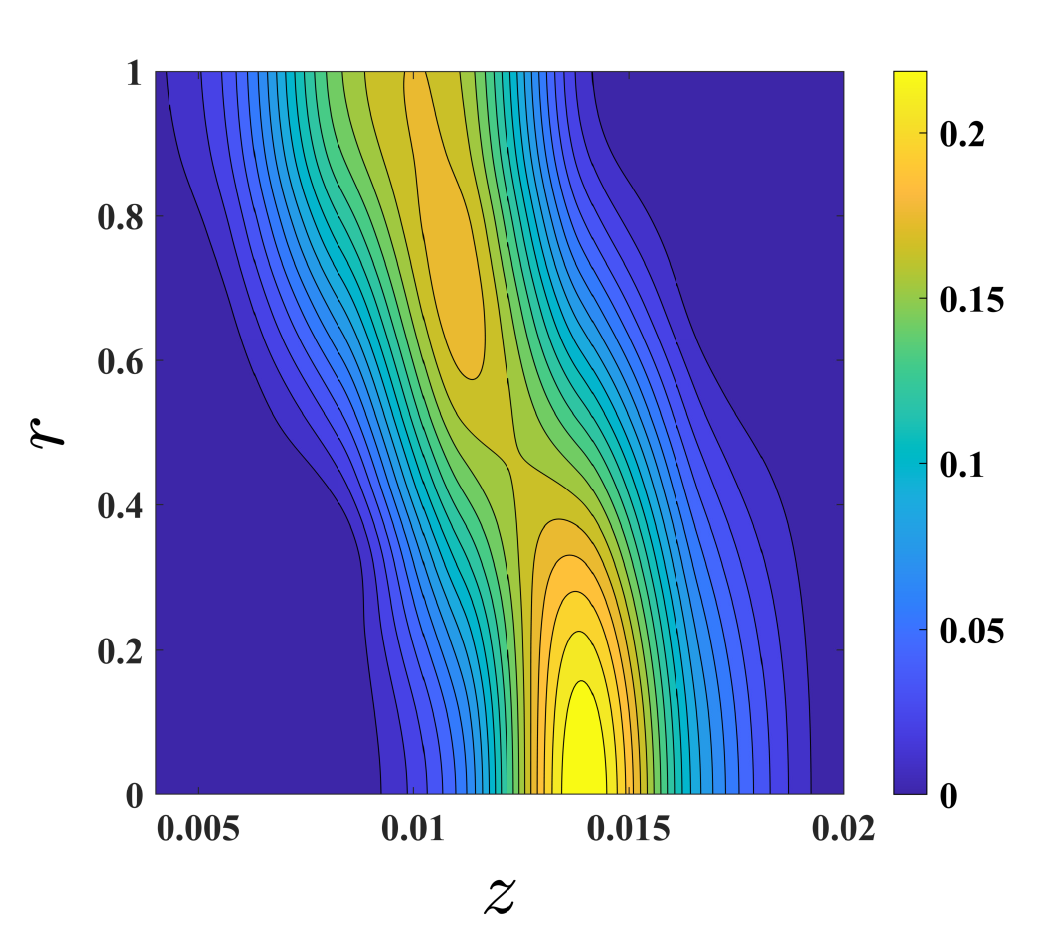}
		$t=0.1$
		\label{fig:16a}
	\end{subfigure}\vspace{1em}
		\begin{subfigure}{0.4\linewidth}
		\raggedright
        \text{(b)}
		\includegraphics[width=\linewidth]{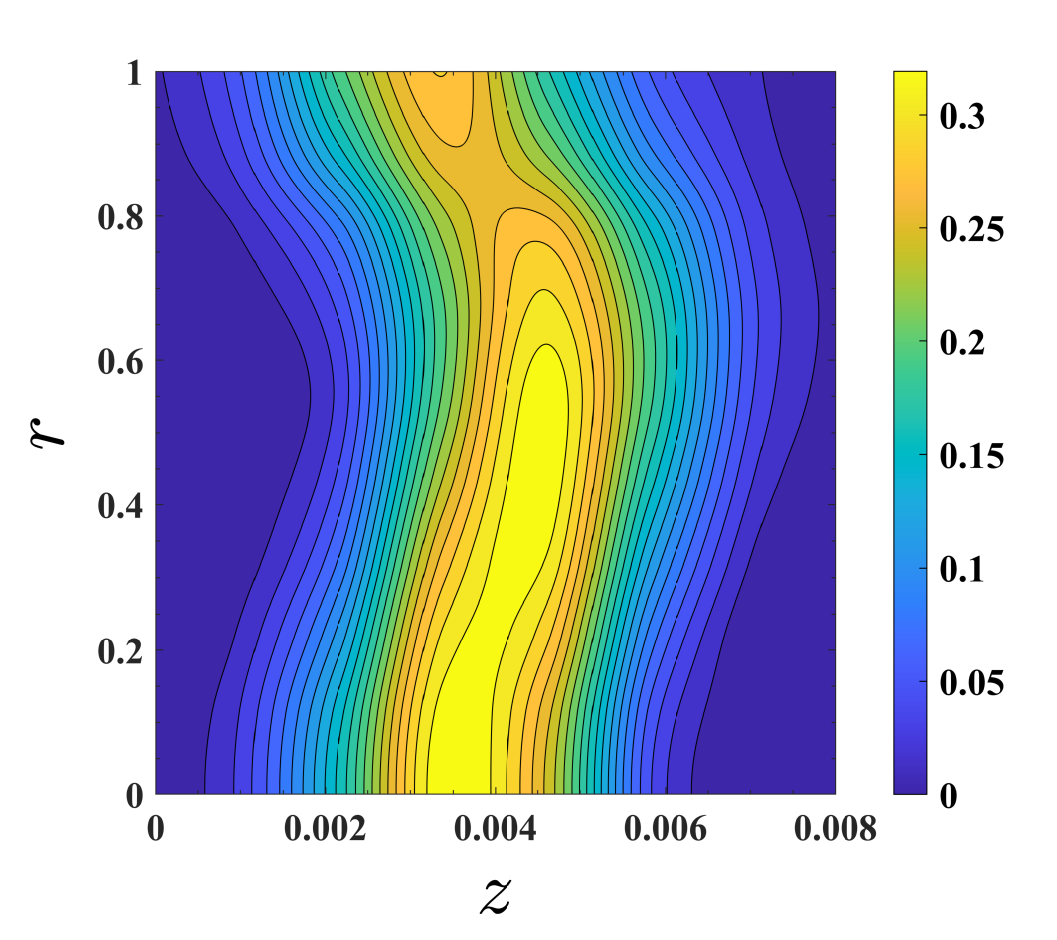}
		\label{fig:16b}
	\end{subfigure}
    \begin{subfigure}{0.4\linewidth}
    	\raggedright
        \text{(c)}
		\includegraphics[width=\linewidth]{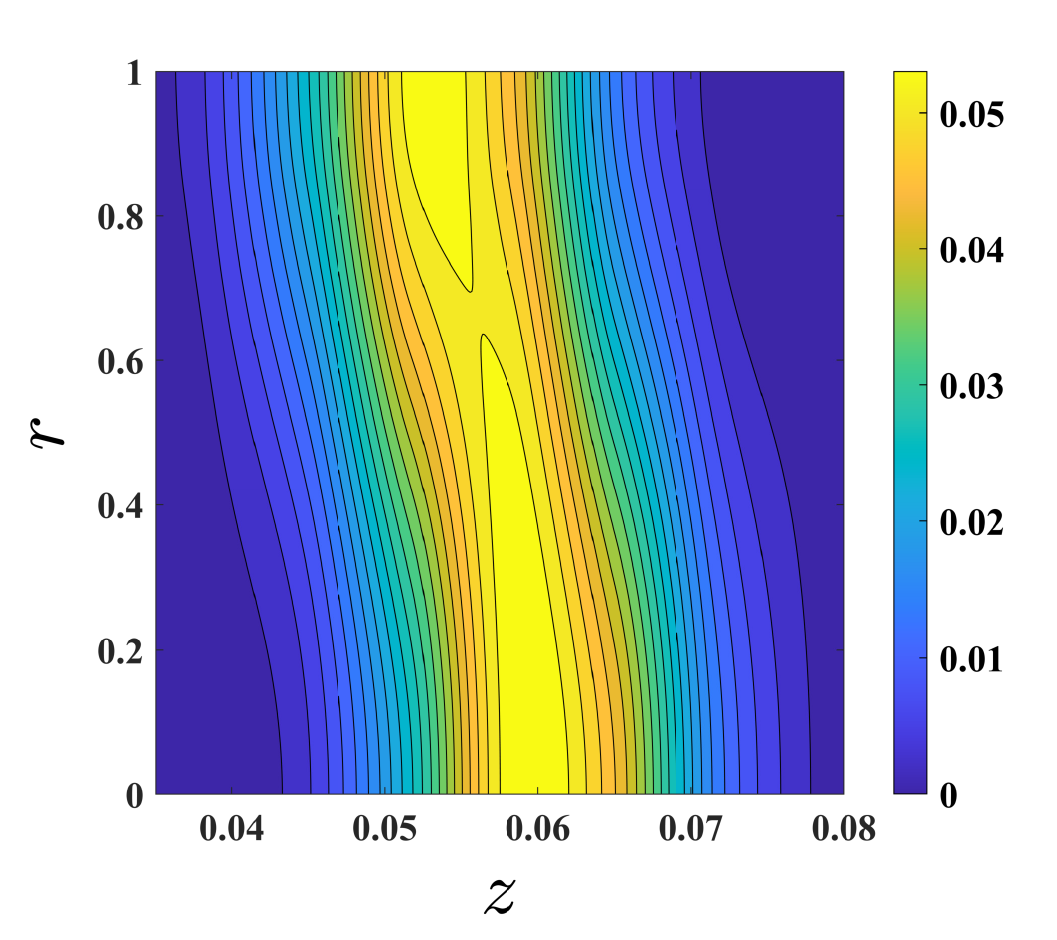}
		$t=0.5$
		\label{fig:16c}
	  \end{subfigure}\vspace{1em}
       \begin{subfigure}{0.4\linewidth}
		\raggedright
        \text{(d)}
		\includegraphics[width=\linewidth]{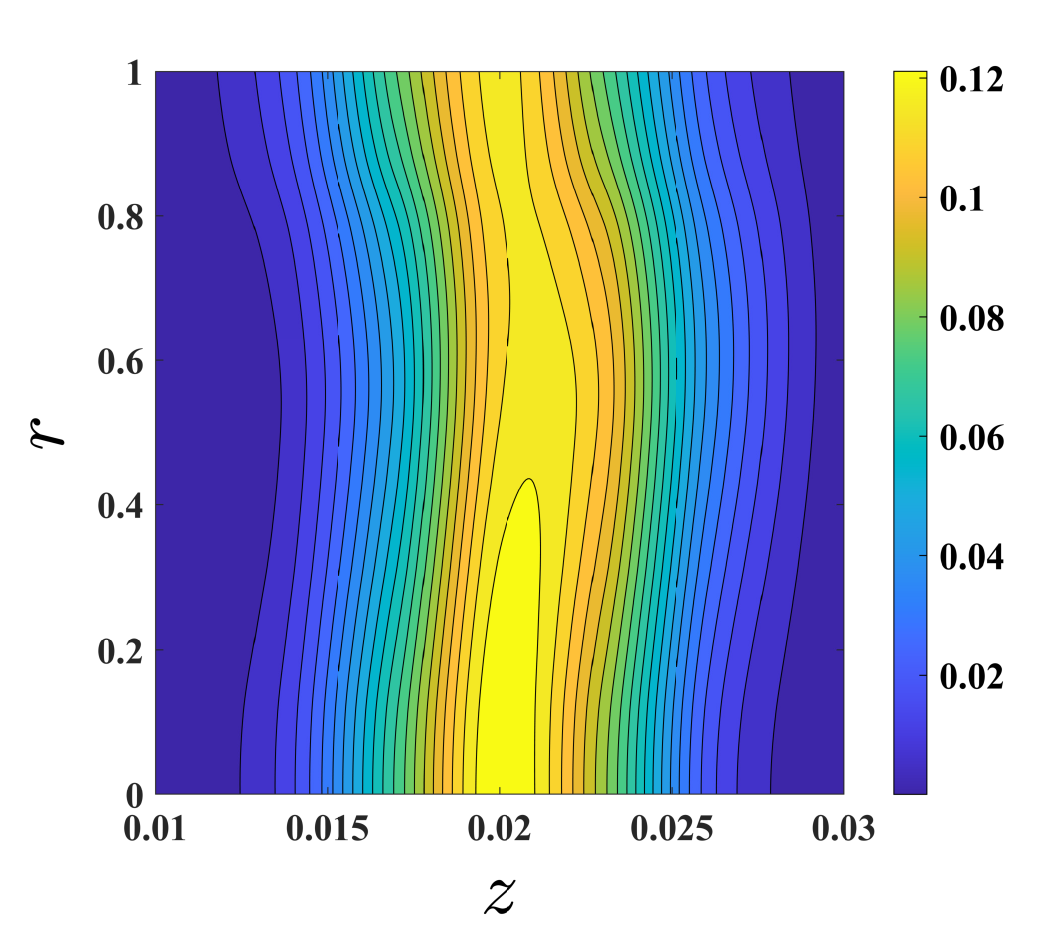}
		\label{fig:16d}
	  \end{subfigure}
       \begin{subfigure}{0.4\linewidth}
       	\raggedright
        \text{(e)}
		\includegraphics[width=\linewidth]{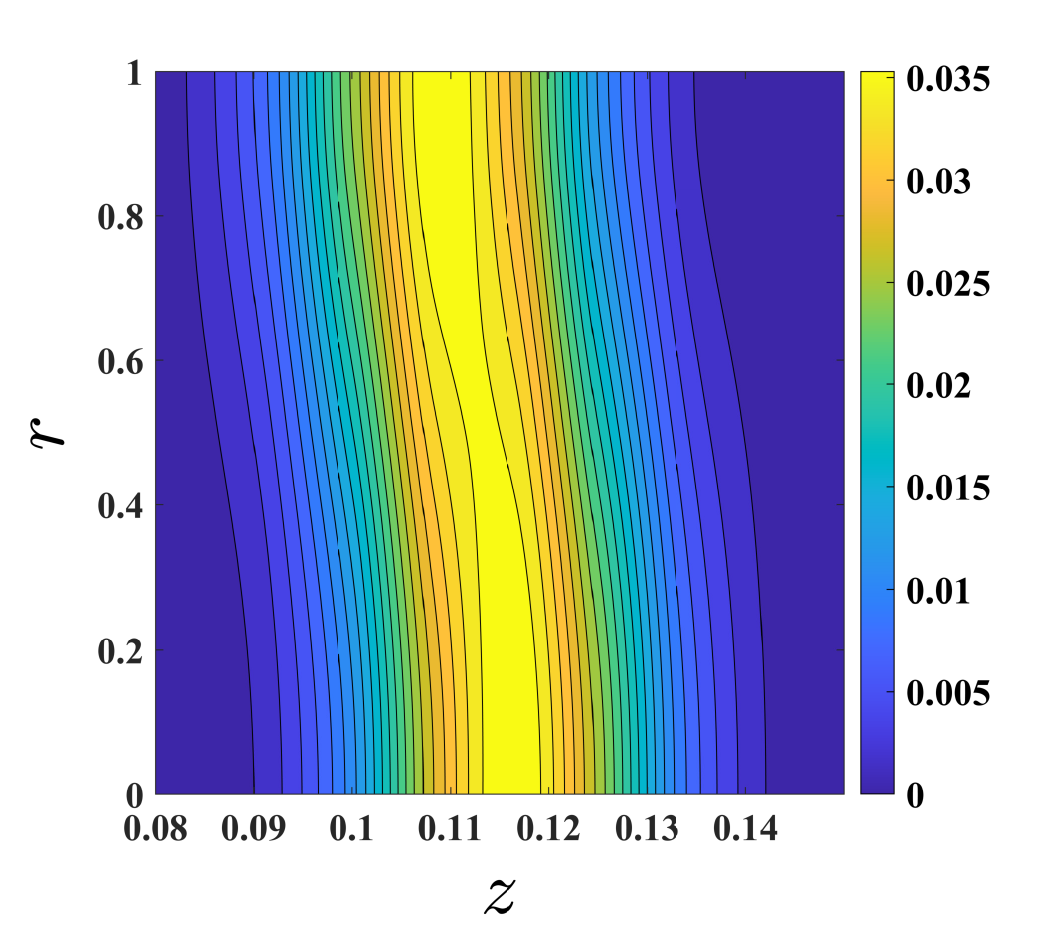}
		$t=1$
		\label{fig:16e}
	  \end{subfigure}
     \begin{subfigure}{0.4\linewidth}
		\raggedright
        \text{(f)}
		\includegraphics[width=\linewidth]{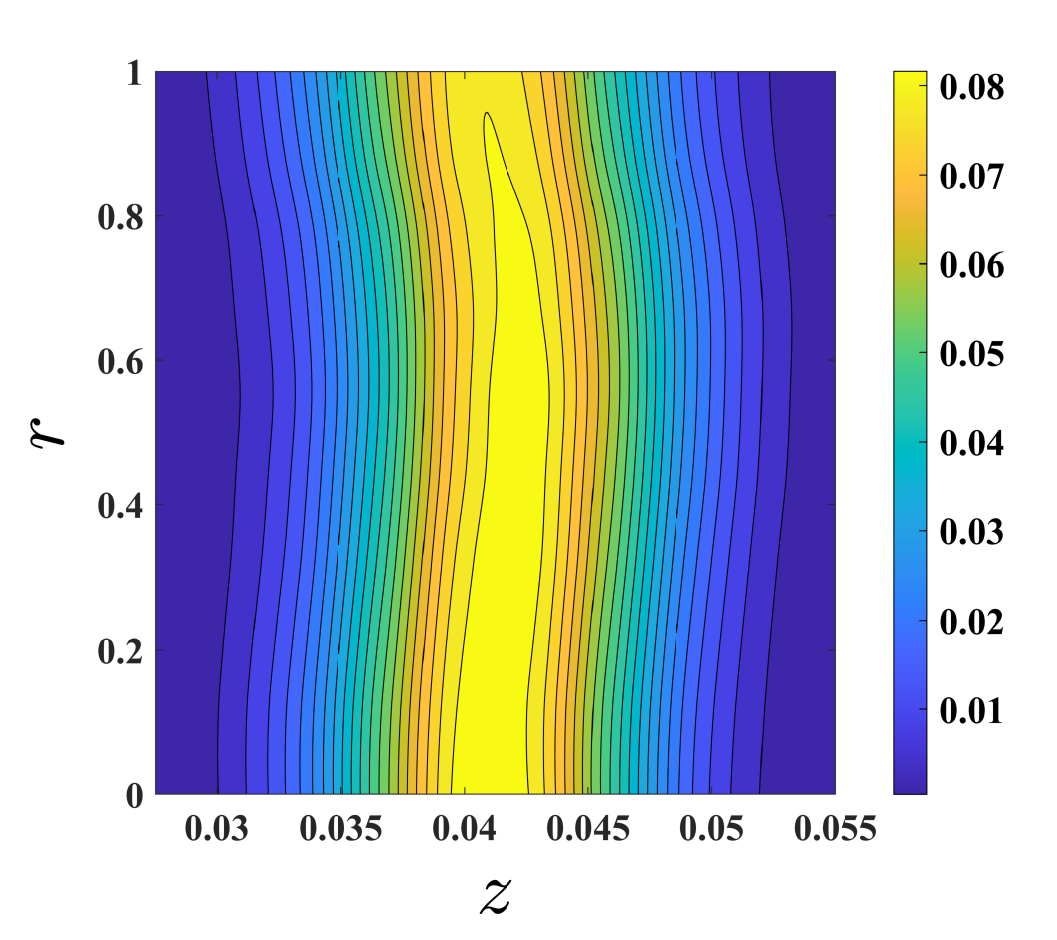}
		\label{fig:16f}
	  \end{subfigure}
\caption{ Time evolution of spatial concentration $C$ for axial (a,c,e) and micro-rotational velocity (b,d,f) with $\beta = 0.01$, $m=10$, $\gamma=0.5$, $\lambda=0.5$, $N_c=0.1$, $n_1=0.5$, $n_2 = 4$, $M=0.5$, $\phi=0.69$, $\xi=0.5$, $\lambda_e=1.2$, $P=-1$ and $Pe = 1000$.}
\label{fig:16}
\end{figure}

Figure \ref{fig:16} depicts the solute concentration distribution in a tube at times $t = 0.1$, $t = 0.5$, and $t = 1$, for both axial and micro-rotational velocity, capturing the distribution in both the axial and radial directions. The solute's spreading is regulated by convection in the axial direction and diffusion in the radial direction. 

Additionally, in the radial direction, solute absorption at the boundary plays a significant supportive role in altering the concentration distribution over time. The absorption parameter $\beta$ reflects the rate at which the solute is removed from the system at $r = 1$. Although this is a tiny amount for a slow rate of absorption, it causes a notable decrease in concentration along the tube wall. Physically, this implies that even weak absorption can create a concentration gradient in the radial direction, leading to lower solute retention near the boundary.

For the axial velocity, the solute cloud moves faster in the axial direction and diffuses well in the fluid. On the other hand, in the case of micro-rotational velocity, due to backflow, the concentration gradient becomes significantly larger near the tube wall at small times. But as we increase the value of $t$, the concentration of solute mixes quite slowly compared to the axial velocity. From figure \ref{fig:16}, for axial velocity, the concentration distribution becomes more downstream compared to that for micro-rotational velocity.

\begin{figure}
	\centering
	\begin{subfigure}{0.4\linewidth}
		\raggedright
        \text{(a)}
		\includegraphics[width=\linewidth]{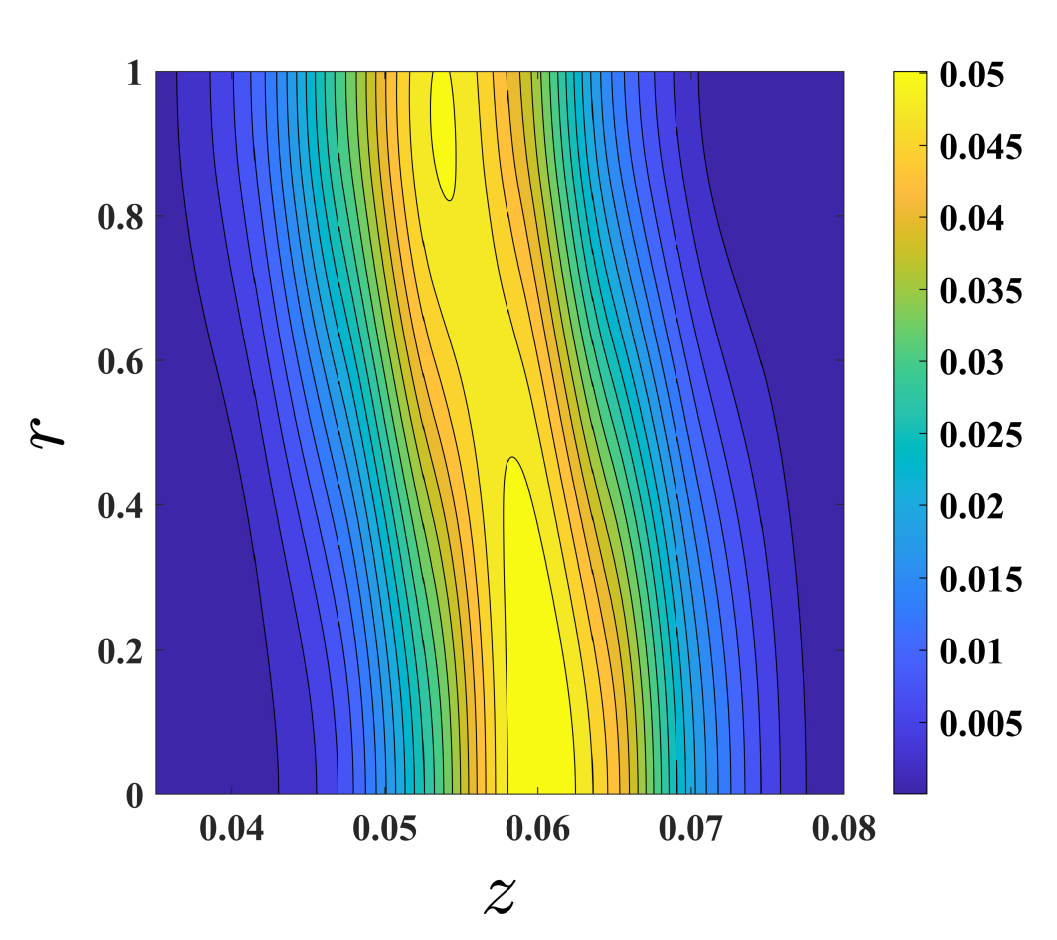}
		$\beta=0.1$
		\label{fig:17a}
	\end{subfigure}\vspace{1em}
		\begin{subfigure}{0.4\linewidth}
		\raggedright
        \text{(b)}
		\includegraphics[width=\linewidth]{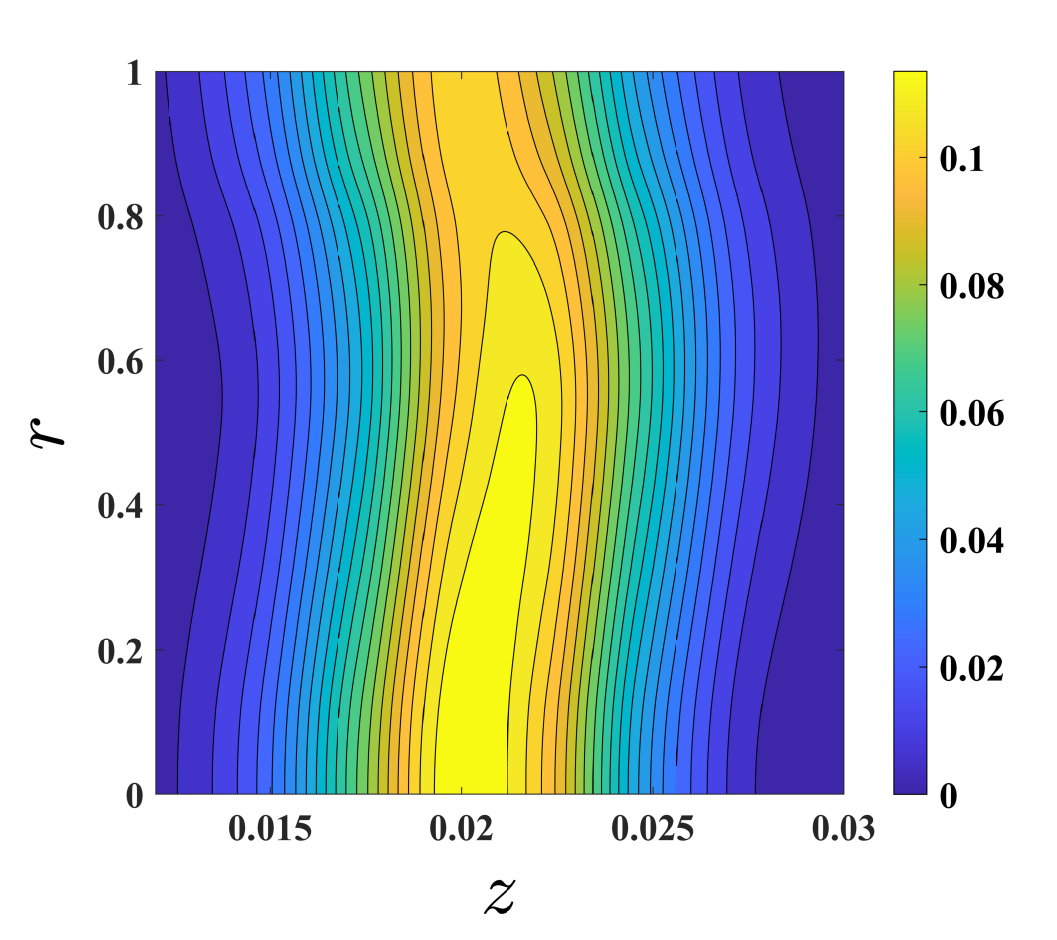}
		\label{fig:17b}
	\end{subfigure}
    \begin{subfigure}{0.4\linewidth}
		\raggedright
        \text{(c)}
		\includegraphics[width=\linewidth]{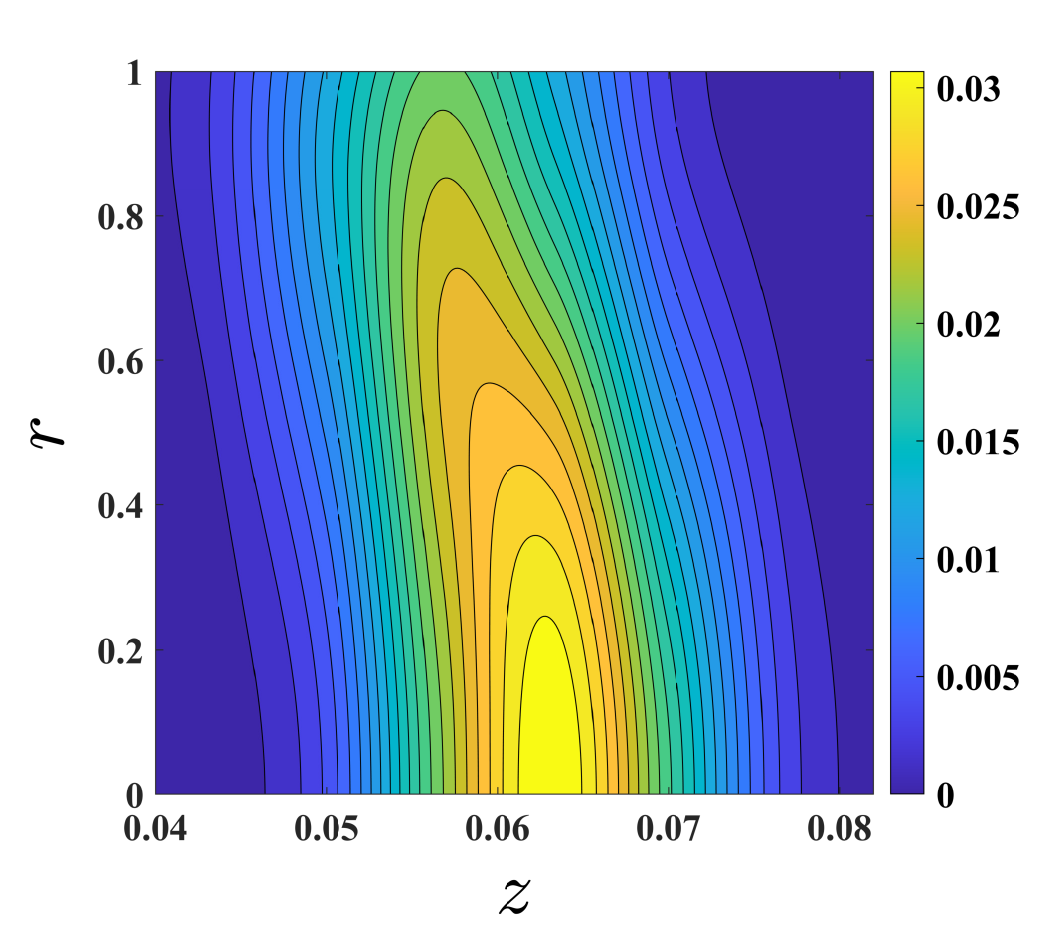}
		$\beta=1$
		\label{fig:17c}
	\end{subfigure}\vspace{1em}
    \begin{subfigure}{0.4\linewidth}
		\raggedright
        \text{(d)}
		\includegraphics[width=\linewidth]{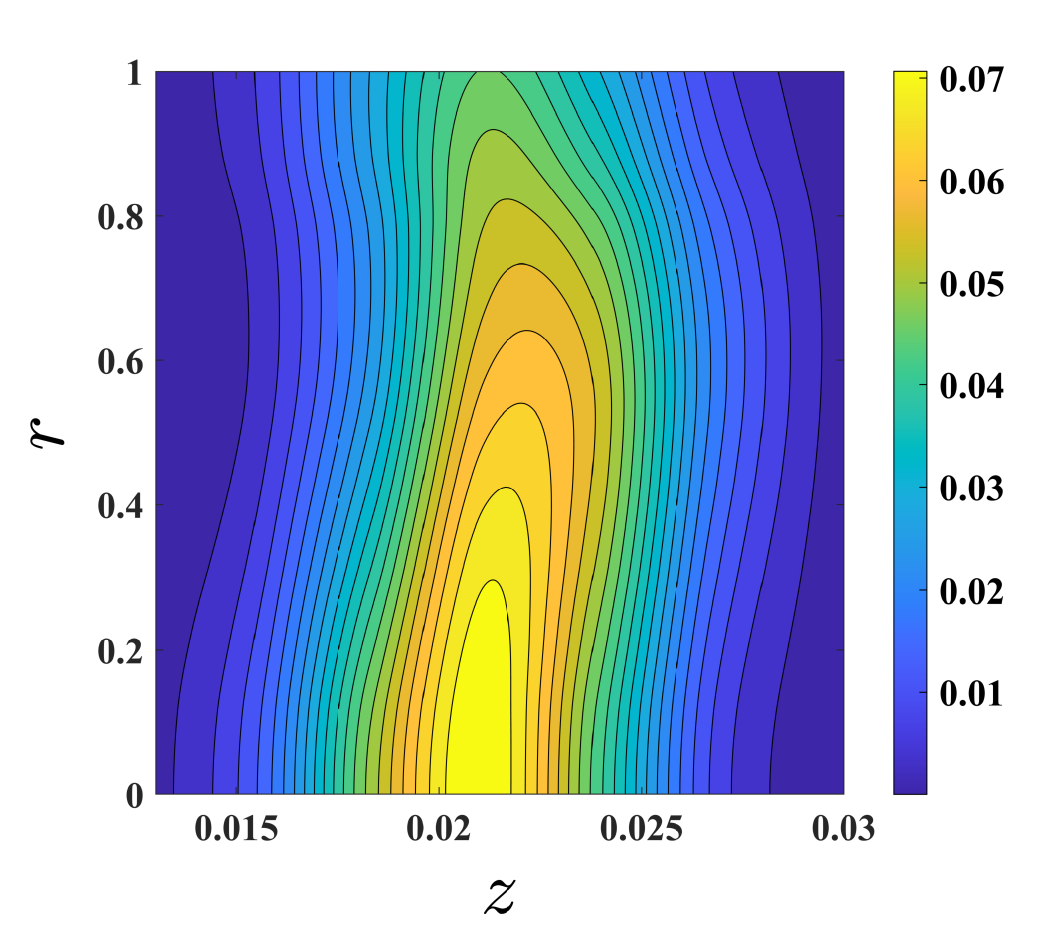}
		\label{fig:17d}
	\end{subfigure}
    \begin{subfigure}{0.4\linewidth}
	    \raggedright
        \text{(e)}
		\includegraphics[width=\linewidth]{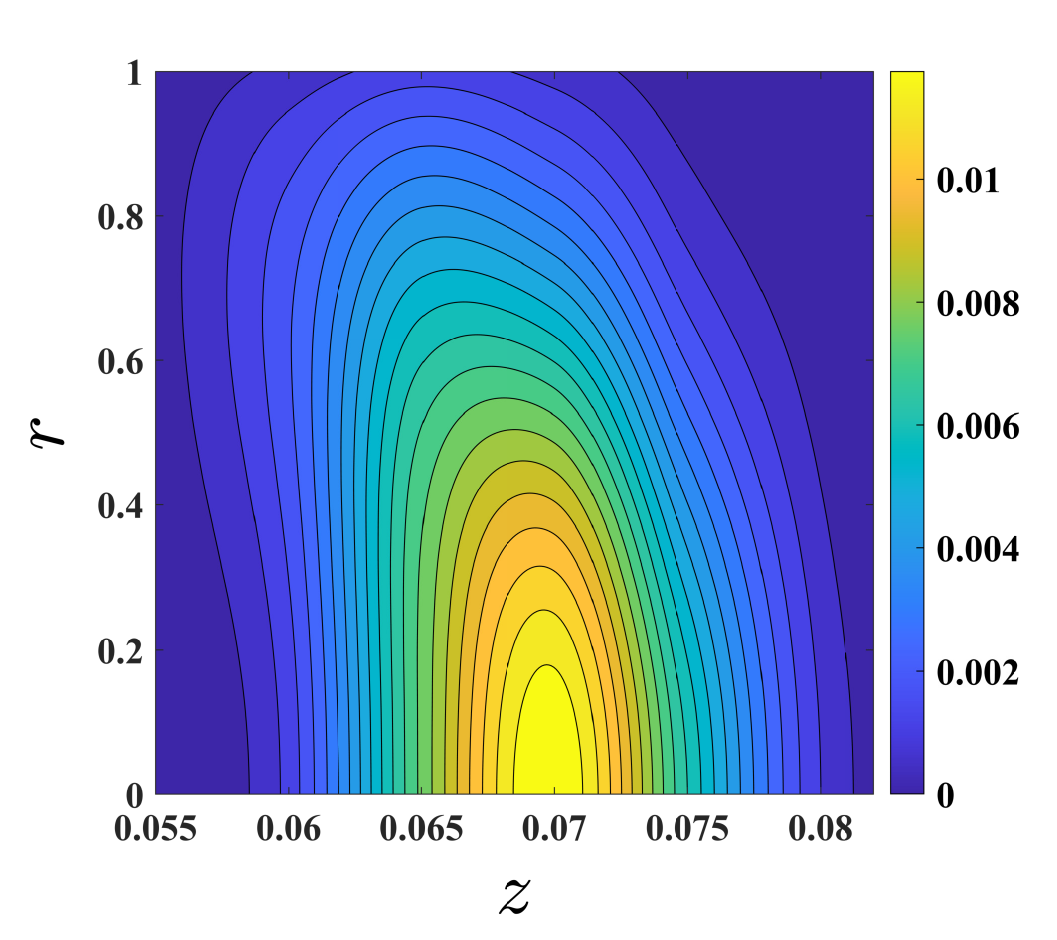}
		$\beta=10$
		\label{fig:17e}
	\end{subfigure}
    \begin{subfigure}{0.4\linewidth}
		\raggedright
        \text{(f)}
		\includegraphics[width=\linewidth]{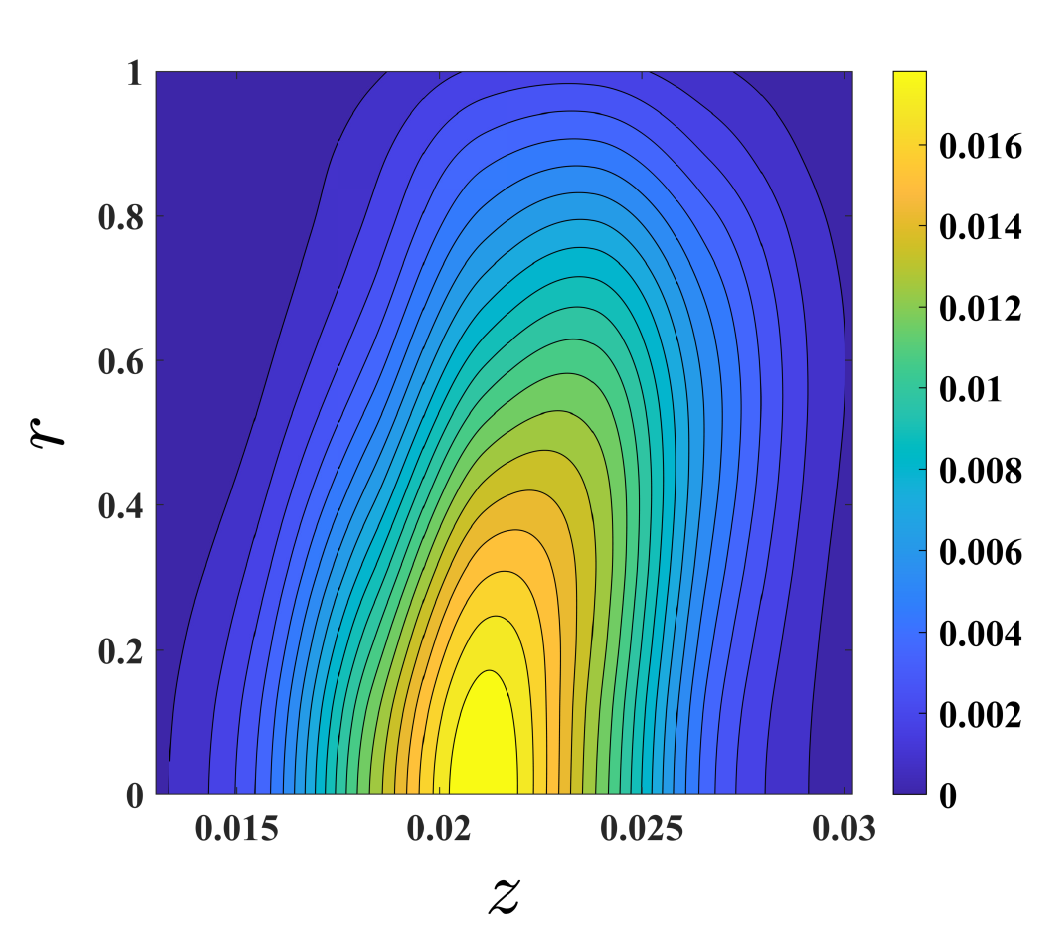}
		\label{fig:17f}
	  \end{subfigure}
	
\caption{ Variation of spatial concentration $C$ for different values of absorption parameter $\beta$ for axial (a,c,e) and micro-rotational velocity (b,d,f) with $t = 0.5$, $m=10$, $\gamma=0.5$, $\lambda=0.5$, $N_c=0.1$, $n_1=0.5$, $n_2 = 4$, $M=0.5$, $\phi=0.69$, $\xi=0.5$, $\lambda_e=1.2$, $P=-1$ and $Pe = 1000$.}
\label{fig:17}
\end{figure}

Figure \ref{fig:17} represents the variation of the solute concentration for different values of $\beta$ for both the axial and the micro-rotational velocities. As the parameter $\beta$ increases, it significantly affects the concentration distribution for both the axial velocity and micro-rotational velocity profiles. In both cases, we can see that for small values of $\beta$, the concentration is significantly higher. That is, the concentration of solute is more prominent in the system. But if we increase the value of $\beta$, the solute is depleted from the system, which leads to a lower concentration of solute near the tube wall region.  

\section{Conclusion}\label{Sec:Conclusion}

In this study, a theoretical framework has been developed for generalized solute dispersion in magnetically influenced two-fluid flow through a porous layered tube with an absorptive wall. The coupled axial and microrotational velocity fields were obtained from the micropolar–Newtonian flow model, and the resulting transport coefficients and concentration distributions were determined using the generalised dispersion approach of Sankarasubramanian and Gill. The analytical predictions were independently validated through Brownian dynamics simulations, demonstrating excellent agreement for the temporal evolution of the zeroth and first axial transport moments. The influence of the governing flow parameters on the coupled axial and microrotational velocity fields, and consequently on the transport coefficients and solute concentration distributions, has been systematically characterised.

It is observed that as the Hartmann number increases, both the axial and micro-rotational velocities decrease, and the magnitude of the convection and dispersion coefficients also decreases, causing solutes to accumulate in certain regions and leading to an increase in the mean concentration. It is noted that two important parameters, the micropolar parameter and coupling number, significantly dictate the fluid flow and, consequently, the solute dispersion. As the micropolar parameter increases, the microrotational velocity rises, enhancing the diffusion of solute over time. Conversely, increasing the coupling number results in lower velocities, which slows the transport and diffusion of solute. An increase in the Darcy resistivity parameter reduces the axial velocity and convection, while the computed effective dispersion coefficient initially increases and subsequently approaches an asymptotic value. The resulting mean concentration distribution shows reduced downstream displacement as the porous resistance increases.

The observations of the mean concentration profiles indicate that the peak of the mean concentration profiles for both velocities rises for high values of Hartmann and the coupling number as they move along the axial direction of the tube. But in the case of the microrotational velocity, it decreases for high values of the micropolar parameter. Finally, we studied the spatial concentration profiles of the solute. Due to the presence of the wall absorption parameter, the solute is consumed more rapidly from the system, and it reaches a steady state as time progresses.

This study has important implications for biofluid mechanics, especially in modelling blood flow and solute transport in arteries and vessels. The micropolar fluid model used here captures essential rheological features of complex biological fluids, accounting for micro-rotation and non-Newtonian effects that are critical in physiological flow systems. The presence of a magnetic field can be modelled as a drug-targeting tool in a rheological situation. The presence of Brinkman and Darcy porous layers at the boundaries introduces additional flow resistance and modifies the shear and diffusion characteristics, reflecting the realistic interactions often found in engineered and biological layered tubes. The excellent agreement between the analytical solution and the Brownian dynamics simulations further demonstrates the reliability of the proposed theoretical framework. The findings offer valuable insights into blood flow modelling, where magnetic fields, vessel wall porosity, and cellular absorption critically influence haemodynamics and solute transport, supporting the development of improved biomedical flow models based on micropolar rheology.

It should be noted that the transport quantities obtained using the microrotational velocity profile are intended as a comparative theoretical measure of the influence of rotational microstructure within the generalised dispersion framework. A fully coupled transport model incorporating explicit translation--rotation kinematics would constitute an interesting extension of the present analysis.

\begin{acknowledgments}
Mr. Sohel Ahmed sincerely acknowledges the Council of Scientific and Industrial Research (CSIR), India, for financial support under grant number 09/1219(16556)/2023-EMR-I. Dr Nanda Poddar gratefully acknowledges Research Ireland for financial support through the Government of Ireland Postdoctoral Fellowship (Project No. GOIPD/2024/226) for this research work.
\end{acknowledgments}

\section*{ Funding.}{ \textbf{SA} gratefully acknowledges the Council of Scientific and Industrial Research (CSIR), India, for financial support through grant number 09/1219(16556)/2023-EMR-I for his PhD study. \textbf{NP} acknowledges the support of Research Ireland through the Government of Ireland Postdoctoral Fellowship. }

\section*{ Authors' contributions:} SA: Conceptualization, Formal analysis, Investigation, Methodology, Software, Visualization, Writing – original draft, Writing – review \& editing; NP: Conceptualization, Formal analysis, Investigation, Methodology, Resources, Software, Supervision, Validation, Visualization, Writing – original draft, Writing – review \& editing; JR: Conceptualization, Formal analysis, Methodology,  Supervision, Writing – review \& editing; KKM: Supervision, Writing – review \& editing; NM: Methodology, Software, Supervision, Validation.

\section*{Declaration of Interests :} The authors declare that they have no competing interests.

\section*{Ethics approval}
Ethics approval was not required because this study is entirely
theoretical and does not involve human participants, animals, or
patient data.

\section*{Data Availability Statement}
The analytical and Brownian dynamics simulation codes developed in MATLAB for this study are available from the corresponding author upon reasonable request. No experimental datasets were generated or analysed in this study.

\section*{APPENDIX A: DIMENSIONLESS VELOCITY DISTRIBUTION}\label{Sec:Appendix_A}
To get the dimensionless forms of the continuity, linear and angular momentum equations, we define the following scaled variables:
\begin{align}\label{eq:A1}\tag{A1}
& v_i=\frac{ {v}_i'}{ {v_0}},~~z=\frac{ {z'}}{ R},~~r=\frac{ {r'}}{ R},~~ {r}_i=\frac{ {r}_i'}{ R},~~p=\frac{ {p'}\, R}{ {\mu'}\,{v_0}},~~w=\frac{ {w'}~ R}{ {v_0}},~~n_j^2=\frac{ R^2}{ {k}_j'},~~\lambda=\frac{ {\mu'}}{ {\mu}_1'}, \nonumber\\
&\sigma=\frac{ {\sigma'}}{ {\sigma'}_1},~~M= {B}_0' R\sqrt{\frac{ {\sigma'}}{ {\mu'}}},~~M_1=M\sqrt{\frac{\lambda}{\sigma}},~~i \in \{1,2,3,4\} \mbox{ and } j \in \{1,2\},\nonumber
\end{align}

where $v_0$ is the characteristic velocity, $R$ is the radius of the tube, $\lambda$ is the viscosity ratio, $\sigma$ is the conductivity ratio, $M$ is the Hartmann number, and $M_1$ is the modified Hartmann number.

Using the above-mentioned scaled variables, the equations in different fluid regions are as follows:
 \begin{equation}\label{eq:A2}\tag{A2}
\left(1-N_c\right)\lambda P-L^2 v_1-N_c L^2\psi+\left(1-N_c\right)M_1^ 2v_1=0 ~~\mbox{in}~~\boldsymbol{\Omega_1},
\end{equation}
\begin{equation}\label{eq:A3}\tag{A3}
\frac{d}{d r}\left[L^2\psi\right]-\frac{m^2}{(2-N_c)}\frac{d v_1}{d r}-\frac{2m^2}{(2-N_c)}~\frac{d \psi}{d r}=0 ~~\mbox{in}~~\boldsymbol{\Omega_1},
\end{equation}

\begin{equation}\label{eq:A4}\tag{A4}
L^2 v_2-M^2 v_2=P ~~\mbox{in}~~\boldsymbol{\Omega_2},
\end{equation}

\begin{equation}\label{eq:A5}\tag{A5}
L^2 v_3 -\frac{(n_1^2+M^2)}{\lambda_e ^2}~v_3=\frac{P}{\lambda_e ^2} ~~\mbox{in}~~\boldsymbol{\Omega_3},
\end{equation}
\begin{equation}\label{eq:A6}\tag{A6}
(n_2^2+M^2)v_4=-P ~~\mbox{in}~~\boldsymbol{\Omega_4},
\end{equation}\\
where $N_c=\frac{ {\kappa'}}{ {\mu}_1'+ {\kappa'}}$,~$m^2=\frac{ R^2~ {\kappa'}}{ {\gamma'}}\frac{(2 {\mu}_1'+ {\kappa'})}{( {\mu}_1'+ {\kappa'})}$,~$L^2\equiv\frac{1}{r}~\frac{d}{d r}\left(r~\frac{d}{d r}\right)$,~and $w=\frac{d \psi}{d r}$. Since pressure $p$ is a function of $z$ only, we take the pressure gradient\, $\frac{\partial p}{\partial z}=\frac{d p}{d z}=P$ (constant). The parameter $N_c$ is called the coupling number; $m$ is called the micropolar parameter, which represents the spinning effect of microparticles in a micropolar fluid; $\lambda_e=\sqrt{\frac{ {\mu'}_e}{ {\mu'}}}$ is the effective viscosity ratio parameter and $L^2$ is taken as a differential operator.

Now, solving equations \eqref{eq:A2}--\eqref{eq:A6}, we have obtained the following derivations for linear and angular velocities in select regions.
The linear velocity of micropolar fluid in $\boldsymbol{\Omega_1}$ is derived as
\begin{equation}\label{eq:A7}\tag{A7}
v_1(r)=c_1I_0(\alpha r)+c_2K_0(\alpha r)+c_3I_0(\beta r)+c_4K_0(\beta r) - \frac{P \lambda}{M_1^2}.
\end{equation}
The micro-rotational velocity of micropolar fluid in $\boldsymbol{\Omega_1}$ is obtained by
\begin{multline}\label{eq:A8}\tag{A8}
w(r)=\frac{A}{r}+\frac{1}{N_c \alpha \beta}\left[\beta(\alpha^2+(N_c-1)M_1^2)(-c_1I_1(\alpha r)+c_2K_1(\alpha r))\right]\\
+\frac{1}{N_c \alpha \beta}\left[\alpha(\beta^2+(N_c-1)M_1^2)(-c_3I_1(\beta r)+c_4K_1(\beta r))\right].
\end{multline}
The Newtonian fluid velocities in $\boldsymbol{\Omega_2}$, $\boldsymbol{\Omega_3}$ and $\boldsymbol{\Omega_4}$ are derived to
\begin{equation}\label{eq:A9}\tag{A9}
v_2(r)=c_5I_0(M r)+c_6K_0(Mr)-\frac{P}{M^2},
\end{equation}
\begin{equation}\label{eq:A10}\tag{A10}
v_3(r)=c_7I_0(Sr)+c_8K_0(Sr)-\frac{P}{(n_1^2+M^2)}
\end{equation}
\begin{equation}\label{eq:A11}\tag{A11}
\mbox{and }\, v_4(r)=-\frac{P}{(n_2^2+M^2)},
\end{equation}
respectively, where $S^2=\frac{(n_1^2+M^2)}{{\lambda_e}^2}$, $A$, $c_1$ $c_2$, $c_3$, $c_4$, $c_5$, $c_6$, $c_7$, and $c_8$ are arbitrary constants.

\bibliography{reference}
\end{document}